\documentclass[11pt]{article}
\usepackage{amsfonts,amsmath,amssymb}
\usepackage{enumerate}
\usepackage{hyperref}
\usepackage{bbm}
\usepackage{nicefrac}
\usepackage[all]{xy}
\usepackage{graphicx}
\usepackage{bm}

\usepackage{upgreek}

\usepackage{booktabs}

\newcommand{\be}{\begin{equation}}
\newcommand{\ee}{\end{equation}}

\newcommand{\bea}{\begin{eqnarray}}
\newcommand{\eea}{\end{eqnarray}}

\newcommand{\bes}{\begin{subequations}}
\newcommand{\ees}{\end{subequations}}

\newcommand{\cM}{{\cal M}}
\newcommand{\cN}{{\cal N}}
\newcommand{\cA}{{\cal A}}
\newcommand{\cB}{{\cal B}}
\newcommand{\cC}{{\cal C}}
\newcommand{\cH}{{\cal H}}

\newcommand{\cV}{{\cal V}}

\def\sst#1{{\scriptscriptstyle #1}}

\def\0{{\sst{(0)}}}
\def\1{{\sst{(1)}}}
\def\2{{\sst{(2)}}}
\def\3{{\sst{(3)}}}
\def\4{{\sst{(4)}}}
\def\5{{\sst{(5)}}}
\def\6{{\sst{(6)}}}
\def\7{{\sst{(7)}}}
\def\8{{\sst{(8)}}}

\def\cA{{{\cal A}}}

\def\tcA{{{\tilde{\cal A}}}}
\def\tcC{{{\tilde{\cal C}}}}

\usepackage{multirow}
\usepackage{rotating}

\allowdisplaybreaks

\begin{document}

\makeatletter
\renewcommand{\theequation}{\thesection.\arabic{equation}}
\@addtoreset{equation}{section}
\makeatother

\begin{titlepage}

\begin{flushright}
NIKHEF-2015-027 \\
CPHT-RR025.0815 
\end{flushright}

\vspace{25pt}

   \begin{center}
   \baselineskip=16pt
   \begin{Large}\textbf{
Dyonic ISO(7) supergravity and the duality hierarchy}
   \end{Large}

\vspace{25pt}
		
{Adolfo Guarino$^{1}$ \, and \, Oscar Varela$^{2,3}$}
		
\vspace{25pt}

	\begin{small}

	{\it ${}^1$Nikhef Theory Group, Science Park 105, 1098 XG Amsterdam, The Netherlands } \\
	%aguarino@nikhef.nl

	\vspace{15pt}

         {\it ${}^2$	Center for the Fundamental Laws of Nature,\\
	Harvard University, Cambridge, MA 02138, USA } \\
	%ovarela@physics.harvard.edu

	\vspace{15pt}

         {\it ${}^3$Centre de Physique Th\'eorique, Ecole Polytechnique,\\CNRS UMR 7644, 91128 Palaiseau Cedex, France} \\

	\end{small}

\vskip 50pt

\end{center}

\begin{center}
\textbf{Abstract}
\end{center}

\begin{quote}

Motivated by its well defined higher dimensional origin, a detailed study of $D=4$ $\cN=8$ supergravity with a dyonically gauged $\textrm{ISO}(7) = \textrm{SO}(7) \ltimes \mathbb{R}^7$ gauge group is performed. We write down the Lagrangian and describe the tensor and duality hierarchies, focusing on an interesting subsector with closed field equations and supersymmetry transformations. We then truncate the $\cN=8$ theory to some  smaller sectors with $\cN=2$ and $\cN=1$ supersymmetry and SU(3), G$_2$ and SO(4) bosonic symmetry. Canonical and superpotential formulations for these sectors are given, and their vacuum structure and spectra is analysed. Unlike the purely electric ISO(7) gauging, the dyonic gauging displays a rich structure of vacua, all of them AdS. We recover all previously known ones and find a new $\cN=1$ vacuum with SU(3) symmetry and various non-supersymmetric vacua, all of them stable within the full $\cN=8$ theory.

\end{quote}

\vfill

\end{titlepage}

\tableofcontents

%%%%%%%%%%%%%%%%%%%%%%%%%%%%%%%%%%%%%%%%%%%%%%%%%%%%%%%%%%%%%%%%%%%%%%%%%%

%\newpage

\section{Introduction}
\label{sec:introduction}

Maximal gauged supergravity in four dimensions often admits continuous or discrete symplectic deformations that respect $\cN=8$ supersymmetry and the gauge group \cite{Dall'Agata:2012bb,Dall'Agata:2014ita}. The simplest type of deformation introduces a dependence on a dimensionless parameter $\,c\,$ in the gauging-dependent couplings of the theory. The covariant derivatives, for example, acquire a new coupling to the magnetic vectors proportional to $\,c\,$,
\begin{equation}
\label{DSO8_intro}
D = d - g \, (\cA^{\Lambda} - c \, \tilde{\cA}_{ \Lambda}) \ ,
\end{equation}
thus leading to a dyonic gauging. The role of this parameter, in a passive picture, is to tune the electric/magnetic symplectic frame prior to introducing the gauging. In the ungauged limit, $\,c\,$ can be set to zero without loss of generality by a symplectic transformation. At finite gauge coupling $g$, however, electric/magnetic duality is broken and the theory typically becomes sensitive to the symplectic frame specified by $c$. Various aspects of this deformation for different gauge groups have now been studied, including its effect on the vacuum structure \cite{Dall'Agata:2012bb,DallAgata:2011aa,Borghese:2012qm,Borghese:2012zs,Borghese:2013dja}, on domain-wall \cite{Guarino:2013gsa,Tarrio:2013qga,Pang:2015mra} and black hole solutions \cite{Anabalon:2013eaa,Cremonini:2014gia,Lu:2014fpa}, or on inflationary models \cite{Dall'Agata:2012sx,Kodama:2015iua}.

An immediate question is whether these $\cN=8$ dyonic gaugings descend from higher dimensions. This was recently answered positively when the gauge group is chosen to be $\textrm{ISO}(7)_{c} \equiv \textrm{CSO}(7,0,1)_{c}$ $\equiv  \textrm{SO}(7) \ltimes \mathbb{R}_{c}^7$ \cite{Guarino:2015jca}. Here and often in the following, we have followed the notation of  \cite{Dall'Agata:2012bb} and have sticked in a subscript $c$ to denote that ISO(7) (more precisely, only its seven translations) is gauged dyonically. In \cite{Guarino:2015jca, Guarino:2015vca} we showed that $D=4$ $\cN=8$ ISO(7)-dyonically-gauged supergravity arises as a consistent truncation of massive type IIA supergravity \cite{Romans:1985tz} on the six-sphere, with the magnetic coupling constant $m \equiv gc$ identified upon reduction with the Romans mass, $\hat F_\0 = m$. All solutions of the $D=4$ theory uplift to solutions of massive type IIA by the consistency of the truncation. In particular, its vacua (all known ones are AdS) give rise to AdS$_4$ backgrounds of massive type IIA string theory. Quantitative evidence was also given in \cite{Guarino:2015jca} that these AdS$_4$ vacua are dual to the simplest type of Chern-Simons theories with a single gauge group and adjoint matter \cite{Schwarz:2004yj}. The answer to the question of the higher-dimensional origin of these dyonic gaugings is of course gauge group dependent. Arguments have been recently given \cite{Lee:2015xga} against an M-theory origin of the dyonic deformation \cite{Dall'Agata:2012bb} of the SO(8) gauging \cite{deWit:1982ig}.

The distinct higher-dimensional origin of the dyonic ISO(7) gauging singles it out and makes it worth of further detailed investigation. This is what we set up to do in this paper from a purely four-dimensional perspective, leaving further research on the precise connection with ten dimensions for separate publications. Various aspects of the ISO(7) gauging have already been studied. The purely electric, $\,c=0\,$ (\textit{i.e.}~$m=0$), ISO(7)-gauged theory was constructed long ago \cite{Hull:1984yy} from the SO(8)-gauged theory \cite{deWit:1982ig} by a  limiting procedure that implements the In\"on\"u-Wigner contraction from SO(8) to ISO(7) directly in the supergravity. The symplectic deformations corresponding to various gauge groups, including ISO(7), were studied in \cite{Dall'Agata:2014ita}. The ISO(7)$_c$ family of gaugings was found to be discrete, containing only two members: the purely electric $c=0$ theory  \cite{Hull:1984yy}, and the dyonic $c \neq 0$ theory. All non-vanishing values of $c$ lead to equivalent theories \cite{Dall'Agata:2014ita}.

The basic formalism to deal with generic gaugings of $D=4$ $\cN=8$ supergravity has been laid out in \cite{deWit:2007mt}, see also \cite{deWit:2005ub}. The gauging is encoded in an embedding tensor that governs both the non-Abelian coupling of the vectors to themselves and to the rest of the supergravity fields, and the embedding of the gauge group into the global U-duality group, E$_{7(7)}$. Gaugings that involve minimal couplings to the magnetic vectors in a given symplectic frame necessarily require the presence of two-form potentials. These appear both sourcing the field strengths of the vectors and in new topological terms in the Lagrangian, without upsetting the count of degrees of freedom. More generally, a larger set of $p$-form potentials, $p=1, \ldots, 4$, of a so-called tensor hierarchy \cite{deWit:2008ta,deWit:2008gc} can be considered. These include all vectors in the theory, a larger set of two-forms than a given gauging would typically require, and three- and four-forms, all of them in irreducible E$_{7(7)}$ representations. Except for the four-form potentials, the fields up the tensor hierarchy are definitely dynamical as they typically cannot be gauged away: they do carry degrees of freedom, albeit not independent ones. Indeed, bringing the metric and scalars into the picture, the higher rank forms can be Hodge-dualised into (of course, dynamical) combinations of scalars and their derivatives. The $\cN=8$ tensor hierarchy equipped with these dualisations has been referred to as the `duality hierarchy' \cite{Bergshoeff:2009ph}. See \cite{Hartong:2009az,Huebscher:2010ib} for the hierarchies in less supersymmetric contexts. In this paper we will specify the Lagrangian for the ISO(7)$_c$ gaugings following the embedding tensor formalism  \cite{deWit:2007mt}. We will also be interested in the duality hierarchy \cite{Bergshoeff:2009ph}, paying particular attention to a subsector with closed field equations and supersymmetry transformations. This subsector arises upon suitable restriction of the full E$_{7(7)}$-covariant duality hierarchy. Although it is only SL(7)-covariant, rather than E$_{7(7)}$, this subsector is still $\cN=8$.

We will also study the vacuum structure of the ISO(7)$_c$ gaugings. More concretely, we provide a systematic classification of the  critical points of the scalar potential that preserve at least SU(3) and at least a particular SO(4) within $\textrm{SO}(7) \subset \textrm{ISO}(7)$. We do this by working out the truncations of the $\cN=8$ theory to the SU(3)- and that particular SO(4)-invariant sectors, and then extremising the resulting potentials. Although the G$_2$-invariant sector is contained within the SU(3) sector, we find it useful to provide a separate treatment for it too. All these sectors are supersymmetric and, as a crosscheck on our calculations, we cast them in the corresponding $\cN=2$ or $\cN=1$ canonical form. We provide explicit parameterisations for the scalars in these subsectors. This allows us to give the location of the critical points in the full scalar space E$_{7(7)}/$SU(8) relative to those parameterisations. This result is new even for the critical points that were already known (see below), which had been found using a method \cite{Dibitetto:2011gm,DallAgata:2011aa} whose power resides, precisely, in its being insensitive to their actual location.

Quite surprisingly, the vacuum structure of the electric and dyonic ISO(7) gaugings turns out to be very different. In fact, while the former has no known vacua, the latter displays a rich (AdS) vacuum structure. Some of these critical points were already known, including points with $\cN=1$, G$_2$  \cite{Borghese:2012qm}, and $\cN=3$, SO(4)  \cite{Gallerati:2014xra} symmetry, and non-supersymmetric points with SO(7), SO(6) \cite{DallAgata:2011aa} and G$_2$ \cite{Borghese:2012qm} symmetry. Among the non-supersymmetric points, only the latter is stable, at least within the full $\cN=8$ theory. Our classification recovers all these extrema and finds new ones with $\cN=2$, $\textrm{SU}(3) \times \textrm{U}(1)$ symmetry (which we already reported on in \cite{Guarino:2015jca}), a point with SU(3), $\cN=1$ symmetry, and stable non-supersymmetric points with SU(3) and SO(4) symmetry. Some, but not all, of these points have counterparts in either the electric \cite{deWit:1982ig} or dyonic \cite{Dall'Agata:2012bb} SO(8) gauging, with the same residual supersymmetry, bosonic symmetry and mass spectrum. See table \ref{Table:SU3&SO4Points} for a summary of the known critical points of the dyonic ISO(7) supergravity.

\begin{table}[t!]
\centering
%\small
%\tabcolsep=0.15cm
\footnotesize{
\scalebox{0.95}{
\begin{tabular}{cccccc}
\hline
\\[-2.5mm]
SUSY  	& bos. sym. & $M^2L^2$ & stability & ref.
\\[2pt]
\hline
\\[-10pt]
$ \cN=3 $ &	 SO(4)  &	$3(1 \pm \sqrt{3})^{(1)}   \ , \;   (1 \pm \sqrt{3})^{(6)}   \ , \;   -\tfrac{9}{4}^{(4)} \ , \;   -2^{(18)}  \ , \; -\tfrac{5}{4}^{(12)}  \ , \;  0^{(22)} $ & yes & \cite{Gallerati:2014xra} \\[2pt]
&  &  $( 3 \pm \sqrt{3})^{(3)} \ , \; \tfrac{15}{4}^{(4)} \ , \; \tfrac{3}{4}^{(12)}   \ , \;  0^{(6)}$
\\[10pt]
$ \cN=2 $ &U(3)  &	$(3 \pm \sqrt{17})^{(1)}  \ , \;   -\tfrac{20}{9}^{(12)} \ , \;   -2^{(16)}  \ , \; -\tfrac{14}{9}^{(18)} \ , \; 2^{(3)}  \ , \;  0^{(19)} $ & yes & \cite{Guarino:2015jca} , [\textrm{here}] \\[2pt]
&  &  $4^{(1)}  \ , \; \tfrac{28}{9}^{(6)} \ , \; \tfrac{4}{9}^{(12)} \ , \; 0^{(9)}$
\\[10pt]
$ \cN=1 $ &	G$_2$ &	$(4 \pm \sqrt{6})^{(1)} \ , \; -\tfrac{1}{6} (11 \pm \sqrt{6} )^{(27)} \ , \;  0^{(14)} $  & yes & \cite{Borghese:2012qm} \\[2pt]
&  &  $\tfrac{1}{2} ( 3 \pm \sqrt{6})^{(7)} \ , \;  0^{(14)}$
\\[10pt]
$ \cN=1 $ &	SU(3)  &	$(4 \pm \sqrt{6})^{(2)}  \ , \;   -\tfrac{20}{9}^{(12)} \ , \;   -2^{(8)}  \ , \; -\tfrac{8}{9}^{(12)} \ , \; \tfrac{7}{9}^{(6)}  \ , \;  0^{(28)} $ & yes  & [\textrm{here}] \\[2pt]
&  &  $6^{(1)}  \ , \; \tfrac{28}{9}^{(6)} \ , \; \tfrac{25}{9}^{(6)} \ , \; 2^{(1)}   \ , \; \tfrac{4}{9}^{(6)} \ , \; 0^{(8)}$
\\[10pt]
$ \cN=0 $ &	SO(7)$_+$  &	$6^{(1)} \ , \; -\tfrac{12}{5}^{(27)} \ , \;  -\tfrac{6}{5}^{(35)} \ , \;  0^{(7)} $ & no & \cite{DallAgata:2011aa} \\[2pt]
&  &  $\tfrac{12}{5} ^{(7)} \ , \;  0^{(21)}$
\\[10pt]
$ \cN=0 $ &	SO(6)$_+$  &	$6^{(2)} \ , \; -3^{(20)}  \ , \;   -\tfrac{3}{4}^{(20)}  \ , \;  0^{(28)} $  & no &  \cite{DallAgata:2011aa} \\[2pt]
&  &  $6^{(1)} \ , \; \tfrac{9}{4} ^{(12)} \ , \;  0^{(15)}$
\\[10pt]
$ \cN=0 $ &	G$_2$  &	$6^{(2)} \ , \;  -1^{(54)} \ , \;   0^{(14)} $ & yes & \cite{Borghese:2012qm} \\[2pt]
&  &  $3^{(14)}   \ , \;  0^{(14)}$
\\[10pt]
$ \cN=0 $ &	SU(3)  &	\textrm{see} (\ref{new_SU3_1_spectrum}) & yes & [\textrm{here}] \\[2pt]
&  & \textrm{see} (\ref{new_SU3_1_spectrum_vec})
\\[10pt]
$ \cN=0 $ & 	SU(3)  &	\textrm{see} (\ref{new_SU3_2_spectrum}) & yes & [\textrm{here}] \\[2pt]
&  & \textrm{see} (\ref{new_SU3_2_spectrum_vec})
\\[10pt]
$ \cN=0 $ &	SO(4)  &	\textrm{see} (\ref{new_SO4_spectrum}) & yes & [\textrm{here}] \\[2pt]
&  & \textrm{see} (\ref{new_SO4_spectrum_vec})
\\[5pt]
\hline
\end{tabular}}
\caption{All critical points of $D=4$ $\cN=8$ dyonically-gauged-ISO(7) supergravity, that preserve at least SU(3) and at least a certain SO(4) (see section \ref{sec:SO(4)-sector}) within $\textrm{SO}(7) \subset \textrm{ISO}(7)$. All points are AdS. For each point it is indicated the residual supersymmetry and bosonic symmetry, the scalar (upper row) and vector (lower row) mass spectra with the corresponding multiplicities, its stability and the reference where it was first found. See tables \ref{Table:SU3Points} and \ref{Table:SO4Points} for their location in scalar space and for their cosmological constants.}}
\label{Table:SU3&SO4Points}
\end{table}

In section~\ref{sec:ISO(7)}, we construct the ISO(7)$_c$ theory using the embedding tensor formalism, and specify the bosonic Lagrangian, an $\cN=8$ subsector of the duality hierarchy and the supersymmetry transformations. In the rest of the paper we flesh out some interesting subsectors with less supersymmetry and bosonic symmetry: see sections \ref{sec:SU(3)-sector}, \ref{sec:G2-sector} and \ref{sec:SO(4)-sector} for discussions of the SU(3), G$_2$ and an SO(4)-invariant sectors, respectively. Canonical supersymmetric formulations are given and the critical points of the scalar potential in these sectors are computed. Four appendices close the paper. The first two offer further discussion. Appendix \ref{app:Z2xSO(3)-sector} contains the truncation of the $\cN=8$ theory to yet another subsector, with $\,\cN=1\,$ supersymmetry and  $\,\mathbb{Z}_{2} \times \textrm{SO}(3)$ bosonic symmetry, relevant to non-geometric type IIA orientifold reductions. Appendix \ref{app:MTheoryonSE7} comments on the relation of the SU(3)-invariant sector of the ISO(7)$_c$ theory to the $\,\cN=2\,$ supergravity that arises from consistent truncation of M-theory on an arbitrary Sasaki-Einstein manifold. The last two are technical:  appendix \ref{app:ConstN=8} gives some details of the construction of the $\cN=8$ ISO(7)$_c$ theory, while appendix \ref{app:ScalarMatrix} gives explicit parameterisations for the supergravity scalar kinetic matrix in the invariant sectors discussed in the main text.

\section{Maximal supergravity with dyonic ISO(7) gauging}
\label{sec:ISO(7)}

We will now present the $D=4$ $\cN=8$ supergravity theory with a dyonically-gauged ISO(7) gauge group, focusing on its bosonic sector. We review the embedding tensor and the field content, including the tensor hierarchy, in section \ref{subsec:FCET}. An interesting subsector of the latter is discussed in \ref{sec:TruncTensorHierarchy}. The bosonic Lagrangian and supersymmetry transformations can be found in \ref{subsec:LST} and \ref{sec:susyvars}. See also appendix \ref{app:ConstN=8} for some details of the construction of the theory from the general formalism of \cite{deWit:2005ub,deWit:2007mt}.

\subsection{Tensor hierarchy and ISO(7) embedding tensor} 
\label{subsec:FCET}

The bosonic field content of maximal supergravity in four dimensions includes the vielbein $e_\mu{}^\alpha$, scalars that parameterise a coset representative ${\cal V}_\mathbb{M}{}^{ij}$, $\mathbb{M} = 1 , \ldots , 56$, $i = 1 , \ldots , 8$, of $\textrm{E}_{7(7)}/\textrm{SU}(8)$ and vectors $\cA^\mathbb{M}$, $\mathbb{M} = 1 , \ldots , 56$, in the fundamental representation of the U-duality group $\textrm{E}_{7(7)}$, with two-form field strengths $\cH_\2^\mathbb{M}$. In the presence of magnetic charges, as it will be the case in this work, a set of two-form potentials\footnote{The flat, SO$(1,3)$ index $\alpha$ on $e_\mu{}^\alpha$ should not cause any confusion with the E$_{7(7)}$ adjoint index on $\cB_\alpha$. Note also that $D=4$ vectors and two-form potentials were denoted with straight, rather than calligraphic, characters in \cite{Guarino:2015jca}.} $\cB_\alpha$,  $\alpha =1 , \ldots, 133$, in the adjoint of $\textrm{E}_{7(7)}$ and with three-form field strengths $\cH_{\3  \alpha}$, is generically required by gauge invariance \cite{deWit:2005ub,deWit:2007mt}. A gauging-dependent projection of the two-form potentials $\cB_\alpha$ typically enters the $D=4$ Lagrangian and the field strengths $\cH_\2^\mathbb{M}$ of the vectors. 

More generally, these $\bm{56}$ vectors $\cA^\mathbb{M}$ and $\bm{133}$ two-forms $\cB_\alpha$ are the first two sets of fields in an E$_{7(7)}$-covariant tensor hierarchy  \cite{deWit:2008ta,deWit:2008gc} that further includes $\bm{912}$ three-form potentials $\cC_{\alpha}{}^{\mathbb{M}}$ with four-form field strengths ${\cH_{\4 \alpha}}^\mathbb{M}$, and $\bm{133} + \bm{8645}$ four-form potentials. Like for the lower rank forms, certain gauging-dependent projections of the three-form potentials $\cC_{\alpha}{}^{\mathbb{M}}$ enter the three-form field strengths $\cH_{\3  \alpha}$, and so on. Obviously, not all the fields in the tensor hierarchy carry independent degrees of freedom: the higher rank forms can be dualised into scalars and their derivatives. This was discussed at length in \cite{Bergshoeff:2009ph}, where the tensor hierarchy equipped with these dualisations was dubbed the `duality hierarchy'. It is possible to write a generic $\cN=8$ gauged supergravity Lagrangian that includes higher-rank fields in the E$_{7(7)}$ tensor hierarchy \cite{Bergshoeff:2009ph}. This Lagrangian reduces, after imposing the duality relations, to the conventional Lagrangian \cite{deWit:2007mt} containing only the metric, scalars, vectors and the two-forms switched on by magnetic gaugings. In section \ref{subsec:LST} we will write the Lagrangian for the dyonic ISO(7) gauging in the formulation of \cite{deWit:2007mt}, although we will still find it useful to consider, in section \ref{sec:TruncTensorHierarchy}, a (restricted) duality hierarchy containing forms of higher rank.

To conclude this summary of the $\cN=8$ field content, recall that the fermionic sector contains the gravitino $\psi_\mu^i$ and spin $1/2$ fields $\chi^{ijk}$, in the $\overline{\textbf{8}}$ and $\overline{\textbf{56}}$ of the R-symmetry group SU(8), respectively. Both fermions are chiral, {\it e.g.}, $\gamma_5 \,  \psi_\mu^i = \psi_\mu^i$, with $\gamma_5 = i \gamma_0\gamma_1\gamma_2\gamma_3$ the chirality operator and $\gamma_\alpha$ the Cliff$(1,3)$ matrices. Recall that, in four dimensions, charge conjugation reverses the fermion chirality. Following convention, we denote negative chirality spinors with lower SU(8) indices, $\gamma_5 \,  \psi_{\mu i}= -\psi_{ \mu i}$.

In order to formulate the ISO(7) gauging, it is natural to branch out the above $\textrm{E}_{7(7)}$-covariant bosonic field content into representations of SL(7), given that  ISO(7) is contained in E$_{7(7)}$ through the chain\footnote{We will not keep track of charges under the SO$(1,1)$ that extends SL(7) into GL(7) in the chain (\ref{eq:chainE7}). This SO$(1,1)$ does not play a role in the gauged ISO$(7)_c$ theory. See nevertheless (\ref{eq:Branchings}), (\ref{eq:Branchings_2}).}
\begin{eqnarray} \label{eq:chainE7}
\textrm{ISO(7)} \equiv \textrm{SO}(7) \ltimes \mathbb{R}^7
 \, \subset \,  
 \textrm{SL}(7) \ltimes \mathbb{R}^7  
 \, \subset \,  
 \textrm{GL}(7) \ltimes \mathbb{R}^7  
 \, \subset \,  
 \textrm{SL}(8) \subset \textrm{E}_{7(7)} \; .
\end{eqnarray}
For this purpose, we find it useful to introduce fundamental SL(8) indices $A,B = 1 , \ldots , 8$, and a collective index  $\Lambda \equiv [AB]=1, \ldots ,28$. For SL(7), we only need to introduce fundamental indices, $I = 1 , \ldots , 7$. The $\bm{56}$ vectors, for example, branch as
\begin{eqnarray} 
\label{vectors56}
\cA^\mathbb{M} = ( \cA^\Lambda \,,\, \tilde{\cA}_\Lambda )  = ( \cA^{AB} \,,\, \tilde{\cA}_{AB} )  = ( \cA^{IJ} ,  \cA^{I}  \,,\,    \tilde{\cA}_{IJ} ,   \tilde{\cA}_{I} ) \ .
\end{eqnarray}
We have dropped the `8' label in $\cA^{I8} $ and  $\tilde{\cA}_{I8}$, and have put tildes on the magnetic vectors. Although the tildes are redundant with the lower position of the indices, we find this emphatic notation visually useful. Similarly, the $\bm{133}$ two-form, $\cB_\alpha$, and $\bm{912}$ three-form, $\cC_\alpha{}^\mathbb{M}$, potentials branch as well into SL(7) representations:  see equation (\ref{eq:Branchings}) for the relevant decompositions.

In $\cN=8$ supergravity, gaugings are completely specified by the embedding tensor $\,{\Theta_{\mathbb{M}}}^{\alpha}\,$ \cite{deWit:2007mt}. This determines the embedding of the gauge group into the $\textrm{E}_{7(7)}$ duality group. Linear constraints enforce $\,{\Theta_{\mathbb{M}}}^{\alpha}\,$ to lie generically in the \textbf{912} of $\textrm{E}_{7(7)}$, and quadratic constraints (see equation (\ref{QC})) ensure the consistency of the gauging \cite{deWit:2005ub,deWit:2007mt}. Fixing the gauge group to be ISO(7), the linear constraint reduces the embedding tensor to lie in the $\bm{28} + \bm{1}$ of SL(7) and the quadratic constraint allows for the following non-vanishing components of  $\,\Theta_\mathbb{M}{}^\alpha\, = ( \,\Theta_\Lambda{}^\alpha \, , \, \Theta^{\Lambda \, \alpha} \, )$ only:
\begin{equation}
\label{Theta-compSL7}
\Theta_{[IJ]\phantom{K}L}^{\phantom{[IJ]}K} = 2\, \delta_{[I}^{K} \, \delta_{J]L} 
\hspace{5mm} , \hspace{5mm}
\Theta_{[I8] \phantom{8}  K}^{\phantom{[I8]} 8} = -\delta_{I K}
%\hspace{5mm} , \hspace{5mm}
%{\Theta^{[IJ] K}}_{L} = 0
\hspace{5mm} \textrm{ and } \hspace{5mm}
{\Theta^{[I8] \, 8}}_{K} = c \, \delta^{I}_{K} \ ,
\end{equation}
see \cite{DallAgata:2011aa}. Here, $c$ is an arbitrary real constant. It was shown in \cite{Dall'Agata:2014ita} that all non-vanishing values of $\,c\,$ lead to equivalent theories up to a rescaling of the gauge coupling $g$. Therefore, for $g \neq 0$, there exist two possible ISO(7) gaugings of $D=4$ $\cN=8$ supergravity \cite{Dall'Agata:2014ita}: $c=0$ and $c \neq 0$. The first two components in (\ref{Theta-compSL7}), associated to the $\bm{28}$, couple to the electric vectors, while the last component, related to the singlet, couples to the magnetic vectors. Strictly speaking, only SO(7) singlets enter (\ref{Theta-compSL7}). In particular, in the first two components, only the singlet in the decomposition of the $\bm{28}$ of SL(7) under SO(7) is involved, and is realised as a Kronecker delta with two lower indices. We nevertheless find it useful to refer to the electric, $\Theta_\Lambda{}^\alpha$, and magnetic, $ \Theta^{\Lambda \, \alpha}$, components of (\ref{Theta-compSL7}) as the $\bm{28}$ and singlet of SL(7), respectively.

The physical difference between the $c=0$ and $c \neq 0$ ISO(7) gaugings is most easily seen by looking at the covariant derivatives. Denoting by $g$ the (electric) gauge coupling and introducing a magnetic gauge coupling $m$ through
\begin{equation} \label{mcg}
m \equiv g \, c  \; , 
\end{equation}
the covariant derivatives induced by the ISO(7) embedding tensor (\ref{Theta-compSL7}) are
\begin{equation}
\label{DISO7}
D = d  \, -    g \,   \cA^{IJ}  \, {t_{[I}}^{K} \, \delta_{J]K}  +  \big(  g \,  \delta_{IJ} \, \cA^{I}  -  m  \, \tcA_{J}  \big) \, {t_{8}}^{J} \ .
\end{equation}
The $\cA^{IJ}$ terms can be equivalently written using the $\bm{48} $  SL(7) generators ${t_{I}}^{J} - \tfrac17 \, t_K{}^K \, \delta_{I}^J  $. These, together with the $\bm{7}^\prime$ generators  $t_8{}^J$, generate the $\textrm{SL}(7) \ltimes \mathbb{R}^7$ subgroup of E$_{7(7)}$ in (\ref{eq:chainE7}). See (\ref{Gener63}),  (\ref{Gener70}) for the expressions of the  E$_{7(7)}$ generators $(t_\alpha)_\mathbb{M}{}^\mathbb{N}$ in the fundamental representation, in the SL(8) basis. In agreement with the table on page 37  of \cite{deWit:2007mt}, the embedding tensor components in the $\bm{28}$ couple the $\bm{21}^\prime$ electric vectors $\cA^{IJ}$ to the $\bm{48}$ generators ${t_{I}}^{J} - \tfrac17 \, t_K{}^K \, \delta_{I}^J  $, and the $\bm{7}^\prime$ electric vectors $\cA^I$ to the $\bm{7}^\prime$ generators $t_8{}^J$, while the singlet component of the embedding tensor couples the $\bm{7}$ magnetic vectors $\tilde{\cA}_I$ to the $\bm{7}^\prime$ generators $t_8{}^J$ whenever $c \neq 0$. The choice $c =0$ in (\ref{Theta-compSL7}) thus leads to the purely electric ISO(7) gauging constructed in \cite{Hull:1984yy} by other methods. For $c\neq 0$, the gauging is dyonic in the symplectic frame where (\ref{Theta-compSL7}) is expressed: the $\mathbb{R}^7$ translations of ISO(7) are gauged dyonically. The rotations SO(7) are only gauged electrically, though: the constraints on the ISO(7) embedding tensor set to zero the $\bm{28}^\prime$  components that would induce a magnetic gauging of SO(7), as well as the $\bm{7}^\prime$, see the table in \cite{deWit:2007mt}. Thus, the $\bm{21}$ magnetic vectors $\tilde{\cA}_{IJ}$ do not participate in the gauging. Observe, finally, that the combinations $\,T_{IJ}\equiv 2 \, {t_{[I}}^{K} \, \delta_{J]K}\,$ and $\,T_{I}\equiv  {t_{8}}^{J}\, \delta_{JI}\,$ in (\ref{DISO7}) correspond to the SO(7) and $\,\mathbb{R}^{7}\,$ generators of the gauge group $\,\textrm{ISO}(7)=\textrm{SO}(7) \ltimes \mathbb{R}^{7}\,$, see (\ref{ISO7_brackets}).

Indices of SL(7) cannot be raised or lowered. For the ISO(7) gauging, these can be identified with SO(7) indices upon contraction with the embedding tensor. Even in this case, we will refrain from raising and lowering them with the SO(7) metric $\delta_{IJ}$.

%%%%%%%%%%%%%%%%%%%%
\subsection{A restricted duality hierarchy}
\label{sec:TruncTensorHierarchy}
%%%%%%%%%%%%%%%%%%%%

In this section, we consider a certain subset of fields in the SL(7)-branched out tensor hierarchy that includes all $\bm{56} \rightarrow (\bm{21}^\prime + \bm{7}^\prime) + ( \bm{21}+ \bm{7}) $ electric and magnetic vectors, but excludes all of the four-forms and most of the SL(7)-covariant two-forms and three-forms that respectively arise in the branching of the $\bm{133}$ and $\bm{912}$ of E$_{7(7)}$ under SL(7). It only includes the two-forms associated to the generators of  $\textrm{SL}(7) \ltimes \mathbb{R}^7$ and the three-forms in the conjugate representation of the electric part of the embedding tensor.  Specifically, we wish to consider the following $\cN=8$ bosonic field content, in SL(7) representations,
\begin{eqnarray} \label{eq:SL7fieldcontent4D}
\bm{1} &  \textrm{metric}:  &  \; ds_4^2 \nonumber \\
\bm{21}^\prime  + \bm{7}^\prime  + \bm{21}+ \bm{7} &  \textrm{coset representatives}:  &   \; {\cal V}^{IJ \, ij} \;, \; {\cal V}^{I8 \, ij}    \;, \;  \tilde{{\cal V}}_{IJ}{}^{ij} \;, \; \tilde{{\cal V}}_{I8}{}^{ij}  \; ,   \nonumber \\
\bm{21}^\prime  + \bm{7}^\prime  + \bm{21}+ \bm{7} &  \textrm{vectors}:  &   \; \cA^{IJ}  \;, \quad \cA^{I}  \;, \quad \;  \tilde{\cA}_{IJ}  \;, \quad \tilde{\cA}_{I}    \; ,  \nonumber \\
\bm{48} + \bm{7}^\prime &  \textrm{two-forms}:  &   \; \cB_{I}{}^J   \;, \quad \cB^I   \; , \nonumber \\
\bm{28}^\prime &  \textrm{three-forms}:  &   \; \cC^{IJ}  \; ,
\end{eqnarray}
along with the fermions $\psi_\mu^i$ and $\chi^{ijk}$ in the $\overline{\textbf{8}}$ and $\overline{\textbf{56}}$ of SU(8).  Note that $\cA^{IJ} \equiv \cA^{[IJ]}$, but $\cC^{IJ} \equiv \cC^{(IJ)}$. The vectors $\cA^{IJ}$ and $\cA^I$ can alternatively be considered to lie respectively in the adjoint and fundamental of SO(7), as they must for the $\textrm{ISO}(7) = \textrm{SO}(7) \ltimes \mathbb{R}^7$ gauging. The representations shown for the coset representatives correspond to their SL(7)  indices $I = 1, \ldots, 7$. Unlike for the vectors and two-forms, we have kept the label `8'  in them that comes from the branching (\ref{eq:chainE7}) through SL(8). Their antisymmetric upper (lower) indices $ij$ label the   $\,\bm{\overline{28}}\,$ ($\bm{28}$)  of SU(8). 

Considering the field content (\ref{eq:SL7fieldcontent4D}) requires some justification, since it contains more fields than necessary to write the ISO(7)-gauged Lagrangian in the formulation of \cite{deWit:2007mt} (see section \ref{subsec:LST}), yet does not include all the fields in the full tensor hierarchy. The relevance of this field content will only become apparent when we discuss the full embedding of the ISO(7) gauging into type IIA \cite{Guarino:2015vca}. It is nevertheless still possible to justify the self-consistency of the field content (\ref{eq:SL7fieldcontent4D}) from a purely four-dimensional perspective. As we will next show, for the $gm \neq 0$ ISO(7) gauging, (\ref{eq:SL7fieldcontent4D})  defines a consistent subsector of the full E$_{7(7)}$ duality hierarchy \cite{Bergshoeff:2009ph}, in the conventional sense. Namely, the Bianchi identities of the $p$-forms, $p=1,2,3$ in (\ref{eq:SL7fieldcontent4D}), the duality relations that these forms satisfy together with the metric and scalars, their equations of motion and supersymmetry variations, all close among themselves. This restricted field content preserves, of course, full $\cN=8$ supersymmetry since we are also keeping the $\bm{\overline{8}}$ gravitini. The rest of this subsection will be devoted to show the closure of the Bianchi identities and duality relations, while the closure of the supersymmetry variations will be verified in section  \ref{sec:susyvars}. 

In order to show the closure of the Bianchi identities, we first compute the field strengths of the $p$-form potentials in (\ref{eq:SL7fieldcontent4D}) specified by the ISO$(7)_c$ gauging (\ref{Theta-compSL7}). The two-form field strengths of the vectors are given by 
\begin{equation}
\label{eqO:2FormFieldStrengths}
\begin{array}{lll}
{\cal H}^{IJ}_\2 &=& d \cA^{IJ} - g \,  \delta_{KL} \, \cA^{IK} \wedge \cA^{LJ} \ , \\[7pt]
{\cal H}^I_\2 &=& d \cA^I -g \, \delta_{JK} \, \cA^{IJ} \wedge \cA^K +\tfrac12 m \, \cA^{IJ} \wedge \tilde \cA_J + m \,  \cB^I \ ,\\[2mm]
\tilde{{\cal H}}_{\2 IJ}  &=&  d \tilde{\cA}_{IJ} + g \,  \delta_{K[I} \, \cA^{KL}  \wedge \tilde{\cA}_{J]L} +  g \,  \delta_{K[I} \, \cA^{K}  \wedge \tilde{\cA}_{J]}  - m  \,  \tilde{\cA}_{I}  \wedge \tilde{A}_{J} +2 g \, \delta_{K[I} \, \cB_{J]}{}^K  \  ,  \\[2mm]
\tilde{{\cal H}}_{\2 I}  &=& d \tilde{\cA}_I  - \tfrac12 g \,  \delta_{IJ}  \, \cA^{JK} \wedge \tilde{\cA}_K + g \, \delta_{IJ} \,  \cB^J  \ ,
\end{array}
\end{equation}
the three-form field strengths of the two-form potentials are
\begin{equation}
\label{eqO:3FormFieldStrengths}
\begin{array}{lll}
{\cal H}_{\3 I}{}^J &=&  D \cB_I{}^J  + \tfrac12  \cA^{JK} \wedge d \tilde{\cA}_{IK} + \tfrac12  \cA^{J} \wedge d \tilde{\cA}_{I}  + \tfrac12  \tilde{\cA}_{IK} \wedge d \cA^{JK} + \tfrac12  \tilde{\cA}_{I} \wedge d \cA^{J}   \\[2mm] 
&&  - \tfrac 12 g \, \delta_{KL} \, \cA^{JK} \wedge \cA^{LM} \wedge \tilde{\cA}_{IM}  - \tfrac 12 g \, \delta_{KL} \, \cA^{JK} \wedge \cA^{L} \wedge \tilde{\cA}_{I}   \\[2mm]
&&  + \tfrac 16 g \, \delta_{IK} \, \cA^{JL} \wedge \cA^{KM} \wedge \tilde{\cA}_{LM}   - \tfrac 13 g \, \delta_{IK} \, \cA^{(J} \wedge \cA^{K)L} \wedge \tilde{\cA}_{L}  \\[2mm] 
&&    - \tfrac12  m  \,  \cA^{JK} \wedge \tilde{\cA}_I \wedge \tilde{\cA}_K - 2 g \, \delta_{IK} \, \cC^{JK}  - \tfrac17 \, \delta_I^J \,  (\textrm{trace})
   \ , \\[4mm]
{\cal H}_{\3}^I   &=&  D \cB^I  -\tfrac12 \cA^{IJ}  \wedge d \tilde{\cA}_J   -\tfrac12  \tilde{\cA}_J \wedge d \cA^{IJ}  +\tfrac12 g \,  \delta_{JK} \, \cA^{IJ} \wedge \cA^{KL} \wedge \tilde{\cA}_L \; ,
\end{array}
\end{equation}
and the four-form field strengths of the three-form potentials read
\begin{equation}
\label{eqO:4FormFieldStrengths}
\begin{array}{lll}
{\cal H}_{\4}^{IJ} & = & D \cC^{IJ}     - {\cal H}_\2^{K(I} \wedge \cB_K{}^{J)} +  {\cal H}_\2^{(I} \wedge \cB^{J)}  - \tfrac{1}{2} m \, \cB^I \wedge \cB^J - \tfrac16  \cA^{K(I} \wedge \tilde{\cA}_{KL} \wedge d \cA^{J)L}    \\[2mm]
&& + \tfrac16  \cA^{IK} \wedge  \cA^{JL}  \wedge d \tilde{\cA}_{KL} -  \tfrac16  \cA^{K(I} \wedge \tilde{\cA}_{K} \wedge d \cA^{J)} -  \tfrac13  \cA^{K(I} \wedge \cA^{J)} \wedge d \tilde{\cA}_{K}  \\[2mm]
&& -  \tfrac16 \cA^{(I} \wedge \tilde{\cA}_{K} \wedge d \cA^{J)K}     -\tfrac16 g\, \delta_{KL} \, \cA^{K(I} \wedge {\cA}^{J)M} \wedge \cA^{LN} \wedge \tilde{\cA}_{MN}    \\[2mm]
&&  + \tfrac16 g\, \delta_{KL} \, \cA^{K(I} \wedge {\cA}^{J)} \wedge \cA^{LM} \wedge \tilde{\cA}_{M}  - \tfrac16 g\, \delta_{KL} \, \cA^{K(I} \wedge {\cA}^{J)M} \wedge \cA^{L} \wedge \tilde{\cA}_{M} \\[2mm]
&&  - \tfrac18 m\,  \cA^{IK} \wedge {\cA}^{JL} \wedge \tilde{\cA}_{K} \wedge \tilde{\cA}_{L}      \ .
\end{array}
\end{equation}
Following (\ref{DISO7}), in (\ref{eqO:3FormFieldStrengths})--(\ref{eqO:4FormFieldStrengths}) we have defined the covariant derivatives
\begin{equation}
\label{eqO:covariantDerFormPotis}
\begin{array}{lll}
D \cB_I{}^J  & \equiv & d \cB_I{}^J   - g \,  \delta_{KL} \, \cA^{JK} \wedge \cB_I{}^L  - g \,  \delta_{IK} \, \cA^{KL} \wedge \cB_L{}^J   - g \,  \delta_{IK} \, \cA^K \wedge \cB^J   +m \,  \tilde{\cA}_I \wedge \cB^J  \\[2pt]
&& -  \tfrac17 \, \delta_I^J \,  (\textrm{trace}) \ ,  \\[6pt]
D \cB^I   & \equiv & d \cB^I  - g \, \delta_{JK} \,  \cA^{IJ} \wedge \cB^K \ ,  \\[6pt]
D \cC^{IJ} & \equiv & d \cC^{IJ}   +2 g \,  \delta_{KL} \, \cA^{K(I} \wedge \cC^{J)L} \ .
\end{array}
\end{equation} 
We have obtained the two- and three-form field strengths (\ref{eqO:2FormFieldStrengths}), (\ref{eqO:3FormFieldStrengths}) by particularising to the ISO$(7)_c$ embedding tensor (\ref{Theta-compSL7}) the generic expressions \cite{deWit:2005ub,Bergshoeff:2009ph} dictated by the $D=4$ embedding tensor formalism (see appendix~\ref{app:ConstN=8}). On the other hand, we obtained the four-form field strength (\ref{eqO:4FormFieldStrengths}) from the IIA truncation formulae \cite{Guarino:2015vca}. This expression is also compatible with that dictated by the $D=4$ embedding tensor formalism, see appendix B of \cite{Bergshoeff:2009ph}. Note the pure Yang-Mills form of the electric field strengths  ${\cal H}^{IJ}_\2$, in agreement with the purely electric gauging of the SO(7) subgroup of ISO(7) when $g \neq 0$. The electric field strengths ${\cal H}_\2^{I}$ contain the contribution expected from the semidirect action of the electric SO(7) rotations on the electric abelian translations $\mathbb{R}^7$, plus contributions of the magnetic vectors $\,\tcA_{I}\,$ and the two-forms $\,\cB^{I}\,$ due to the dyonic gauging when $ m \neq 0$.

Introducing, from (\ref{DISO7}) with the generators in the appropriate representation, the following covariant derivatives of the two-form field strengths
\begin{equation}
\label{DH2}
\begin{array}{lll}
D {\cal H}_\2^{IJ} & \equiv &  d {\cal H}_{\2}^{IJ} - 2 \, g \, \delta_{KL} \, \cA^{K[I} \wedge {\cal H}_\2^{J]L} \ , \\[6pt]
D {\cal H}_\2^I &  \equiv & d {\cal H}_\2^I - g \, \delta_{JK} \cA^{IJ} \wedge \cH_\2^K + g \, \delta_{JK} \, \cA^J \wedge \cH_\2^{IK}  - m \, \tilde{\cA}_J \wedge \cH_\2^{IJ} \ , \\[6pt]
D \tilde{{\cal H}}_{\2 IJ} &  \equiv & d \tilde{{\cal H}}_{\2 IJ}   + 2 \, g \,  \delta_{K[I} \cA^{KL} \wedge \tilde{\cH}_{\2 J]L}  + 2 \, g \, \delta_{K[I} \cA^{K} \wedge \tilde{\cH}_{\2 J]}   - 2 \, m \, \tcA_{[I} \wedge \tilde{\cH}_{\2 J]}                       \ , \\[6pt]
D \tilde{{\cal H}}_{\2 I}  &  \equiv &  d \tilde{{\cal H}}_{\2 I} - g \, \delta_{IJ} \, \cA^{JK} \wedge \tilde{\cH}_{\2 K} \ ,
\end{array}
\end{equation}
and of the three-form field strengths, 
\begin{equation}
\label{DH3}
\begin{array}{lll}
D {\cal H}_{\3 I}{}^J & \equiv &  d {\cal H}_{\3 I}{}^J - g \, \delta_{KL} \, \cA^{JK} \wedge \cH_{\3 I}{}^L- g \, \delta_{IK} \, \cA^{KL} \wedge \cH_{\3 L}{}^J - g \, \delta_{IK} \, \cA^{K} \wedge \cH_{\3}^J  \\[4pt]
&& + \, m \,  \tilde{\cA}_I \wedge \cH_\3^J   - \tfrac17 \, \delta_I^J \,  (\textrm{trace}) \ ,  \\[8pt]
D {\cal H}_\3^I &  \equiv & d {\cal H}_\3^I   - g \, \delta_{JK} \, \cA^{IJ} \wedge \cH_{\3}^K \ , 
\end{array}
\end{equation}
we find that the Bianchi identities corresponding to the form potentials in (\ref{eq:SL7fieldcontent4D}) can be written as
\begin{equation}
\label{eq:TruncBianchis}
\begin{array}{l}
D {\cal H}_{\2}^{IJ} = 0 \hspace{2mm} , \hspace{2mm}
D {\cal H}_\2^{I} = m \, \cH^{I}_{\3} \hspace{2mm} , \hspace{2mm}
D {\tilde{\cal H}}_{\2 IJ} = - 2 \, g \, {\cH_{\3 [I}}^{K}\,\delta_{J]K} \hspace{2mm} , \hspace{2mm}
D {\tilde{\cal H}}_{\2 I} = g \, \delta_{IJ} \,  \cH^{J}_{\3} \ , \\[3mm]
D {\cal H}_{\3 I}{}^J = \cH_\2^{JK} \wedge \tilde{\cH}_{\2 IK}  + \cH_\2^{J} \wedge \tilde{\cH}_{\2 I}   -2 g\, \delta_{IK} \, \cH_\4^{JK} - \tfrac17 \, \delta_I^J \,  (\textrm{trace})   \ , \\[3mm]
D {\cal H}_\3^{I} = -\cH_\2^{IJ} \wedge \tilde{\cH}_{\2 J}  \hspace{3mm} , \hspace{5mm}
D {\cal H}_{\4}^{IJ} \equiv 0 \ .
\end{array}
\end{equation}
The Bianchi identities (\ref{eq:TruncBianchis})  indeed close among themselves, as we wanted to show. An equivalent way of phrasing this is that  (\ref{eq:TruncBianchis})  defines a free differential algebra (FDA) which is a sub-FDA of the FDA defined by the Bianchi identities of the full tensor hierarchy. 

We now turn to discuss the closure of the field content (\ref{eq:SL7fieldcontent4D}) under Hodge duality. Closure is really automatic: the magnetic two-form field strengths are dual to scalar-dependent combinations of the electric two-form field strengths; the three-form field strengths are dual to scalar dependent combinations of covariant derivatives of scalars; the four-form field strengths are dual to combinations of scalars; and all vectors and scalars have been retained in (\ref{eq:SL7fieldcontent4D}). It is nevertheless useful to write the explicit duality relations. For the vectors and two-form potentials, these  have been given in \cite{deWit:2005ub,Bergshoeff:2009ph}, while for the three-form potentials the duality relations have been given in \cite{Bergshoeff:2009ph}. In particular, the four-form field strengths are dual to the derivative of the scalar potential (see (\ref{V_generalRewrite}) below) with respect to the embedding tensor. 

In order to write the duality relations, we need to introduce two scalar-dependent symmetric matrices, $\cM$, real, and $\cN$, complex, respectively E$_{7(7)}$- and SL(8)-covariant. The former is the square of the E$_{7(7)}/$SU(8) coset representative, $\mathcal{M} \,\,=\,\, \mathcal{V} \, \mathcal{V}^{t}$, and is also related to the real and imaginary parts of the latter,
\begin{equation}
\label{N_Matrix}
\mathcal{N}_{\Lambda\Sigma} = \mathcal{R}_{\Lambda\Sigma} + i \, \mathcal{I}_{\Lambda\Sigma} \ ,
\end{equation}
where $\mathcal{I}_{\Lambda\Sigma}$ is invertible and negative definite. More concretely,
\begin{eqnarray}
\label{Mscalar}
\mathcal{M}_{\mathbb{MN}}= 2 \,  {\cal V_{( \mathbb{M}}}^{ij} \,  {\cal V}_{ \mathbb{N}) \, ij }  \equiv 
\left(
\begin{array}{ll}
\mathcal{M}_{\Lambda \Sigma} & {\mathcal{M}_{\Lambda}}^{\Sigma} \\[2mm]
{\mathcal{M}^{\Lambda}}_{\Sigma} & \mathcal{M}^{\Lambda \Sigma}
\end{array}
\right)
=
\left(
\begin{array}{ll}
-(\mathcal{I}+\mathcal{R}\mathcal{I}^{-1}\mathcal{R})_{\Lambda \Sigma} & {(\mathcal{R}\mathcal{I}^{-1})_{\Lambda}}^{\Sigma} \\[2mm]
{(\mathcal{I}^{-1}\mathcal{R})^{\Lambda}}_{\Sigma} & -(\mathcal{I}^{-1})^{\Lambda \Sigma}
\end{array}
\right) .
\end{eqnarray} 
The inverse of ${\cal M}_{\mathbb{M} \mathbb{N}}$ is 
${\cal M}^{\mathbb{M} \mathbb{N}} = \Omega^{\mathbb{M} \mathbb{P}} \Omega^{\mathbb{N} \mathbb{Q}} {\cal M}_{\mathbb{P} \mathbb{Q}} $,
with $ \Omega^{\mathbb{M} \mathbb{N}}$ the Sp$(56,\mathbb{R})$-invariant matrix.

From \cite{deWit:2005ub,Bergshoeff:2009ph}, we obtain the following duality relations for the $\bm{56}$, $\bm{133}$ and $\bm{912}$ E$_{7(7)}$-covariant two-, three- and four-form field strengths,
\begin{eqnarray} \label{DualityH2}
\tilde{\cH}_{\2 \Lambda} &=& {\cal R}_{\Lambda \Sigma} \, {\cal H}_\2^\Sigma +  {\cal I}_{\Lambda \Sigma} \, *{\cal H}_\2^\Sigma  \; , \\[5pt]
\label{DualityH3}
\cH_{\3 \alpha} &=& \tfrac{1}{12} \,  (t_\alpha)_{\mathbb{M}}{}^\mathbb{P} \, \cM_{\mathbb{N}\mathbb{P}} *D \cM^{\mathbb{M} \mathbb{N}}\; ,  \\[5pt]
\label{DualityH4}
{\cH_{\4 \alpha}}^\mathbb{M} &=& - \tfrac{1}{84} \,  (t_\alpha)_{\mathbb{P}}{}^{\mathbb{R}}  {X_{\mathbb{NQ}}}^{\mathbb{S}}  \mathcal{M}^{\mathbb{MN}}   \Big(  \mathcal{M}^{\mathbb{PQ}}  \mathcal{M}_{\mathbb{RS}}  +   7 \,   \delta^\mathbb{P}_\mathbb{S} \,  \delta^\mathbb{Q}_\mathbb{R}  \Big) \, \textrm{vol}_4 \; .
\end{eqnarray}
Here, $\,{X_{\mathbb{MN}}{}^\mathbb{P} \equiv \Theta_{\mathbb{M}}}^{\alpha} \, {(t_{\alpha})_{\mathbb{N}}}^{\mathbb{P}}\,$, see (\ref{ET}),  is the contraction of the ISO(7) embedding tensor (\ref{Theta-compSL7}) with the generators $ {(t_{\alpha})_{\mathbb{N}}}^{\mathbb{P}}$ of E$_{7(7)}$ in the fundamental representation, see (\ref{Gener63}),  (\ref{Gener70}). The duality relations for the restricted field content (\ref{eq:SL7fieldcontent4D}) simply follow from (\ref{DualityH2})--(\ref{DualityH4}) by branching the adjoint SL(8) index on the vectors as in (\ref{vectors56}), and restricting the E$_{7(7)}$ generators to only those of $\textrm{SL}(7) \ltimes \mathbb{R}^7$: 
{\setlength\arraycolsep{2pt}
\begin{eqnarray}
\label{H2IJDuality}
\tilde{\cH}_{\2 IJ} &=& \tfrac{1}{2} {\cal I}_{[IJ][KL]} \, *{\cal H}_\2^{KL} + {\cal I}_{[IJ][K8]} \, *{\cal H}_\2^{K}  + \tfrac{1}{2} {\cal R}_{[IJ][KL]} \, {\cal H}_\2^{KL} +  {\cal R}_{[IJ][K8]} \, {\cal H}_\2^{K}  \ , \\[7pt]
\label{H2IDuality}
\tilde{\cH}_{\2 I} &=& \tfrac{1}{2} {\cal I}_{[I8][KL]} \, *{\cal H}_\2^{KL} + \, {\cal I}_{[I8][K8]} \, *{\cal H}_\2^{K}  + \tfrac{1}{2} {\cal R}_{[I8][KL]} \, {\cal H}_\2^{KL} + {\cal R}_{[I8][K8]} \, {\cal H}_\2^{K} \  , \\[7pt]
\label{H3IJDuality}
\cH_{\3 I}{}^J &=& \tfrac{1}{12}  (t_I{}^J)_{\mathbb{M}}{}^\mathbb{P}  \, \cM_{\mathbb{N}\mathbb{P}} *D \cM^{\mathbb{M} \mathbb{N}} - \tfrac17 \, \delta_I^J \, (\textrm{trace}) \ , \\[7pt]
\label{H3IDuality}
\cH_{\3 }{}^I &=& \tfrac{1}{12}  (t_8{}^I)_{\mathbb{M}}{}^\mathbb{P} \, \cM_{\mathbb{N}\mathbb{P}} *D \cM^{\mathbb{M} \mathbb{N}} \  , \\[7pt]
\label{H4Duality}
\cH_{\4}^{IJ} &=&  \tfrac{1}{84}  {X_{\mathbb{NQ}}}^{\mathbb{S}} \big( (t_K{}^{(I|})_{\mathbb{P}}{}^{\mathbb{R}}    \mathcal{M}^{|J)K \, \mathbb{N}} +  (t_8{}^{(I|})_{\mathbb{P}}{}^{\mathbb{R}}    \mathcal{M}^{|J)8 \, \mathbb{N}} \big)  \big(  \mathcal{M}^{\mathbb{PQ}}  \mathcal{M}_{\mathbb{RS}}  +   7 \,   \delta^\mathbb{P}_\mathbb{S} \,  \delta^\mathbb{Q}_\mathbb{R}  \big)  \textrm{vol}_4  \ . \,\, \qquad 
\end{eqnarray}
}In (\ref{H4Duality}), only components $\cM^{\Lambda \mathbb{N}}$ in the notation of (\ref{Mscalar}), and not $\cM_\Lambda{}^{ \mathbb{N}}$, are contracted with the  $\textrm{SL}(7) \ltimes \mathbb{R}^7$ generators. The combination of these duality relations with the Bianchi identities (\ref{eq:TruncBianchis}) reproduces  a subset of the equations of motion: see section \ref{ScalarCriticalPoints}.

Extensions of the duality hierarchy (\ref{eq:SL7fieldcontent4D}) may be considered that are still smaller than the full E$_{7(7)}$ hierarchy. A natural extension includes, besides the  $\bm{28}^\prime$ $\cC^{IJ}$ three-form potentials in (\ref{eq:SL7fieldcontent4D}) conjugate to the electric embedding tensor, also the SL(7)-singlet three-form potential $\tilde{\cC}$ conjugate to the singlet magnetic component of the embedding tensor. Consistency then requires that the singlet two-form potential $\,\cB\,$ that renders $\cB_I{}^J$ traceful is also retained. 
%
%This `pure trace' two-form $\,\cB\,$ has field strength
%%
%\begin{equation}
%\begin{array}{lll}
%{\cal H}_{\3} &=&  D \cB  + \tfrac12  \cA^{IJ} \wedge d \tilde{\cA}_{IJ} + \tfrac12  \cA^{I} \wedge d \tilde{\cA}_{I}  + \tfrac12  \tilde{\cA}_{IJ} \wedge d \cA^{IJ} + \tfrac12  \tilde{\cA}_{I} \wedge d \cA^{I}   \\[2mm] 
%%
%&&  - \tfrac 12 g \, \delta_{KL} \, \cA^{IK} \wedge \cA^{LM} \wedge \tilde{\cA}_{IM}  - \tfrac 12 g \, \delta_{KL} \, \cA^{IK} \wedge \cA^{L} \wedge \tilde{\cA}_{I}   \\[2mm]
%%
%&&  + \tfrac 16 g \, \delta_{IK} \, \cA^{IL} \wedge \cA^{KM} \wedge \tilde{\cA}_{LM}   - \tfrac 13 g \, \delta_{IK} \, \cA^{(I} \wedge \cA^{K)L} \wedge \tilde{\cA}_{L}  \\[2mm] 
%%
%&&    - \tfrac12  m  \,  \cA^{IK} \wedge \tilde{\cA}_I \wedge \tilde{\cA}_K - 2 g \, \delta_{IJ} \, \cC^{IJ} - 14 \, m \, \cC \ ,
%\end{array}
%\end{equation}
%%
%with
%%
%\begin{equation}
%\label{eqO:covariantDerB_singlet}
%\begin{array}{lll}
%D \cB & \equiv & d \cB   - g \,  \delta_{IJ} \, \cA^J \wedge \cB^I   + m \,  \tilde{\cA}_I \wedge \cB^I  \ .
%\end{array}
%\end{equation} 
%
%
The extension of the Bianchi identities (\ref{eq:TruncBianchis}) to also include these singlets reads
\begin{equation}
\label{DH_singlets}
\begin{array}{lll}
D {\cal H}_{\3} &=& \cH_\2^{IJ} \wedge \tilde{\cH}_{\2 IJ}  + \cH_\2^{I} \wedge \tilde{\cH}_{\2 I}   -2 g\, \delta_{IJ} \, \cH_\4^{IJ}  - 14 \, m \, \tilde{\cH}_{\4} \ , \\[2mm]
D {\tilde{\cH}}_{\4} & \equiv & 0 \ ,
\end{array}
\end{equation}
while their duality relations are, from (\ref{DualityH3}) and  (\ref{DualityH4}),
\begin{equation}
\label{H4Duality_singlet}
\begin{array}{lll}
\cH_{\3} &=&  -\tfrac{1}{12}  (t_8{}^8)_{\mathbb{M}}{}^\mathbb{P}  \, \cM_{\mathbb{N}\mathbb{P}} *D \cM^{\mathbb{M} \mathbb{N}} \ ,\\[2mm]
\tilde{\cH}_{\4} &=& \tfrac{1}{84} \,  {X_{\mathbb{NQ}}}^{\mathbb{S}}   (t_8{}^{K})_{\mathbb{P}}{}^{\mathbb{R}}    {\mathcal{M}_{8K}}^{\mathbb{N}}  \mathcal{M}^{\mathbb{PQ}}  \mathcal{M}_{\mathbb{RS}}  \,  \textrm{vol}_4 \ .
\end{array}
\end{equation}
We have used $\,{t_{I}}^{I}=-{t_{8}}^{8}\,$ and $\,\textrm{Tr}({t_{I}}^{J} \, {t_{8}}^{K})=\textrm{Tr}({t_{8}}^{J} \, {t_{8}}^{K})=0\,$ to simplify the results. For $\,\tilde{\cH}_{\4}\,$ in (\ref{H4Duality_singlet}), components  $\cM_\Lambda{}^{ \mathbb{N}}$, and not $\cM^{\Lambda \mathbb{N}}$, in the notation of (\ref{Mscalar}), are now contracted with the $\mathbb{R}^7$ generators, opposite to what happened for $\cH_{\4}^{IJ}$ in (\ref{H4Duality}). Although the singlet $\,\tilde{\cC}\,$ does not play a role in the restricted duality hierarchy (\ref{eq:SL7fieldcontent4D}), its dualised field strength $\tilde{\cH}_{\4}$  in (\ref{H4Duality_singlet}) is still crucial to recover the scalar potential, as we will show in the next subsection. The significance of this asymmetric role of $\,\tcC\,$ for the  massive type IIA embedding of dyonic ISO(7) supergravity will be discussed in \cite{Guarino:2015vca}.

\subsection{Bosonic Lagrangian} 
\label{subsec:LST}

We will now write the Lagrangian of $\cN=8$ dyonically gauged ISO(7) supergravity, focusing on the bosonic terms. While it is possible to write a Lagrangian that includes higher rank fields in the E$_{7(7)}$ tensor hierarchy (or in the restricted hierarchy (\ref{eq:SL7fieldcontent4D})) supplemented by duality relations \cite{Bergshoeff:2009ph}, we will instead write a Lagrangian in the formulation of \cite{deWit:2007mt}. The latter includes, besides the metric and scalars, only some of the vectors and two-forms in  (\ref{eq:SL7fieldcontent4D}). More concretely, the Lagrangian can be expressed in terms of the $\bm{21}^\prime + \bm{7}^\prime$ electric vectors $ \cA^\Lambda = ( \, \cA^{IJ} , \cA^I \,  ) $ and their field strengths $\,\cH_{\2}^{\Lambda} = ( \, \cH_{\2}^{IJ} \, , \,  \cH_{\2}^{I}\, )\,$, the $\bm{7}$ magnetic vectors  $\tilde{\cA}_I$ and their field strengths $\tilde{\cH}_{\2 I}$, and the $\bm{7}^\prime$ two-form potentials $\cB^I$.

The bosonic Lagrangian of $\cN=8$ dyonically gauged ISO(7) supergravity is
{\setlength\arraycolsep{2pt}
\begin{eqnarray} 
\label{BosLag}
{\cal L} &=&  R \, \textrm{vol}_4  -\tfrac{1}{48} D {\cal M}_{\mathbb{M}\mathbb{N}} \wedge * D {\cal M}^{\mathbb{M}\mathbb{N}}  +\tfrac12 \, {\cal I}_{\Lambda \Sigma} \, \cH_{(2)}^\Lambda \wedge * \cH_{(2)}^\Sigma + \tfrac12 \, \mathcal{R}_{\Lambda \Sigma} \, \cH_{(2)}^\Lambda \wedge  \cH_{(2)}^\Sigma   \\[4pt]
&& - V  \textrm{vol}_4  - m  \left[  \cB^I \wedge \big( \tilde{\cal H}_{\2 I} - \tfrac{g}{2}  \delta_{IJ}  \cB^{J}  \big) 
- \tfrac14 \,  \tilde \cA_I \wedge \tilde \cA_J \wedge \big(  d \cA^{IJ} + \tfrac{g}{2}  \, \delta_{KL} \, \cA^{IK} \wedge \cA^{JL} \big)  \right]  . \nonumber 
\end{eqnarray}
}See appendix \ref{app:ConstN=8} for some details of its derivation. The second line of (\ref{BosLag}) is entirely due to the ISO(7) gauging. It contains, on the one hand, a scalar potential,
\begin{equation}
\begin{array}{ccl}
\label{V_generalRewrite}
V & = & \dfrac{g^{2}}{168} \,  {X_{\mathbb{MP}}}^{\mathbb{R}}  {X_{\mathbb{NQ}}}^{\mathbb{S}}  \mathcal{M}^{\mathbb{MN}}   \Big(  \mathcal{M}^{\mathbb{PQ}}  \mathcal{M}_{\mathbb{RS}}  +   7 \,   \delta^\mathbb{P}_\mathbb{S} \,  \delta^\mathbb{Q}_\mathbb{R}  \Big) \ ,
\end{array}
\end{equation}
with the $X$-tensor  in (\ref{ET}) particularised for the ISO$(7)_c$ embedding tensor (\ref{Theta-compSL7}). Upon using (\ref{mcg}), this scalar potential contains pieces in $g^2$, $gm$ and $m^2$. On the other hand, the second line of (\ref{BosLag}) contains some topological terms whenever $m \neq 0$. Note, in particular, the topological mass $ gm \,  \delta_{IJ} \, \cB^I \wedge \cB^J$, which generalises a similar term in ${\cN=2}$ compactifications of massive type IIA on Calabi-Yau \cite{Louis:2002ny}. In the first line, the only contributions from the gauging appear in the covariant derivatives (\ref{DISO7}) and the gauging-modified field strengths of the electric vectors given in (\ref{eqO:2FormFieldStrengths}). In the SL(7) symplectic frame we are using, the scalar-dependent matrices $\cM$, $ \mathcal{R}$ and $ \mathcal{I}$ given in (\ref{Mscalar}) are independent of the gauging and, in particular, of the dyonically-gauging parameter $c = m/g$. 

The generic $\bm{912}$ four-form field strengths (\ref{DualityH4}) and the scalar potential (\ref{V_generalRewrite}) are related through the embedding tensor via 
\begin{eqnarray}
\label{H4/Potential1}
\Theta_\mathbb{M}{}^\alpha \, { \cH_{\4 \alpha}}^\mathbb{M} = - 2 \, V \, \textrm{vol}_4 \; .
\end{eqnarray}
Combining (\ref{H4Duality}) and the second equation in (\ref{H4Duality_singlet}), it is easy to show that this relation simplifies for the ISO$(7)_c$ gauging to
\begin{eqnarray}
\label{H4/Potential2}
 g \, \delta_{IJ} \, \cH_{\4}^{IJ} + m \, \tilde{\cH}_{\4} = - 2 \, V \, \textrm{vol}_4 \; . \; 
\end{eqnarray}
In particular, the dualisation of both four-forms $\cH_{\4}^{IJ}$ and $\tilde{\cH}_{\4} $ contains terms linear in $g$ and $m$; only when combined through (\ref{H4/Potential2}) is the quadratic dependence of $V$ on $g$ and $m$ reproduced.

The theory (\ref{BosLag}) with (\ref{DISO7}),  (\ref{eqO:2FormFieldStrengths}) admits three different smooth limits of the coupling constants $g$ and $m$. In the limit $m \rightarrow 0$, $ g\neq 0$, Hull's purely electric ISO(7) gauging \cite{Hull:1984yy} is recovered. This theory arises from consistent truncation of massless type IIA supergravity on $S^6$ \cite{Hull:1988jw}. The limit $g \rightarrow 0$, $ m \neq 0$ corresponds to a purely magnetic gauging of a nilpotent extension of  $\textrm{U}(1)^6 \times \mathbb{R}$ with 21 non-compact central charges. This theory arises as a $T^6$ truncation of massive type IIA. Finally, the $g \rightarrow 0$, $ m \rightarrow 0$ limit yields the ungauged $\cN=8$ supergravity \cite{Cremmer:1979up}, which is well known to arise from $D=11$ supergravity on $T^7$ \cite{Cremmer:1979up},  or massless type IIA on $T^6$.

\subsection{SO(7)-covariant critical point conditions} \label{ScalarCriticalPoints}

The combination of the duality relations with the Bianchi identities of the $\cN=8$ tensor hierarchy gives rise to the vector equations of motion and (projections of) the scalar equations of motion \cite{Bergshoeff:2009ph}. In the restricted duality hierarchy (\ref{eq:SL7fieldcontent4D}), all vectors were retained. Accordingly, the duality conditions (\ref{H2IJDuality}), (\ref{H2IDuality}) reproduce all of the vector equations of motion, as derived from the Lagrangian (\ref{BosLag}), upon substitution into (the first line of) the Bianchi identities (\ref{eq:TruncBianchis}). In contrast, not all of the three-form field strengths of the full hierarchy were retained in (\ref{eq:SL7fieldcontent4D}). Thus, it is interesting to enquire to which scalar equations of motion are their Bianchi identities related to when combined with the dualisation conditions. As we will now show, these are related to the equations of motion of the proper (parity even) scalars of E$_{7(7)}/$SU(8). We will focus on maximally symmetric solutions for which the scalar equations of motion reduce to the extremisation conditions for the scalar potential $V$.

For this particular discussion, we will incorporate the singlet three-form $\cH_\3$ in (\ref{DH_singlets}), (\ref{H4Duality_singlet}) along with the three-forms $\cH_{\3 I}{}^J$, $\cH_{\3}^I$ of the restricted duality hierarchy (\ref{eq:SL7fieldcontent4D}). Substituting the duality relations (\ref{H3IJDuality})--(\ref{H4Duality}), (\ref{H4Duality_singlet}) into the Bianchi identities (\ref{eq:TruncBianchis}), (\ref{DH_singlets}), we obtain a set of $ \bm{1} + \bm{48} + \bm{7}^\prime$ equations, in representations of SL(7). From the discussion of \cite{Bergshoeff:2009ph} adapted to our context, these correspond to the projections to the generators of $\textrm{GL}(7) \ltimes \mathbb{R}^7 \subset  \textrm{E}_{7(7)}$ of the equations of motion of the E$_{7(7)}/$SU(8) scalars. Further branching into representations of SO(7) and restricting to zero tensors and constant scalars (thus, critical points of $V$), these projections become
\begin{eqnarray}
\label{Extreme1}
\bm{1} : & & \,\,\, \big( g \, \delta_{IJ} \, \cH_{\4}^{IJ}  + 7 \, m \, \tilde{\cH}_{\4}  \,  \big)|_{0}  = 0 \; , \\[6pt]
\label{Extreme27}
\bm{27}: & & \big( \cH_\4^{IJ}   - \tfrac17 \, \delta^{IJ} \,  \delta_{KL} \, \cH_\4^{KL} \big) |_{0}  = 0 \;  ,  \\[6pt]
\label{Extreme21}
\bm{21}: & & \textrm{identically zero} \;  ,  \\[6pt]
\label{Extreme7}
\bm{7}: & & \textrm{identically zero} \;  ,
\end{eqnarray}
where $|_0$ denotes evaluation at a critical point of $V$. In these equations, we have used the four-form field strengths $ \cH_{\4}^{IJ}$ and $\tilde{\cH}_{\4} $ as shorthand for the scalar functions on the r.h.s.~of the duality relations (\ref{H4Duality}), (\ref{H4Duality_singlet}). Equations (\ref{Extreme21}), (\ref{Extreme7}) correspond to projections to the $\bm{21} + \bm{7}$ generators of the gauge group ISO(7). They turn out to be identically zero, in agreement with the scalar potential being invariant under the gauge group. 

Although originally obtained as projections, the SO(7)-covariant equations (\ref{Extreme1})--(\ref{Extreme7}) are in fact in one-to-one correspondence with extremisation conditions with respect to definite scalars. The singlet equation (\ref{Extreme1}) corresponds\footnote{We thank Gianluca Inverso for pointing out to us this interpretation of eq.(\ref{Extreme1}).} to the extremisation condition with respect to the SO$(1,1)$ dilaton that extends SL(7) into GL(7). Equation (\ref{Extreme27}), in the symmetric traceless of SO(7), corresponds to the extremisation of the potential with respect to the 27 scalars of SL$(7)/$SO(7). The $\bm{7}$ scalars of $\mathbb{R}^7 \subset \textrm{ISO}(7)$ are St\"uckelberg and therefore do not enter the scalar potential, hence they do not give rise to extremisation conditions. Put together, equations (\ref{Extreme1}), (\ref{Extreme27}), (\ref{Extreme7}) thus correspond to the conditions of extremisation of the potential $\,V\,$ with respect to the 35 (parity even) scalars\footnote{Alternatively, these scalars can be viewed as parameterising the coset $\textrm{SL}(8)/\textrm{SO}(8) \subset \textrm{E}_{7(7)}/\textrm{SU}(8)$.} of $\textrm{GL}(7) \ltimes \mathbb{R}^7 / \textrm{SO}(7)  \subset \textrm{E}_{7(7)}/\textrm{SU}(8)$. Note, incidentally, that these equations also depend typically on the (parity odd) pseudoscalars. Finally, the $\bm{21}$ identities (\ref{Extreme21}) can be reinterpreted as being trivial in that the corresponding compact SO(7) scalars have been modded out from the coset  E$_{7(7)}/$SU(8).

\subsection{Supersymmetry transformations} \label{sec:susyvars}

We conclude our characterisation of ISO(7)-dyonically-gauged $\cN=8$ supergravity with the supersymmetry transformations. The only effects of the gauging on the supersymmetry variations of the ungauged theory occur in the fermion variations, through the gauging-modified field strengths of the vectors and new (`shift') scalar-dependent terms. The supersymmetry variations of the bosons are the same in gauged and in ungauged supergravity. We will nevertheless find it useful to spell out the  supersymmetry variations of the SL(7)-covariant bosonic fields in (\ref{eq:SL7fieldcontent4D}) to show that they only involve fields within the same set. See  \cite{deWit:2007mt} for the $\cN=8$ supersymmetry transformations of the fermions.

The $\cN=8$ supersymmetry transformations for the vectors $A^\mathbb{M}$  are linear in the scalar coset representative\footnote{The actual $\,\textrm{E}_{7(7)}/\textrm{SU}(8)\,$ coset element is a $\,56 \times 56\,$ matrix of the form $\,{\cal V}_\mathbb{M}{\,}^{\underline{\mathbb{N}}}\,$. This coset representative $\,{\mathcal{V}_{\mathbb{M}}}^{\,\underline{\mathbb{N}}}\,$ is in a \textit{mixed} basis in which the global (not underlined) and the local (underlined) indices are taken in the SL(8) and SU(8) basis, respectively. As a result, one has the decomposition $\,{{\mathcal{V}_{\mathbb{M}}}^{\underline{\mathbb{N}}}=({\mathcal{V}_{\mathbb{M}}}^{ij},\mathcal{V}_{\mathbb{M} \, ij})}\,$, with ${\,\mathcal{V}_{\mathbb{M} \, ij}=({\mathcal{V}_{\mathbb{M}}}^{ij})^{*}\,}$, together with $\,{\mathcal{V}_{\mathbb{M}}}^{ij}=({\tilde{\mathcal{V}}_{\Lambda}}^{\,\,\,\,ij},\mathcal{V}^{ \Lambda \, ij})\,$ and $\,\mathcal{V}_{\mathbb{M} \, ij}=(\tilde{\mathcal{V}}_{\Lambda \, ij},{\mathcal{V}^{ \Lambda}}_{ij})\,$. The change of basis between the SL(8) and SU(8) basis is given in terms of SO(8)-invariant real tensors $\,[\Gamma_{AB}]^{ij}\,$, namely,
\begin{equation} 
\label{eq:SL8SU8}
{R_{\mathbb{M}}}^{\,\mathbb{N}}  =\frac{1}{2 \sqrt{2}} \,\,  [\Gamma_{AB}]^{ij} \,\,  \otimes \,\,
\left(
\begin{array}{cc}
1   &  1 \\
-i  &  i 
\end{array}\right) \ ,
\end{equation}
where $\,A=1, \ldots ,8\,$ is a fundamental SL(8) index and $\,i=1, \ldots ,8\,$ is a fundamental SU(8) index. This is compatible with identifying the $\textbf{8}$ of SL(8) with the $\textbf{8}_{v}$ of SO(8) and the $\textbf{8}$ of SU(8) with the chiral $\textbf{8}_{s}$ of SO(8). The same change of basis (\ref{eq:SL8SU8}) applies to local (underlined) indices.} 
$\,{\cal V}_\mathbb{M}{}^{ij}\,$ \cite{Cremmer:1979up}, which also sits in the $\bm{56}$ of $\textrm{E}_{7(7)}$. The transformations of the $\bm{133}$ two-form potentials $\cB_\alpha$ were worked out in \cite{deWit:2007mt}, where they were shown to be quadratic in the coset representative. In order to write the variations of the $\bm{48} + \bm{7}^\prime$ two-forms in (\ref{eq:SL7fieldcontent4D}) we will only have to branch the result of \cite{deWit:2007mt} accordingly and select these SL(7) representations. The $\cN=8$ supersymmetry variations of the $\bm{912}$ three-forms $\cC_{\alpha}{}^{\mathbb{M}}$ have not appeared in the literature\footnote{Results are known for less than maximal supersymmetry: the supersymmetry transformations of the three-form potentials for $\cN=1$ and $\cN=2$ hierarchies have been computed in \cite{Hartong:2009az}  and \cite{Huebscher:2010ib}, respectively.}. We conjecture these variations to be 
{\setlength\arraycolsep{0pt}
\begin{eqnarray} \label{eq:susyThreeForms}
&& \delta \cC_{\mu \nu \rho \, \alpha}{}^\mathbb{M} = (t_\alpha)_\mathbb{R}{}^\mathbb{P} \, \Omega^{\mathbb{R} \mathbb{Q}} \,  \Omega^{\mathbb{M} \mathbb{N}} \Big( \tfrac{4i}{7} {\cal V}_{\mathbb{N} \, jl}  \, {\cal V}_{\mathbb{P}}{}^{lk}   \, {\cal V}_{\mathbb{Q} \, ik} \, \bar{\epsilon}^i \, \gamma_{[\mu \nu} \, \psi_{\rho]}^j 
 -i \tfrac{\sqrt{2}}{ 6} \,   {\cal V}_{\mathbb{N}}{}^{hi}  \, {\cal V}_{\mathbb{P} \, [ij|}   \,  {\cal V}_{\mathbb{Q} \, | kl]} \, \bar{\epsilon}_h \, \gamma_{\mu \nu \rho} \, \chi^{jkl} \nonumber \\[4pt]
&& \qquad  \qquad\qquad\qquad \qquad \qquad \;+ \textrm{h.c.}   \Big)   +3 \, \cB_{[\mu \nu| \alpha} \, \delta \cA^\mathbb{M}_{|\rho]}
-  \,  (t_\alpha)_\mathbb{R}{}^\mathbb{P} \, \Omega_{\mathbb{P} \mathbb{N}}  \, \cA^\mathbb{M}_{[\mu} \, \cA^\mathbb{R}_{\nu}  \, \delta  \cA^\mathbb{N}_{\rho]} \; ,
\end{eqnarray}
}up to a possible symmetrisation in $ \bar{\epsilon}^{(i} \, \gamma_{[\mu \nu} \, \psi_{\rho]}^{j)}$. As usual, $\gamma_{\mu_1 \ldots \mu_p} = e_{\mu_1}{}^{\alpha_1} \ldots  e_{\mu_p}{}^{\alpha_p} \gamma_{\alpha_1 \ldots \alpha_p}$. This conjecture passes several consistency checks. Being cubic in the coset representative, (\ref{eq:susyThreeForms}) follows the pattern of the variations of the vectors and two-forms. The terms in $\cB_{[\mu \nu| \alpha} \, \delta \cA^\mathbb{M}_{|\rho]}$  and  $\cA^\mathbb{M}_{[\mu} \, \cA^\mathbb{R}_{\nu}  \, \delta  \cA^\mathbb{N}_{\rho]}$ match the corresponding terms of the $\cN=1$ and $\cN=2$ three-form transformations  \cite{Hartong:2009az,Huebscher:2010ib}. The truncation of (\ref{eq:susyThreeForms}) to one of the singlets in the decomposition (\ref{eq:Branchings}) of the $\bm{912}$ under SL(7) coincides with the supersymmetry variation of the type IIA Ramond-Ramond three-form in the $\textrm{SO}(1,3) \times \textrm{SL}(7)$-covariant reformulation of type IIA supergravity of \cite{Guarino:2015vca}. Here, we will instead be interested in the $\bm{28}^\prime$ SL(7) components of (\ref{eq:susyThreeForms}). As we will show in \cite{Guarino:2015vca}, this too can be reproduced from consistent truncation of massive type IIA on $S^6$.

We can therefore specify the supersymmetry variations of the bosonic fields (\ref{eq:SL7fieldcontent4D}) in our conventions as follows. The vielbein and scalar coset representatives transform as
\begin{equation}
\label{SUSY-Vielbeine}
\begin{array}{llll}
\delta {e_{\mu}}^{\alpha} &=&   \tfrac{1}{4} \,  \, \bar{\epsilon}_{i} \, \gamma^{\alpha} \, {\psi_{\mu}}^{i} + \tfrac{1}{4} \, \bar{\epsilon}^{i} \, \gamma^{\alpha} \, \psi_{\mu i}  & , \\[3mm]
\delta {\mathcal{V}_{\mathbb{M}}}^{ij} &=&  \tfrac{1}{ \sqrt{2}} \,  \mathcal{V}_{\mathbb{M} \, kl} \, \left( \bar{\epsilon}^{[i} \, \chi^{jkl]}  \, + \, \tfrac{1}{4!} \, \varepsilon^{ijklmnpq} \, \bar{\epsilon}_{m} \, \chi_{npq} \right) & , 
\end{array}
\end{equation}
with the fundamental E$_{7(7)}$ index $\mathbb{M}$ on the coset representative branched out into SL(7) representations according to (\ref{eq:Branchings}). For reference from \cite{Guarino:2015vca}, we do branch out  the supersymmetry variations of the vectors under SL(7):
\begin{equation}
\label{VectorsBranchingSL7}
\begin{array}{llrl}
 \delta \cA_\mu{}^{IJ} &=&  i\,  \cV^{IJ}{}_{ij} \left(  \, \bar{\epsilon}^i  \psi_\mu{}^j + \tfrac{1}{2\sqrt{2}} \,  \bar{\epsilon}_k  \gamma_\mu \chi^{ijk}\right) + \textrm{h.c.} & , \\[3mm]
\delta \cA_\mu{}^{I} &=&  i\,  \cV^{I8}{}_{ij} \left(  \, \bar{\epsilon}^i  \psi_\mu{}^j + \tfrac{1}{2\sqrt{2}}  \,  \bar{\epsilon}_k  \gamma_\mu \chi^{ijk}\right) + \textrm{h.c.} & , \\[3mm]
\delta \tilde{\cA}_{\mu \, IJ} &=&  -i\,  \tilde{\cV}_{IJ \, ij} \left(  \, \bar{\epsilon}^i  \psi_\mu{}^j + \tfrac{1}{2\sqrt{2}} \,  \bar{\epsilon}_k  \gamma_\mu \chi^{ijk}\right) + \textrm{h.c.} & ,  \\[3mm]
\delta \tilde{\cA}_{\mu \, I} &=&  -i\,  \tilde{\cV}_{I8 \, ij} \left(  \, \bar{\epsilon}^i  \psi_\mu{}^j + \tfrac{1}{2\sqrt{2}}  \,  \bar{\epsilon}_k  \gamma_\mu \chi^{ijk}\right) + \textrm{h.c.} & .
\end{array}
\end{equation}

The supersymmetry transformations of the $ \bm{48} + \bm{7}^\prime $ two-forms read
{\setlength\arraycolsep{1pt}
\begin{eqnarray} \label{susytensors4dSL7bis}
\delta \cB_{\mu \nu \, J}{}^I & = &   \Big[ - \tfrac{2}{3} \big( {\cal V}^{IK}{}_{jk} \,  \tilde{{\cal V}}_{JK}{}^{ik} + {\cal V}^{I8}{}_{jk} \,  \tilde{{\cal V}}_{J8}{}^{ik}  +  \tilde{{\cal V}}_{JK \, jk}{} \,  {\cal V}^{IK}{}^{ik}   +  \tilde{{\cal V}}_{J8 \, jk}{} \,  {\cal V}^{I8}{}^{ik} \big)  \bar{\epsilon}_{i} \gamma_{[\mu} \psi_{\nu]}^j \nonumber \\
&& \quad -  \tfrac{\sqrt{2}}{3} \, \big(  {\cal V}^{IK}{}_{ij} \,  \tilde{{\cal V}}_{JK \, kl}  +  {\cal V}^{I8}{}_{ij} \,  \tilde{{\cal V}}_{J8 \, kl}   \big) \,  \bar{\epsilon}^{[i} \gamma_{\mu \nu} \chi^{jkl]} +\textrm{h.c.} \Big]  \nonumber \\
&&  \quad  +  \big( \cA_{[\mu}^{IK} \, \delta \tilde{\cA}_{\nu] JK} + \cA_{[\mu}^{I} \, \delta \tilde{\cA}_{\nu] J} +  \tilde{\cA}_{[\mu| \, JK} \,  \delta  \cA_{|\nu]}{}^{IK}+  \tilde{\cA}_{[\mu| \, J} \,  \delta  \cA_{|\nu]}{}^{I} \big) 
-\tfrac17 \,  \delta_J^I \,  (\textrm{trace}) \; , \nonumber  \\[12pt]
\delta \cB_{\mu \nu}{}^I & = & \Big[  \tfrac{2}{3}  \big( {\cal V}^{IJ}{}_{jk} \,  \tilde{{\cal V}}_{J8}{}^{ik} +  \tilde{{\cal V}}_{J8 \, jk}{} \,  {\cal V}^{IJ}{}^{ik} \big)  \, \bar{\epsilon}_{i}  \gamma_{[\mu} \psi_{\nu]}^j + \tfrac{\sqrt{2}}{3} \,  {\cal V}^{IJ}{}_{ij} \,  \tilde{{\cal V}}_{J8 \, kl}  \,  \bar{\epsilon}^{[i} \gamma_{\mu \nu} \chi^{jkl]} +\textrm{h.c.} \Big]   \nonumber \\
&&  \quad -  \big( \cA_{[\mu}^{IJ} \, \delta \tilde{\cA}_{\nu] J} +  \tilde{\cA}_{[\mu| \, J} \,  \delta  \cA_{|\nu]}{}^{IJ} \big) \; . 
\end{eqnarray}
}Note the same pattern of SL(7) indices in coset and vector contributions: $\cV^{IK} \, \tilde{\cV}_{JK}$ parallels $\cA^{IK} \, \delta  \tilde  \cA_{JK}$, etc. Finally, the variations that follow from (\ref{eq:susyThreeForms})  for the $\bm{28}^\prime$ three-forms are
{\setlength\arraycolsep{1pt}
\begin{eqnarray} \label{susy3forms4dSL7bis}
\delta \cC_{\mu \nu \rho}{}^{IJ} & = &  \Big[ -  \tfrac{4i}{7} \,\Big( 
 \cV^{K(I}{}_{jl} \big(  \cV^{J)L \, lk} \,  \tilde \cV_{KL \, ik}  + \tilde \cV_{ KL}{}^{ lk} \,  \cV^{J)L}{}_{ ik}   \big) \nonumber \\
&& \qquad  \quad  + \, \cV^{K(I}{}_{jl} \big(  \cV^{J)8 \, lk} \,  \tilde \cV_{K8 \, ik}  + \tilde \cV_{K8}{}^{ lk} \,  \cV^{J)8}{}_{ ik}   \big) \nonumber \\
&& \qquad  \quad + \, \cV^{(I|8}{}_{jl} \big(  \cV^{|J)K \, lk} \,  \tilde \cV_{K8 \, ik}  + \tilde \cV_{K8}{}^{ lk} \, \cV^{|J)K}{}_{ik}  \big) \Big)  \, \bar{\epsilon}^{i}  \gamma_{[\mu \nu} \psi_{\rho]}^j  \nonumber \\
&& + \, i \tfrac{\sqrt{2}}{3}  \,\Big( 
 \cV^{K(I|}{}^{hi}  \, \cV^{|J)L}{}_{[ij|} \,  \tilde \cV_{KL}{}_{|kl]}   + \,  \cV^{K(I}{}^{hi} \,  \cV^{J)8}{}_{[ij|} \,  \tilde \cV_{K8}{}_{|kl]}  \\ 
&& \qquad  \quad  + \, \cV^{(I|8}{}^{hi}  \, \cV^{|J)K}{}_{[ij|} \,  \tilde \cV_{K8}{}_{|kl]}   \Big)  \, \bar{\epsilon}_{h}  \gamma_{\mu \nu \rho } \chi^{jkl} \; + \textrm{h.c.} \Big] \nonumber \\
&& - \, 3 \,  \Big(  \cB_{[\mu \nu| K}{}^{(I} \, \delta \cA^{J)K}_{|\rho]} +   \cB_{[\mu \nu}{}^{(I} \, \delta \cA^{J)}_{\rho]}   \Big) \nonumber \\
&& + \,  \cA^{K(I}_{[\mu} \big(  \cA^{J)L}_\nu \,  \delta \tilde \cA_{\rho] KL}  + \tilde \cA_{\nu KL} \, \delta \cA^{J)L}_{\rho]}   \big)  + \cA^{K(I}_{[\mu} \big(  \cA^{J)}_\nu \,  \delta \tilde \cA_{\rho] K}  + \tilde \cA_{\nu K} \, \delta \cA^{J)}_{\rho]}   \big) \nonumber \\
&& \qquad  \quad + \,   \cA^{(I}_{[\mu} \big(  \cA^{J)K}_\nu \,  \delta \tilde \cA_{\rho] K}  + \tilde \cA_{\nu K} \, \delta \cA^{J)K}_{\rho]}   \big)  \; . \nonumber
\end{eqnarray}
}Again, the SL(7) structure of indices in coset and vector contributions is the same. 

Equations (\ref{SUSY-Vielbeine})--(\ref{susy3forms4dSL7bis}) show that the supersymmetry variations of the bosonic fields (\ref{eq:SL7fieldcontent4D}) close among themselves and into the fermions $\psi^i_\mu$, $\chi^{ijk}$. In turn, the supersymmetry variations of the fermions close into scalars and field strengths of vectors, all of which were retained in (\ref{eq:SL7fieldcontent4D}). This shows the consistency of the subsector (\ref{eq:SL7fieldcontent4D}) of the duality hierarchy, plus fermions, at the level of the supersymmetry variations. Finally, the supersymmetry variations of the fields that enter the Lagrangian (\ref{BosLag}) involve fields that appear as well in the Lagrangian. Note, however, that the fields entering the Lagrangian can still source the supersymmetry variations of fields not entering the Lagrangian, \textit{e.g,} the terms $ \cA_{[\mu}^{I} \, \delta \tilde{\cA}_{\nu] J} +  \tilde{\cA}_{[\mu| \, J} \,  \delta  \cA_{|\nu]}{}^{I} -\tfrac17 \,  \delta_J^I \,  (\textrm{trace})$ source the supersymmetry variation $\delta \cB_{\mu \nu \, J}{}^I$.

\begin{center}
\begin{table}[t!]
\renewcommand{\arraystretch}{1.5}
\scalebox{0.77}{
\begin{tabular}{lllllll}
\noalign{\hrule height 1pt}
\textrm{Field}   & SO$(7)$  & SO$(6)$ & $\textrm{G}_{2}$ & $\textrm{SO}(4)=\textrm{SO}(3)_{d} \times \textrm{SO}(3)_{R}$ & SU(3) & \\
\noalign{\hrule height 1pt}
\large{scalars}
&  \textbf{1} + \textbf{7} & $(3 \times)$ \textbf{1} + $(2 \times)$ \textbf{6} & $(2 \times)$ \textbf{1}  & $(4 \times)$ \textbf{(1,1)} + $(4 \times)$ \textbf{(2,2)}   & $(6 \times)$ \textbf{1} + $(4 \times)$ \textbf{3} + $(4 \times)$ $\overline{\textbf{3}}$ \\
&+ \textbf{27}  & + \textbf{20} + \textbf{15} & + $(2 \times)$ \textbf{7} &  + $(2 \times)$ \textbf{(3,3)} + $(2 \times)$ \textbf{(3,1)}  &  + $(2 \times)$ \textbf{6} + $(2 \times)$ $\overline{\textbf{6}}$ \\
&  + \textbf{35}  &   + \textbf{10} + $\overline{\textbf{10}}$ & + $(2 \times)$ \textbf{27} &    + $(2 \times)$ \textbf{(4,2)} + $(2 \times)$ \textbf{(5,1)}  &  + $(2 \times)$ \textbf{8} 
\\[2mm]
\noalign{\hrule height 1pt}
$\cA^{IJ}$ & \textbf{21} & \textbf{15} + \textbf{6}  & \textbf{14} + \textbf{7} &  \textbf{(1,3)} + $(2\times)$ \textbf{(3,1)}  &  $(2 \times)$ \textbf{3} + $(2 \times)$ $\bar{\textbf{3}}$  \\[-2mm]
 & & & & + \textbf{(2,2)} + \textbf{(4,2)} & + \textbf{8} + \textbf{1} \\[2mm]
$\cA^{I}$  & \textbf{7} & \textbf{6} + \textbf{1}  & \textbf{7}  & \textbf{(2,2)} + \textbf{(3,1)} &  \textbf{3} + $\bar{\textbf{3}}$ + \textbf{1} \\[2mm]
$\tcA_{IJ}$ & \textbf{21} & \textbf{15} + \textbf{6}  & \textbf{14} + \textbf{7} &  \textbf{(1,3)} + $(2\times)$ \textbf{(3,1)}  &  $(2 \times)$ \textbf{3} + $(2 \times)$ $\bar{\textbf{3}}$  \\[-2mm]
 & & & & + \textbf{(2,2)} + \textbf{(4,2)} & + \textbf{8} + \textbf{1} \\[2mm]
$\tcA_{I}$  & \textbf{7} & \textbf{6} + \textbf{1}  & \textbf{7}  & \textbf{(2,2)} + \textbf{(3,1)} &  \textbf{3} + $\bar{\textbf{3}}$ + \textbf{1} \\[2mm]
\noalign{\hrule height 1pt}
${\cB_{I}}^{J}$  & \textbf{21} &  \textbf{1} + $(2\times)$ \textbf{6}   & \textbf{14 + 7}  & $(2\times)$ \textbf{(2,2)} + $(2\times)$ \textbf{(4,2)} &  $(2\times)$ \textbf{1} + $(3\times)$ \textbf{3} + $(3\times)$ $\bar{\textbf{3}}$ \\[-2mm]
& \textbf{+ 27} & + \textbf{15} + \textbf{20} &  + \textbf{27} &  + $(2\times)$ \textbf{(3,1)}  + \textbf{(1,3)}   &  + $(2\times)$ \textbf{8} + \textbf{6} + $\bar{\textbf{6}}$ \\[-2mm]
& & & &  + \textbf{(3,3)} + \textbf{(5,1)}  & \\[-2mm]
& & & &  + \textbf{(1,1)} & \\[2mm]
$\cB^{I}$ & \textbf{7} & \textbf{6} + \textbf{1}  & \textbf{7} & \textbf{(2,2)} + \textbf{(3,1)} &  \textbf{3} + $\bar{\textbf{3}}$ + \textbf{1} \\[2mm]
\noalign{\hrule height 1pt}
$\cC^{IJ}$  & \textbf{1 + 27} & $(2 \times)$ \textbf{1}  & \textbf{1} + \textbf{27} & $(2\times)$ \textbf{(1,1)} + \textbf{(2,2)} + \textbf{(3,3)}   & $(2\times)$ \textbf{1} + \textbf{3} + $\bar{\textbf{3}}$    \\[-2mm]
& & + \textbf{20} + \textbf{6} & &  + \textbf{(4,2)} + \textbf{(5,1)}    &  + \textbf{8} + \textbf{6} + $\bar{\textbf{6}}$ \\[2mm]
%$\cC$ & \textbf{1} & \textbf{1} & \textbf{1} &  \textbf{(1,1)} & \textbf{1}  \\[2mm]
\noalign{\hrule height 1pt}
\end{tabular}
}
\caption{Branching rules of the SL(7)-covariant tensor hierarchy (\ref{eq:SL7fieldcontent4D}) for different invariant sectors of the ISO(7)$_{c}$ supergravity. Only singlets are retained in each sector. Following the discussion around (\ref{H4Duality_singlet}), all sectors can be extended to include a singlet two-form $\,\cB\,$ that makes ${\cB_{I}}^{J}$ traceful, and a singlet three-form $\tcC$ dual to the magnetic component of the embedding tensor.}
\label{Table:group_theory}
\end{table}
\end{center}

%\vspace{-40pt}

\section{An $\mathcal{N}=2$ truncation: the SU(3)-invariant sector}
\label{sec:SU(3)-sector}

In the remainder of the paper, we will specify the truncation of the $\cN=8$ theory to various interesting subsectors that preserve $\cN=2$ and $\cN=1$ supersymmetry and SU(3), G$_2$, and an SO(4) subgroup of the ISO(7) bosonic gauge symmetry.  See table \ref{Table:group_theory} for a summary of the field content of these subsectors, and of the SO$(7)$ and SO$(6)$  further subsectors of the SU(3) sector.

We begin by discussing the consistent truncation of the $\cN=8$ theory to its SU(3)-invariant sector. The analog truncation for the purely electric SO(8) gauging \cite{deWit:1982ig} has been studied in \cite{Warner:1983vz,Ahn:2000mf,Bobev:2010ib} and for the dyonic SO(8) gauging \cite{Dall'Agata:2012bb}, in \cite{Borghese:2012zs,Guarino:2013gsa}. This sector corresponds to $\,\cN=2\,$ supergravity coupled to one vector multiplet and one hypermultiplet. The corresponding $2+4$ real scalars take values on a submanifold
\begin{eqnarray} 
\label{ScalManN=2}
\frac{\textrm{SU}(1,1)}{\textrm{U}(1)} \, \times \,   \frac{\textrm{SU}(2,1)}{\textrm{SU}(2) \times \textrm{U}(1)}
\end{eqnarray}
of E$_{7(7)}/$SU(8) which is the product of two well known special K\"ahler (SK) and quaternionic K\"ahler (QK) manifolds. The gauging inherited in this sector from the $\cN=8$  ISO(7)$_{c}$ gauging is an abelian $\,\textrm{U}(1) \times \textrm{SO}(1,1)_{c}\,$ dyonic gauging in the hypersector.  In section \ref{subsec:construction} we construct the Lagrangian of this theory, including an explicit parameterisation for the scalar kinetic terms and potential, and discuss the duality hierarchy in section \ref{subsec:SU3-hierarchy}. We then give a superpotential and the canonical $\cN=2$ formulations of this sector in sections \ref{subsec:SuperPot} and \ref{CanonicalN=2}, respectively. Some further subsectors are discussed in \ref{sec:subsecsSU3} and the vacuum structure  is analysed in \ref{subsec:CriticalPoints}. See also appendix~\ref{app:SU3sectorScalMat} for the explicit expression of the SU(3)-invariant scalar matrix $\cM_{\mathbb{M}\mathbb{N}}$, and appendix \ref{app:MTheoryonSE7} for the relation of this sector to the similar model that arises from consistent truncation of M-theory on an arbitrary Sasaki-Einstein seven-manifold \cite{Gauntlett:2009zw}.

\subsection{Construction and bosonic Lagrangian} 
\label{subsec:construction}

The embedding of the relevant SU(3) into $\textrm{SO}(7)\subset \textrm{ISO}(7)$  can be described by the chain 
\begin{equation}
\label{Embeding_SO3}
\textrm{SO}(7) \supset \textrm{SO}(6) \sim \textrm{SU}(4) \supset \textrm{SU}(3) \ ,
\end{equation}
so that $\,\textbf{7} \rightarrow  \textbf{1} \,+\, \textbf{3} \,+\, \bar{\textbf{3}}\,$. In terms of SL(8) indices, we have a splitting $\,A \rightarrow ( a \,\oplus\, 8) \,\oplus\, (1 \,\oplus\, \hat{a} )$ with $\,a=2,4,6\,$ and $\,\hat{a}=3,5,7\,$, followed by a complexification of the form
\begin{equation}
\label{SU3_complexification}
z_{0} = x_{1} + i \, x_{8}
\hspace{3mm} \textrm{ , } \hspace{3mm}
z_{1} = x_{2} + i \, x_{3}
\hspace{3mm} \textrm{ , } \hspace{3mm}
z_{2} = x_{4} + i \, x_{5}
\hspace{3mm} \textrm{ , } \hspace{3mm}
z_{3} = x_{6} + i \, x_{7} \ ,
\end{equation}
so that SU(3) is realised as a singlet ($z_{0}$) and a triplet $(z_{1,2,3})$ of complex coordinates. When restricted to this sector, the retained bosonic fields take values along the SU(3)-invariant metric\footnote{Indices $i,j$ in this section and in appendix \ref{app:SU3sectorScalMat} are in the fundamental of the SO(6) in the chain (\ref{Embeding_SO3}). These indices should not cause any confusion with the SU(8) indices of the $\cN=8$ coset representative ${\cal V}_{\mathbb{M}}{}^{ij}$ of section \ref{sec:ISO(7)}.}
$\,\delta_{ij}\,$, the two-form $\,J_{ij}\,$, $\,i=2, \ldots ,7\,$, and the complex totally antisymmetric tensor of SU(3) (or equivalently, a complex decomposable three-form $\Omega_{ijk}$, see appendix~\ref{app:SU3sectorScalMat}). In fact, only the scalar matrix $ \cM_{\mathbb{MN}}$ has components along the latter. The fields in the $\cN=8$ duality hierarchy (\ref{eq:SL7fieldcontent4D}) give rise to the following SU(3)-invariant fields: 
\begin{equation}
\label{SU(3)_full_content}
\begin{array}{rrrrl}
\textrm{metric} & : & g_{\mu\nu} \\[3mm]
\textrm{scalars} & : & \cM_{\mathbb{MN}} & \rightarrow & (\chi,\varphi) \,\,\textrm{ and }\,\, (\phi,a,\zeta,\tilde{\zeta}) \,\,\,\,\,\,\, , \,\,\,\,\,\, \textrm{[ see appendix~\ref{app:SU3sectorScalMat}  ]} \\[2mm]
\textrm{vectors} & : & \cA^{I} & \rightarrow  &  \cA^{1} \equiv A^{0} \\[1mm]
        &  & \cA^{IJ} & \rightarrow  &   \cA^{ij} =  A^{1} \, J^{ij} \\[1mm]
        &  & \tcA_{I} & \rightarrow  &  \tcA_{1} \equiv \tilde{A}_{0} \\[1mm]
        &  & \tcA_{IJ} & \rightarrow  &  \tcA_{ij} = \tfrac{1}{3} \tilde{A}_{1} \, J_{ij} \\[3mm]
\textrm{two-forms} & : & \cB^{I} & \rightarrow  &  \cB^{1} \equiv B^{0} \\[1mm]
        &  & {\cB_{I}}^{J} & \rightarrow  &  {\cB_{1}}^{1} = \tfrac{6}{7} \, B_{1} \,\,\,\,\, , \,\,\,\,\,  {\cB_{i}}^{j} = -\tfrac{1}{7} \, B_{1} \, \delta_{i}^{j} + \tfrac{1}{3} B_{2} \, {J_{i}}^{j}  \\[3mm]
\textrm{three-forms} & : & \cC^{IJ} & \rightarrow  &   \cC^{11}  \equiv C^{0}   \,\,\,\,\, , \,\,\,\,\,    \cC^{ij} = C^{1} \, \delta^{ij}  \; , 
%        &  & \cC & \rightarrow  &  C 
\end{array}
\end{equation}
in agreement with the number of singlets in the last column of table~\ref{Table:group_theory}. The real scalars $\,(\chi,\varphi)\,$ and $\,(\phi,a,\zeta,\tilde{\zeta})\,$ respectively parameterise each factor of the scalar manifold (\ref{ScalManN=2}). The superscript $\Lambda = 0 , 1$ on the electric vectors $A^\Lambda$ labels them as the graviphoton and the vector in the vector multiplet, respectively, and similarly for their magnetic counterparts $\tilde{A}_\Lambda$. The superscripts or subscripts on the two- and three-forms are just labels with no further meaning. The vectors $A^0$, $\tilde{A}_0$ gauge dyonically the SO$(1,1)_c$ generated by $\,T_{1}\,$ in (\ref{linear_comb_t}), while $A^1$ gauges electrically the U(1) generated by $\,T_{23}+T_{45}+T_{67}\,$ in (\ref{linear_comb_t}). Along with the metric and the six scalars, only $A^0$, $A^1$, $\tilde{A}_0$, their field strengths and $B^0$ enter the SU(3)-invariant bosonic Lagrangian, see (\ref{L_full_SU3}). Finally, the branching of the gravitini, in the $\bm{\overline{8}}$ of SU(8), under this SU(3) produces two singlets, in agreement with the $\,\mathcal{N}=2\,$ supersymmetry of this sector.

We can construct an explicit parameterisation of the scalar manifold (\ref{ScalManN=2}) of this sector as follows. We first identify the generators (\ref{Gener63}), (\ref{Gener70}) of E$_{7(7)}$ that are invariant under the SU(3), (\ref{Embeding_SO3}), (\ref{eq:chainE7}), under consideration. These are
\begin{equation}
\label{Gener_SU3}
\begin{array}{cclc}
g_{1} &=& {t_{2}}^{2} + {t_{4}}^{4} + {t_{6}}^{6} + {t_{3}}^{3} + {t_{5}}^{5} + {t_{7}}^{7} - 3 \, ({t_{1}}^{1} + {t_{8}}^{8})  & , \\[2mm]

g_{2} &=& g^{(-)}_{2} + g^{(+)}_{2} =  ({t_{1}}^{8}) + ({t_{8}}^{1}) & , \\[2mm]

g_{3} &=&  {t_{1}}^{1} - {t_{8}}^{8} & , \\[2mm]

g_{4} &=&  g^{(-)}_{4} + g^{(+)}_{4} = (t_{4567} + t_{2367} + t_{2345} ) + (t_{1238} + t_{1458} + t_{1678} ) & ,  \\[2mm]

g_{5} &=&  g^{(-)}_{5} + g^{(+)}_{5} =  ( t_{1246}  - t_{1257}  - t_{1347}- t_{1356}  ) + ( t_{8357} - t_{8346} - t_{8256} - t_{8247}) & ,  \\[2mm]

g_{6} &=&  g^{(-)}_{6} + g^{(+)}_{6} = ( t_{3571} - t_{3461} - t_{2561} - t_{2471} ) +  (t_{8246} - t_{8257} - t_{8347} - t_{8356} )   & ,
\end{array}
\end{equation}
where $g_1$, $g_3$ are Cartan generators and a subscript $(\pm)$ indicates a positive or negative root. The exponentiations
\begin{equation}
\label{coset_SU3}
\mathcal{V}_{\textrm{SK}}  = e^{-12 \, \chi \, g^{(+)}_{4}} \, e^{\frac{1}{4} \, \varphi \, g_{1}} 
\hspace{10mm} \textrm{ and } \hspace{10mm}
\mathcal{V}_{\textrm{QK}}  = e^{a \,  g^{(+)}_{2} - \, 6 \, \zeta \,g^{(+)}_{5} - \,6 \, \tilde{\zeta} \,g^{(+)}_{6} } \, e^{\phi \, g_{3}} \ 
\end{equation}
lead to coset representatives for each factor in (\ref{ScalManN=2}), and the total representative is simply the product\footnote{ \label{FootnoteSL8} This coset is in the SL(8) basis. This is enough for our purposes, since we will not discuss couplings to the fermions. Should one be interested in, for example, restricting the $\cN=8$ supersymmetry variations (\ref{SUSY-Vielbeine})--(\ref{susy3forms4dSL7bis}) to the SU(3)-invariant sector, a rotation (\ref{eq:SL8SU8}) of this coset representative would be needed.} $\,{\mathcal{V} = \mathcal{V}_{\textrm{SK}} \, \mathcal{V}_{\textrm{QK}}}\,$. Finally, the scalar matrix is the quadratic combination $\,{\cM=\mathcal{V} \, \mathcal{V}^{t}}$. See appendix \ref{app:SU3sectorScalMat} for its explicit expression.

With this scalar parameterisation, the Lagragian of the SU(3)-invariant sector can be written as 
\begin{equation}
\label{L_full_SU3}
\begin{array}{lll}
{\cal L} &=&  (R - V) \, \textrm{vol}_4 + \tfrac{3}{2} \left[ d\varphi \wedge * d\varphi +  e^{2 \varphi} \, d\chi \wedge * d\chi \right] \\[2mm]
&+& 2 \, d\phi \wedge * d\phi +  \frac{1}{2} \, e^{2 \phi} \,  [ D\zeta \wedge *  D\zeta  +  D\tilde{\zeta} \wedge * D\tilde{\zeta} ] \\[2mm] 
&+&  \tfrac{1}{2} \, e^{4 \phi} \,   [ Da +  \tfrac{1}{2}  ( \zeta D \tilde{\zeta} - \tilde{\zeta} D \zeta  ) ] \wedge * [ Da +  \tfrac{1}{2}  ( \zeta D \tilde{\zeta} - \tilde{\zeta} D \zeta  )  ]    \\[2mm]
&+& \tfrac{1}{2} \, \mathcal{I}_{\Lambda\Sigma} \, H_{\2}^{\Lambda} \wedge  * H_{\2}^{\Sigma}  + \tfrac{1}{2} \, \mathcal{R}_{\Lambda\Sigma} \, H_{\2}^{\Lambda} \wedge H_{\2}^{\Sigma}  - m \,   B^{0} \wedge  d\tilde{A}_{0}\,-\, \tfrac{1}{2} \, g \, m \,  B^{0}  \wedge  B^{0}   \ ,
\end{array}
\end{equation}
and follows by truncating (\ref{BosLag}) according to (\ref{SU(3)_full_content}). Here, the covariant derivatives are
\begin{equation}
\label{Cov_aB}
Da = da \,+\, g \, A^{0} \, - \, m \, \tilde{A}_{0} 
\hspace{4mm} \textrm{ , } \hspace{4mm}
D\zeta = d\zeta - 3 \, g \,  A^{1} \, \tilde{\zeta} 
\hspace{4mm} \textrm{ , } \hspace{4mm}
D \tilde{\zeta} = d\tilde{\zeta} + 3 \, g \,  A^{1}\, \zeta \,\,\, \ ,
\end{equation} 
and the electric vector field strengths
\begin{equation}
\label{H_modified_Elec_SU3}
H_{\2}^{0}=dA^{0} +  m \, B^{0}
\hspace{8mm} \textrm{ , } \hspace{8mm}
H_{\2}^{1}=dA^{1} \,\,\, \ ,
\end{equation}
follow from (\ref{eqO:2FormFieldStrengths}). The gauge kinetic matrix in (\ref{L_full_SU3}) is obtained from the scalar matrix $\,\cM\,$ through (\ref{Mscalar}). In the scalar parameterisation that we are using here, it explicitly reads
\begin{equation}
\label{NMatrix}
\mathcal{N}_{\Lambda \Sigma} = \mathcal{R}_{\Lambda \Sigma} + i \, \mathcal{I}_{\Lambda \Sigma} = 
\frac{1}{(2\, e^{\varphi } \, \chi +i )}
\left(
\begin{array}{cc}
 -\dfrac{e^{3 \varphi }}{(e^{\varphi } \, \chi -i )^2} & \dfrac{3 \, e^{2 \varphi } \, \chi }{(e^{\varphi} \, \chi -i )} \\[5mm]
 \dfrac{3 \, e^{2 \varphi } \, \chi }{(e^{\varphi } \, \chi -i)} & 3 \,  ( e^{\varphi } \, \chi^2+e^{-\varphi})
\end{array}
\right) \ .
\end{equation}
Note that $\,\mathcal{I}_{\Lambda\Sigma}\,$ is negative definite so that the vector kinetic terms have the correct sign. Finally, the explicit expression of the scalar potential in  (\ref{L_full_SU3}) can be derived from (\ref{V_generalRewrite}) to be 
\begin{equation}
\label{VSU3}
\begin{array}{lll}
V &=& \frac{1}{2} \, g^{2} \left[ e^{4 \phi -3 \varphi } \big(1+e^{2\varphi} \chi^2 \big)^3 -12  \, e^{2 \phi -\varphi }  \big(1+e^{2\varphi} \chi^2 \big) -24 \, e^{\varphi} \right.
\\[6pt]
&& \left. \qquad +  \tfrac34 \, e^{4\phi +\varphi} \big( \zeta^2 + \tilde \zeta^2 \big)^2   \big(1+3 \, e^{2\varphi} \chi^2 \big)+ 3\, e^{4\phi +\varphi} \big( \zeta^2 + \tilde \zeta^2 \big) \chi^2   \big(1+e^{2\varphi} \chi^2 \big) \right.
\\[6pt]
&& \left. \qquad  -3 \, e^{2\phi +\varphi} \big( \zeta^2 + \tilde \zeta^2 \big)   \big(1-3 \, e^{2\varphi} \chi^2 \big)   \right] -    \tfrac12 \, g \, m \, \chi  \, e^{4 \phi+3 \varphi  } \left( 3\big( \zeta^2 + \tilde \zeta^2 \big) + 2\chi ^2 \right)  \\[6pt]
&& + \,  \frac{1}{2} \, m^2 \, e^{4 \phi+3 \varphi  }  \ .
\end{array}
\end{equation}
Out of the six real scalars in this sector, this potential effectively depends on only four. The non-compact St\"uckelber scalar $a$ and the U(1) phase $\beta$ of the complex combination 
\begin{equation}
\label{polar_coord}
\tilde{\zeta}+ i \, \zeta = 2 \, \rho \, e^{i \, \beta} \ ,
\end{equation}
do not enter the potential. As we will discuss in section \ref{subsec:CriticalPoints}, this potential displays a rich structure of critical points, both supersymmetric and non-supersymmetric, when $\,gm \neq 0\,$.

\subsection{Duality hierarchy} 
\label{subsec:SU3-hierarchy}

The duality hierarchy (\ref{eq:SL7fieldcontent4D}) in this sector reduces to the field content (\ref{SU(3)_full_content}), which includes only singlets in the branching of SL(7) under SU(3). The formulae in section \ref{sec:TruncTensorHierarchy} simplify accordingly. The electric vector field strengths have already been given in (\ref{H_modified_Elec_SU3}), and their magnetic counterparts are
\begin{equation}
\label{H2_SU3sector}
\tilde{H}_{\2 0} =  d\tilde{A}_{0} + g B^{0} 
\hspace{8mm} \textrm{ , } \hspace{8mm}
\tilde{H}_{\2 1} =  d\tilde{A}_{1} - 2 g B_{2} \,\,\, \ .
\end{equation}
The field strengths (\ref{eqO:3FormFieldStrengths}) for the two-form potentials $(B^{0} \, ; \, B_{1} , B_{2})$ reduce to
\begin{equation}
\label{H3_SU3sector}
\begin{array}{lll}
H_{\3}^{0} & = & dB^{0} \ ,  \\[1mm]
H_{\3 1} & = & DB_{1} + \tfrac{1}{2} ( A^{0} \wedge d\tilde{A}_{0} +  \tilde{A}_{0} \wedge dA^{0}  - \tfrac{1}{3} A^{1} \wedge d\tilde{A}_{1} -  \tfrac{1}{3} \tilde{A}_{1} \wedge d A^{1} ) + 2 g (C^{1} - C^{0}) \ , \\[1mm]
H_{\3 2} & = & dB_{2} \ , 
\end{array}
\end{equation}
with $\,DB_{1}=dB_{1}-g A^{0} \wedge B^{0}+m \tilde{A}_{0} \wedge B^{0}\,$, and those (\ref{eqO:4FormFieldStrengths}) for the three-form potentials $(C^{0},C^{1})$ read
\begin{equation}
\label{H4_SU3sector}
\begin{array}{lll}
H_{\4}^{0} & = & dC^{0} + H_{\2}^{0} \wedge B^{0} - \tfrac{1}{2} m B^{0} \wedge B^{0} \ ,  \\[1mm]
H_{\4}^{1} & = & dC^{1} - \tfrac{1}{3} H^{1}_{\2} \wedge B_{2} \ .
\end{array}
\end{equation}
The Bianchi identities (\ref{eq:TruncBianchis}) simplify to
\begin{equation}
\label{eq:TruncBianchis_SU3}
\begin{array}{l}
d {H}_\2^{0} = m \, H^{0}_{\3} \hspace{3mm} , \hspace{3mm}
d {H}_{\2}^{1} = 0 \hspace{3mm} , \hspace{3mm}
d {\tilde{H}}_{\2 0} = g \,  H^{0}_{\3} \hspace{3mm} , \hspace{3mm} 
d {\tilde{H}}_{\2 1} = - 2 \, g \, H_{\3 2} \\[3mm] 
d H_\3^{0} = 0  \hspace{3mm} , \hspace{3mm} D H_{\3 1} = H_\2^{0} \wedge \tilde{H}_{\2 0}  - \tfrac{1}{3} H_\2^{1} \wedge \tilde{H}_{\2 1}  + 2 g \, (H^1_\4 - H^0_\4 )  \hspace{2mm} , \hspace{2mm} d H_{\3 2} = 0 \ , \\[3mm]
d H_{\4}^{0} \equiv d H_{\4}^{1} \equiv 0 \ ,
\end{array}
\end{equation}
where $\,D H_{\3 1} =d H_{\3 1} -g A^{0} \wedge H_{\3}^0 +m \tilde{A}_{0} \wedge H_{\3}^0 \,$. These again close among themselves, in agreement with the consistency of the SU(3)-invariant truncation.
 
The duality relations also simplify, and can be written in terms of the explicit scalar parameterisation on this sector given in section \ref{subsec:construction} and appendix \ref{app:SU3sectorScalMat}. The vector/vector duality relations (\ref{H2IDuality}) and (\ref{H2IJDuality}) reduce to
\begin{equation}
\label{duality:vector/vector}
\begin{array}{lll}
\tilde{H}_{\2 0} &=& - \dfrac{e^{3 \varphi } (1+ 3 e^{2 \varphi } \chi ^2)}{(1+e^{2 \varphi } \chi ^2)^2 (1+4 e^{2 \varphi } \chi ^2)} * H_{\2}^{0}
+  \, \chi ^2 \,  \dfrac{3 e^{3 \varphi }}{1+e^{2 \varphi } \chi ^2 ( 5+4 e^{2 \varphi } \chi^2)} * H_{\2}^{1} 
 \\[4mm]
&-& 2 \, \chi ^3 \,  \dfrac{ e^{6 \varphi } }{(1+e^{2 \varphi } \chi ^2)^2 (1+4 e^{2 \varphi } \chi ^2)} \, H_{\2}^{0}
+  \chi \, \dfrac{3 e^{2 \varphi }  (1+2 e^{2 \varphi } \chi ^2)}{1+e^{2 \varphi } \chi ^2 ( 5+4 e^{2 \varphi } \chi^2)} \, H_{\2}^{1}  \ ,  \\[5mm]
\tilde{H}_{\2 1} &=&  \chi ^2 \, \dfrac{3 e^{3 \varphi }}{1+e^{2 \varphi } \chi ^2 ( 5+4 e^{2 \varphi } \chi^2)}  \, * H_{\2}^{0} - e^{-\varphi} \, \dfrac{3 (1 + e^{2 \varphi } \chi ^2)}{1 + 4 e^{2 \varphi } \chi ^2} \, * H_{\2}^{1} \\[4mm]
&+&  \chi  \, \dfrac{3e^{2 \varphi }  (1+2 e^{2 \varphi } \chi ^2)}{1+e^{2 \varphi } \chi ^2 ( 5+4 e^{2 \varphi } \chi^2)}  \, H_{\2}^{0}  + \chi \, \dfrac{6 (1 + e^{2 \varphi } \chi ^2)}{1 + 4 e^{2 \varphi } \chi ^2} H_{\2}^{1} \ ,
%
%
%   KEEP THIS ONE FOR FUTURE !!!
%
%\tilde{\cH}_{\2 mn} &=& - \chi ^2 \, \dfrac{e^{3 \varphi }}{1+e^{2 \varphi } \chi ^2 ( 5+4 e^{2 \varphi } \chi^2)} \, %J_{mn} \, * \cH_{\2}^{1} +  \chi  \, \dfrac{e^{2 \varphi }  (1+2 e^{2 \varphi } \chi ^2)}{1+e^{2 \varphi } \chi ^2 ( %5+4 e^{2 \varphi } \chi^2)} \, J_{mn} \, \cH_{\2}^{1}\\[4mm]
%&-& e^{-\varphi} \left( \frac{1}{2} \, \chi ^2 \dfrac{e^{2\varphi }}{1 + 4 e^{2 \varphi } \chi ^2} \, J_{pq} J_{mn}  -  
%\delta_{mp}\, \delta_{nq} \right) * \cH_{\2}^{pq} \\[4mm]
%&+& \chi \left( \frac{1}{2}  \dfrac{ (1 + 2 e^{2 \varphi } \chi ^2)}{1 + 4 e^{2 \varphi } \chi ^2} \, J_{pq} J_{mn}  -  
%\delta_{mp}\, \delta_{nq} \right) \cH_{\2}^{pq} \ .
%
\end{array}
\end{equation}
the duality relations (\ref{H3IDuality}), (\ref{H3IJDuality}) for the three-form field strengths simplify as
\begin{equation}
\label{duality:two-form/scalar}
\begin{array}{lll}
H_{\3}^{0} &=& - e^{4 \phi} \, * \big(   Da  +\tfrac{1}{2}\, (\zeta \, D\tilde{\zeta} - \tilde{\zeta}  \, D\zeta)     \big) \ , \\[3mm]
{H_{\3 1}} &=&  * \big[  2(d \varphi - e^{2 \varphi} \chi d\chi) - 2 d\phi  + a e^{4 \phi}  \big( Da  +\tfrac{1}{2} (\zeta  D\tilde{\zeta} - \tilde{\zeta}   D\zeta) \big)   + \tfrac{1}{2} e^{2\phi} ( \zeta D\zeta +  \tilde{\zeta} D\tilde{\zeta} )  \big]   , \\[3mm]
H_{\3 2} &=& \tfrac{1}{2} * \Big[ e^{2\phi}  (\zeta \, D\tilde{\zeta} - \tilde{\zeta}  \, D\zeta) + \tfrac{1}{2} \, (\zeta^2 + \tilde{\zeta}^2) \,  e^{4 \phi}  \big( Da  +\tfrac{1}{2}\, (\zeta \, D\tilde{\zeta} - \tilde{\zeta}  \, D\zeta) \big)  \Big]  \ ,
\end{array}
\end{equation}
and the duality relations  (\ref{H4Duality}) for the four-form field strengths give rise to
\begin{equation}
\label{duality:three-form/ET}
\begin{array}{lll}
H_{\4}^{0} &=&  \Big[  \tfrac{1}{2} \, g \,  \big(1+e^{2\varphi} \chi^2 \big)  \Big( 12 \, e^{2 \phi - \varphi } - 2  \,  e^{4 \phi - 3\varphi }  \big(1+e^{2\varphi} \chi^2 \big)^2  - 3 e^{4 \phi +\varphi } \chi^2 \big( \zeta^2 +\tilde{\zeta}^2 \big) \Big)  \\[4pt]
&& \,\,\, + \, m\, e^{4 \phi + 3\varphi } \,  \chi^3 \Big]  \, \textrm{vol}_{4} \ , \\[3mm]
H_{\4}^{1} &=& \Big[ \tfrac{1}{2} \, g \Big( 8 \, e^{\varphi } + 2 \, e^{2 \phi - \varphi }  \big(1+e^{2\varphi} \chi^2 \big) + e^{2\phi +\varphi} \big( \zeta^2 + \tilde \zeta^2 \big)   \big(1-3 \, e^{2\varphi} \chi^2 \big)    \\[4pt]
&&  \qquad \, \, - \, \tfrac12 \, e^{4\phi +\varphi} \big( \zeta^2 + \tilde \zeta^2 \big) \chi^2   \big(1+e^{2\varphi} \chi^2 \big)
 -  \tfrac14 \, e^{4\phi +\varphi} \big( \zeta^2 + \tilde \zeta^2 \big)^2   \big(1+3 \, e^{2\varphi} \chi^2 \big) \Big)  \\[4pt]
&& \,\,\, + \, \tfrac14 \, m \, e^{4 \phi + 3\varphi } \, \chi \,   \big( \zeta^2 + \tilde \zeta^2 \big)  \Big]  \, \textrm{vol}_{4} \ .
\end{array}
\end{equation}
For later reference, we also give the SU(3)-invariant truncation of the duality relation (\ref{H4Duality_singlet}) for the four-form field strength $\tilde{H}_{\4} \equiv \tilde{\cH}_{\4} $ of the singlet three-form potential $\,\tilde{C}\equiv \tcC\,$ related to the magnetic component of the embedding tensor. It reads
\begin{equation} \label{H4magneticDual}
\begin{array}{lll}
\tilde{H}_{\4} &=&  \frac{1}{2}  \, g \, \chi  \, e^{3 \varphi +4 \phi } \big(3 (\zeta ^2+\tilde{\zeta}^2)+2 \chi ^2 \big) \, \textrm{vol}_{4} -  m \, e^{3 \varphi +4 \phi } \, \textrm{vol}_{4} \ .
\end{array}
\end{equation}

These duality relations manifestly show that, in the symplectic frame we are using, the magnetic vectors and the higher rank forms in the tensor hierarchy do not carry independent degrees of freedom, but rather depend on the metric, the electric vector field strengths and the scalars. Alternatively, these relations can be used to transfer independent degrees of freedom within the duality hierarchy. For example, the first relation in (\ref{duality:two-form/scalar}) can be used to dualise the St\"uckelberg scalar $a$ into the two-form $B^0$, so that the latter can be regarded as carrying the independent degrees of freedom. This duality relation can also be obtained by varying the Lagrangian (\ref{L_full_SU3}) with respect to the magnetic graviphoton $\tilde{A}_0$. Solving this duality relation and substituting into (\ref{L_full_SU3}), the following new Lagrangian is obtained:
\begin{equation}
\label{L_full_SU3_dual}
\begin{array}{llll}
\widetilde{\mathcal{L}} &=& (R - V) \, \textrm{vol}_4  + \tfrac{1}{2} \, e^{-4 \phi} \,   H_{\3}^{0}\wedge * H_{\3}^{0} + \tfrac{3}{2} \left[ d\varphi \wedge * d\varphi + e^{2 \varphi} \, d\chi \wedge * d\chi \right]   \\[2mm] 
&+&  2 \, d\phi \wedge * d\phi  + \frac{1}{2} \, e^{2 \phi}  \left[ D\zeta \wedge * D\zeta  +   D\tilde{\zeta} \wedge * D\tilde{\zeta} \right]  \\[2mm]
&+& \tfrac{1}{2} \, \mathcal{I}_{\Lambda\Sigma} \, H_{\2}^{\Lambda} \wedge  * H_{\2}^{\Sigma}  + \tfrac{1}{2} \, \mathcal{R}_{\Lambda\Sigma} \, H_{\2}^{\Lambda} \wedge H_{\2}^{\Sigma}  \\[2mm]
&+&    H_{\3}^{0} \wedge  \, \left[  \,  g \, A^{0}  \, + \, \tfrac{1}{2} \, ( \zeta D \tilde{\zeta} - \tilde{\zeta} D \zeta  )   \, \right] \,-\, \tfrac{1}{2} \, gm \,  B^{0}  \wedge  B^{0} \ .
\end{array}
\end{equation}
The kinetic terms are now expressed in terms of the field strength $H_{\3}^{0}$ of $B^0$ given in (\ref{H3_SU3sector}), and the magnetic vector $\tilde{A}_0$ no longer appears in this Lagrangian. See {\it e.g.} section 4.1 of \cite{Cassani:2012pj} for a discussion in a similar context. In the Lagrangian (\ref{L_full_SU3_dual}), $B^0$ is a propagating massive two-form with conventional kinetic term  $H_{\3}^{0}\wedge * H_{\3}^{0}$ and mass term $B^0 \wedge * B^0$ (coming from the $m \, B^0$ dependence of $H_{\2}^{0}$ in (\ref{H_modified_Elec_SU3})), in addition to the topological mass term $B^0 \wedge B^0$. Lagrangians similar to (\ref{L_full_SU3_dual}) but naturally written in terms of the magnetic field strengths, as in (\ref{duality:vector/vector}), usually appear in dimensional reductions of massive IIA or M-theory to $\cN=2$ supergravity, see {\it e.g.} \cite{Louis:2002ny,Gauntlett:2009zw}. See appendix \ref{app:MTheoryonSE7} for the relation of the SU(3)-invariant sector and the theory of \cite{Gauntlett:2009zw}. See \cite{Bergshoeff:2009ph} for a more general discussion of Lagragians involving higher-rank fields in the $\cN=8$ duality hierarchy.

The duality relations can be also used to relate the four-form field strengths to the potential, as discussed for the full $\cN=8$ theory in section \ref{sec:TruncTensorHierarchy}. With the parameterisation of the scalars in the SU(3) sector that we gave in section \ref{subsec:construction}, the duality relation (\ref{H4/Potential2}) can be explicitly verified in this sector. With the help of (\ref{duality:three-form/ET}) and (\ref{H4magneticDual}), equation (\ref{H4/Potential2}) can be seen to reduce to
\begin{equation} 
\label{DualitySU3H4V}
g \, ( \, H^{0}_{\4} + 6 \, H^{1}_{\4} \, ) + m \, \tilde{H}_{\4} = -2 \, V \, \textrm{vol}_{4} \ ,
\end{equation}
where $H^{0}_{\4}$, $H^{1}_{\4}$ are the field strengths (\ref{H4_SU3sector}) of the three-form potentials $C^0$, $C^1$ in the truncated hierarchy (\ref{SU(3)_full_content}), $\tilde{H}_{\4}$ is the field strength of the three-form potential $\tilde{C}$ related to the magnetic component of the embedding tensor, and $V$ is the scalar potential in the SU(3) sector, given in  (\ref{VSU3}). From (\ref{Extreme1}) and (\ref{Extreme27}), we also find that the following relations hold at every critical point (see section \ref{subsec:CriticalPoints}) of the scalar potential  (\ref{VSU3}):
\begin{equation}
\label{EOM_SU3}
g \, ( \, H^{0}_{\4}|_0 + 6 \, H^{1}_{\4}|_0 \, ) +7 m \, \tilde{H}_{\4}|_0 =  0  \ , \qquad 
H^0_\4 |_0 = H^1_\4 |_0 \ . 
\end{equation}
Recall that $|_0$ and $V_0$ denote evaluation at a critical point. Combining (\ref{DualitySU3H4V}) with the first equation in (\ref{EOM_SU3}) yields $\,V_{0} \,\, \textrm{vol}_{4} = 3 \, m \, \tilde{H}_{\4}|_0\,$. This condition relates the AdS character of the critical points in this sector with a non-vanishing value of the magnetic gauge coupling $m$, provided $\tilde{H}_{\4}|_0 \neq 0 $, which is indeed the case.

\subsection{Superpotential formulation} 
\label{subsec:SuperPot}

Two superpotentials exist \cite{Ahn:2000mf} (see also \cite{Bobev:2010ib}) from which the scalar potential of the SU(3)-invariant sector \cite{Warner:1983vz} of the electric SO(8) gauging \cite{deWit:1982ig} derives. The same statement holds \cite{Borghese:2012zs} for the SU(3) sector of the dyonic SO(8) gauging \cite{Dall'Agata:2012bb}. Here we will show that this is also true for the SU(3)-invariant sector of ISO$(7)_c$ supergravity. See \cite{Ahn:2001by,Ahn:2002qga} for superpotentials in the SO$(7)_+$ and G$_2$ sectors of the electric ISO(7) gauging \cite{Hull:1984vg}. In order to see this following a notation close to \cite{Bobev:2010ib,Borghese:2012zs}, we first introduce coordinates $\,t\,$ and $\,u\,$ on two copies of the upper-half of the complex plane
\begin{equation}
\label{mappingN1Sugra}
t = -\chi + i \, e^{-\varphi}
\hspace{10mm} \textrm{ and } \hspace{10mm}
u =  -\rho + i \, e^{-\phi} \ ,
\end{equation}
with $\, \rho^2=\tfrac{1}{4}\, (\tilde{\zeta}^2 + \zeta^2)\,$ as follows from (\ref{polar_coord}), and then further convert into two copies of the unit-disk via
\begin{equation}
\label{mappingN2Chiral}
z= \frac{t-i}{t+i}
\hspace{8mm} \textrm{ and } \hspace{8mm}
\zeta_{12}= \frac{u-i}{u+i} \ ,
\end{equation}
so that $|z| <1$, $|\zeta_{12}| <1$. In terms of the new complex fields $\,z\,$ and $\,\zeta_{12}\,$, the kinetic terms for the $(\chi,\varphi)$ and $(\rho,\phi)$ scalars in (\ref{L_full_SU3}) can be recast as 
\begin{equation}
\label{LkinUnitDisk}
\tfrac{1}{2} \,  \mathcal{L}^{\textrm{kin}}_{\textrm{scalar}}  =   3 \, \frac{dz \wedge * d\bar{z}}{( 1- |z|^2)^2}   + 4 \,  \frac{d\zeta_{12} \wedge *  d\bar{\zeta}_{12}}{( 1- |\zeta_{12}|^2)^2}  \ . 
\end{equation}

Introducing
\begin{equation}
\label{WN=2-SU3}
\begin{array}{llll}
\mathcal{W} & \equiv &   (1-|z|^{2})^{-\frac{3}{2}} \, (1-|\zeta_{12}|^{2})^{-2}  \\[4mm]
& \times &   \left[ \, g \, \left( \dfrac{7}{8} \,  (1- \zeta_{12})^{4} \, (1+z)^{3} \, + \, 3 \, (\zeta_{12}-z) \, (1+z) \,(1-\zeta_{12})^2 \,(1- z \, \zeta_{12})  \right) \right.  \\[4mm]
&+& \left.  i\, \dfrac{m}{8} \,  (1-\zeta_{12})^{4} \, (1-z)^{3}  \, \right] \ ,
\end{array}
\end{equation}
we find that the scalar potential (\ref{VSU3}) is reproduced through the expression
\begin{equation}
\label{VN=2-SU3}
\tfrac{1}{4} \, V = 2 \, \left[  \frac{4}{3} \,  (1-|z|^{2})^{2} \, \left| \frac{\partial W}{\partial z} \right|^{2} +   (1-|\zeta_{12}|^{2})^{2} \, \left| \frac{\partial W}{\partial \zeta_{12}} \right|^{2}   -3 \, W^{2} \right] \ ,
\end{equation}
with the superpotential $\,W\,$ given in terms of (\ref{WN=2-SU3}) by either $\,W = | {\cal W}_+| \equiv | {\cal W} (z, \zeta_{12})|\,$ or $\,W =  | {\cal W}_-| \equiv | {\cal W} (z, \bar{\zeta}_{12})|\,$. We take this match as a consistency check on our calculation of the potential  (\ref{VSU3})  in section \ref{subsec:construction} with the $\cN=8$ embedding tensor formalism followed by SU(3)-invariant truncation.

All the supersymmetric critical points of the scalar potential  (\ref{VSU3}), as given in table \ref{Table:SU3Points} of section \ref{subsec:CriticalPoints}, are critical points of  $\,|\mathcal{W}_{+}|\,$. The $\cN=2$ point is an extremum of both $\,|\mathcal{W}_{+}|\,$ and $\,|\mathcal{W}_{-}|\,$. Under the map $\,\zeta_{12} \rightarrow \bar{\zeta}_{12}\,$ (or, equivalently, $\rho \rightarrow -\rho$ in (\ref{mappingN1Sugra}), {\it i.e.},  $\beta \rightarrow -\beta $ in (\ref{polar_coord})), the $\cN=1$ points become extrema of $\,|\mathcal{W}_{-}|\,$, rather than $\,|\mathcal{W}_{+}|\,$. Due to the overall $\,c^{1/3}\,$ dependence of the critical points in table~\ref{Table:SU3Points}, there are two asymptotic limits: $\,z=\zeta_{12}=-1\,$ at $\,c \rightarrow 0\,$ ({\it i.e.}, $m \rightarrow 0$, $g \neq 0$)  and $\,z=\zeta_{12}=+1\,$ at $\,c \rightarrow \infty\,$ ({\it i.e.}, $g \rightarrow 0$, $m \neq 0$). These critical points thus disappear for the purely electric and purely magnetic gaugings.

These superpotentials will be useful to holographically study RG flows between different Chern-Simons phases of the D2-brane field theory with at least SU(3) flavour symmetry. We leave this for future work. See \cite{Ahn:2000mf,Bobev:2009ms} for studies of RG flows with at least SU(3) invariance in the M2-brane field theory from electrically gauged SO(8) supergravity, and \cite{Guarino:2013gsa,Tarrio:2013qga} for similar domain wall solutions in dyonic SO(8) supergravity.

\subsection{Canonical $\cN=2$ formulation} 
\label{CanonicalN=2}

As a further crosscheck on our calculations, we will now cast  the SU(3)-invariant Lagrangian (\ref{L_full_SU3}) in $\cN=2$ canonical form, focusing on the special geometry quantities that enter the canonical formulation. The scalar manifold (\ref{ScalManN=2}) is the product of two well-known special K\"ahler and quaternionic K\"ahler manifolds, corresponding to the vector multiplet and hypermultiplet scalars, respectively. The parameterisation (\ref{coset_SU3}) leads to the familar form for the metric on this space that appears in the scalar kinetic terms in the Lagrangian (\ref{L_full_SU3}).  Indices $\,M=1, \ldots, 4\,$, $\,\alpha=1,\ldots, 8\,$, $\,u=1, \ldots , 4\,$ and $\,i=t$ in this subsection \mbox{respectively} correspond to $\,\textrm{Sp}(4,\mathbb{R})\,$ vector indices, $\,\textrm{SU}(2,1)\,$ adjoint indices, $\,{\textrm{SU}(2,1)/\textrm{U}(2)}\,$ curved indices and $\,\textrm{SU}(1,1)/\textrm{U}(1)\,$ curved holomorphic indices -- we denote by  $\,t\,$ the only value that  $\,i\,$ takes on. The index $\,\Lambda=0,1\,$ introduced below (\ref{SU(3)_full_content}) labels, as usual, ``half" the fundamental representation of $\,\textrm{Sp}(4,\mathbb{R})\,$.

Let us start by describing the special K\"ahler geometry of the scalars in the vector \mbox{multiplet}. We find the sections $\,X^M = (X^\Lambda, F_\Lambda)\,$,
\be
\label{holosections}
\begin{array}{lclclclc} 
X^0 = -t^3 & \,\,,\, & X^1= -t & \,,\, &  F_0 = 1 & \,,\, &  F_1= 3 \, t^2 & ,
\end{array}
\ee
which are holomorphic in the coordinate $\,t\,$ (\ref{mappingN1Sugra}) on the upper-half plane realisation of  $\textrm{SU}(1,1)/\textrm{U}(1)\,$, to be the relevant ones for our model. In the symplectic frame in which the Lagrangian (\ref{L_full_SU3}) is written, the sections $\,F_\Lambda\,$ can be obtained from the prepotential
\begin{equation}
\label{Prepot_SU3}
\mathcal{F}=-2\,\sqrt{ X^{0} \, (X^{1})^3} \ ,
\end{equation}
as $\,F_{\Lambda}=\partial \mathcal{F}/\partial X^{\Lambda}\,$. The K\"ahler potential
\begin{eqnarray} 
\label{KahlerPot}
K= -\log \big( i \, \bar{X}^M \Omega_{MN} X^N \big) = - 3 \,\log ( - i(t-\bar{t}) ) 
\hspace{3mm} \textrm{ with } \hspace{3mm}
\Omega_{MN} = \left(\begin{array}{cc}
0 					& \mathbb{I}_2  \\
- \mathbb{I}_2	& 0 				  \\
\end{array} \right)  ,
\end{eqnarray}
gives rise to the metric 
\begin{equation} 
\label{metricfromKPot}
- K_{t \bar{t}} \, dt \, d\bar{t} = - (\partial_{t} \partial_{\bar{t} } K)  \, d t  \, d\bar{t} = \frac{3\, dt \, d\bar{t}}{(t -\bar{t})^2} = -\frac{3}{4} \left[ d \varphi ^2 + e^{2 \varphi} \, d \chi^2\right]   \ ,
\end{equation}
on the relevant scalar kinetic terms in (\ref{L_full_SU3}). The components of the vielbein
\begin{equation}
\begin{array}{llllll}
f_{t}{}^M &=& (f_{t}{}^\Lambda ,f_{t}{}_\Lambda ) &\equiv&  \partial_{t} (e^{K/2} X^M ) \,+\, \tfrac{1}{2} \,  e^{K/2} X^M \partial_{t} K & , \\[2mm]
\bar{f}_{\bar{t}}{}^M &=& (\bar{f}_{t}{}^\Lambda ,\bar{f}_{\bar{t}}{}_\Lambda ) &\equiv&  \partial_{\bar{t}} (e^{K/2} \bar{X}^M ) \,+\, \tfrac{1}{2} \,  e^{K/2} \bar{X}^M \partial_{\bar{t}} K & ,
\end{array}
\end{equation}
explicitly read
\begin{equation}
\label{VMvielbein}
\begin{array}{lllll}
f_{t}{}^0 =  \phantom{-} \dfrac{3 \, t^2 \,  \bar{t}}{[ \, i(t-\bar{t} )\, ]^{5/2}} & \,\, , \,\,&  f_{t}{}^1 = \phantom{-}  \dfrac{2 \, t +  \bar{t}}{[ \, i({t}-\bar{t} ) \, ]^{5/2}}   & , \\[4mm]
f_{t}{}_0 =  - \dfrac{3 }{[ \, i(t -\bar{t} ) \, ]^{5/2}} & \,\, , \,\, & f_{t}{}_1 = - \dfrac{3 \, t \,  (t +2 \, \bar{t})}{[\,  i(t -\bar{t} ) \, ]^{5/2}}  & ,
\end{array}
\end{equation}
and $\,\bar{f}_{\bar{t}}{}^M \equiv (f_{t}{}^M)^{*}\,$. Together with the gauge kinetic matrix $\,{\cal N}_{\Lambda \Sigma}\,$ given in (\ref{NMatrix}), we have verified these quantities to satisfy a number of special geometry identities\footnote{These include:
\begin{eqnarray} 
\label{N=2identities}
& F_\Lambda = {\cal N}_{\Lambda\Sigma} X^\Sigma \; , \qquad
 f_{t}{}_\Lambda = \bar{{\cal N}}_{\Lambda \Sigma} f_{t}{}^\Sigma \; , \qquad
 f_{t}{}^\Lambda \, K^{t  \bar{ t }}  \,   \bar{f}_{ \bar{t}}{}^\Sigma = -\tfrac12 \,  (\mathcal{I}^{-1})^{\Lambda \Sigma}  -e^{K}   \,\bar{X}^\Lambda X^\Sigma \; , \nonumber \\[5pt]
 & X^M \, \Omega_{MN} \,   f_{t}{}^N  = X^M \, \Omega_{MN} \,   \bar{f}_{\bar{t}}{}^N =0 \; , \qquad
X^\Lambda \, {\cal I}_{\Lambda \Sigma}  \,  \bar{X}^\Sigma = -\tfrac12 \, e^{-K}  \; ,  \\[6pt]
 & K_{t \bar{t} } = - i  \, \bar{f}_{\bar{t}}{}^M \, \Omega_{MN} \, f_{t}{}^N   =-2 \, f_{t}{}^\Lambda  \, {\cal I}_{\Lambda\Sigma} \,  \bar{f}_{\bar{t}}{}^\Sigma \; . \nonumber
\end{eqnarray}
}.

Let us now turn to the gauged hypermultiplet. Of the eight Killing vectors $\,k_\alpha\,$ of the quaternionic K\"ahler metric $\,h_{uv}\,$ in (\ref{L_full_SU3}), only
\begin{eqnarray} 
\label{KVs}
k_1 = \partial_a 
\hspace{10mm} \textrm{ and } \hspace{10mm}
k_2 = 3 \,   \big( \zeta \, \partial_{\tilde{\zeta}} \,-\,\tilde{\zeta} \, \partial_{\zeta}\big)  \ ,
\end{eqnarray}
participate in the gauging. As anticipated, these Killing vectors generate an abelian $\textrm{SO}(1,1) \times \textrm{U}(1)$ subgroup of SU$(2,1)$. This gauge group arises from the ISO(7) gauge group of the full $\cN=8$ theory by first breaking $\textrm{ISO}(7) =\textrm{SO}(7) \ltimes \mathbb{R}^7$ to $( \textrm{SO} (6) \ltimes \mathbb{R}^6 ) \times \mathbb{R}$; then, the compact U(1) is the singlet in the branching of the adjoint of SO(6) under SU(3) and $\textrm{SO}(1,1) \sim \mathbb{R}$ is the $\,\mathbb{R}\,$ factor in the direct product. The moment maps corresponding to the isometries (\ref{KVs}) are
\begin{eqnarray} 
\label{MomentMaps}
P_1^x = \Big( 0 \,,\,  0\,,\,  - \tfrac{1}{2} \,  e^{2\phi}  \Big)
\hspace{5mm} \textrm{ and } \hspace{5mm}
P_2^x = 3 \, \Big( - e^\phi \, \tilde\zeta \,,\,   e^\phi  \, \zeta \,,\,    1 -\tfrac14 \,  (\zeta^2 + \tilde\zeta^2) \,  e^{2\phi} \Big)  \ ,
\end{eqnarray}
with $\,x=1,2,3\,$. Finally, the embedding tensor $\,\Theta_M{}^{\alpha} = ( \Theta_\Lambda{}^{\alpha} , \Theta^{\Lambda \, \alpha} )\,$ in this sector follows from the $\,\cN=8\,$ embedding tensor  (\ref{Theta-compSL7}) via the identifications $\,\Theta_{[IJ]\phantom{K}L}^{\phantom{[IJ]}K} \leftrightarrow \Theta_1{}^2\,$, $\,{\Theta_{[I8]\phantom{8}K}^{\phantom{[I8]}8}  \leftrightarrow \Theta_0{}^1}\,$, $\,{{\Theta^{[IJ] K}}_{L} \leftrightarrow \Theta^{12}}\,$ and $\,{{\Theta^{[I8] 8}}_{K} \leftrightarrow \Theta^{01}}\,$, namely
\begin{equation} 
\label{N=2Theta}
\Theta_0{}^1 = 1 
\hspace{5mm} \textrm{ , } \hspace{5mm}
\Theta^{01} = -c 
\hspace{5mm} \textrm{ , } \hspace{5mm}
\Theta_1{}^2 = 1 
\hspace{5mm} \textrm{ and } \hspace{5mm}
\Theta^{12} = 0 \ ,
\end{equation}
with all other components vanishing. Thus, the compact U(1) is gauged electrically only, whereas $\textrm{SO}(1,1) \sim \mathbb{R}$ is gauged dyonically. 

Bringing the definitions (\ref{holosections})--(\ref{VMvielbein}), (\ref{KVs})--(\ref{N=2Theta}), along with the metrics $\,h_{uv}\,$  (which can be read off from (\ref{L_full_SU3})) and $\,K_{t \bar{t}} \,$ in (\ref{metricfromKPot}), to the canonical expression for the $\,\cN=2\,$ scalar potential due to a dyonic gauging in the hypermultiplet sector \cite{deWit:2005ub,Michelson:1996pn},
\begin{eqnarray} 
\label{potN=2}
\tfrac{1}{4} \, V= g^{2} \, \Theta_M{}^\alpha \Theta_N{}^\beta \Big[ 4\, e^K X^M \bar{X}^N  h_{uv} k^u{}_\alpha  k^v{}_\beta
+ P^x_\alpha P^x_\beta  \big( K^{t \bar{t}} f_{t}{}^M \bar{f}_{\bar{t}}{}^N  -3 \, e^K X^M \bar{X}^N \big) \Big] ,
\end{eqnarray}
we exactly reproduce the scalar potential (\ref{VSU3}). We have also verified that the equations \cite{Hristov:2009uj, Louis:2012ux}
\begin{eqnarray} 
\label{N=2Consgen}
X^M \Theta_M{}^ \alpha k^u{}_\alpha = 0 
\hspace{3mm} \textrm{ , } \hspace{3mm}
\epsilon_{xyz} X^M \bar{X}^N \Theta_M{}^\alpha \Theta_N{}^\beta P_\alpha^y P_\beta^z = 0   
\hspace{3mm} \textrm{ , } \hspace{3mm}
f_t{}^M \Theta_M{}^ \alpha P_\alpha^ x = 0 \ ,
\end{eqnarray}
for maximally supersymmetric solutions within this $\,\cN=2\,$ sector reduce to
\begin{eqnarray}
\tilde\zeta = \zeta = 0 
\hspace{3mm} \textrm{ , } \hspace{3mm}
g\, t^3 + m = 0 
\hspace{3mm} \textrm{ , } \hspace{3mm}
g \, \big( \, 4 \, t  + 2 \, \bar{t}  - e^{2\phi} \,  t^2 \,  \bar{t} \, \big) - m \, e^{2\phi} =0 \ .
\end{eqnarray}
For $\,g m \neq 0\,$,  these have the $\,\cN=2\,$, $\,\textrm{SU}(3) \times \textrm{U}(1)$-invariant AdS$_{4}$ critical point in Table~\ref{Table:SU3Points} of section \ref{subsec:CriticalPoints} as their unique solution.

\subsection{Further subsectors}
\label{sec:subsecsSU3}

Let us now briefly discuss some further consistent truncations of the SU(3)-invariant sector which lead, accordingly, to subsectors with smaller field content and larger symmetry. The field contents discussed below agree with those recorded in table \ref{Table:group_theory}.

The field content of the $\,\textrm{G}_{2}$-invariant sector is obtained from (\ref{SU(3)_full_content}) by truncating out all vectors and two-forms, and identifying the three-forms as $\,C^0 = C^1 \equiv C\,$ and the scalars as
\begin{equation}
\label{G2_identification}
\varphi = \phi
\hspace{8mm} \textrm{ , } \hspace{8mm}
\chi = \tfrac{1}{2} \, \tilde{\zeta}
\hspace{8mm} \textrm{ and } \hspace{8mm}
a=\zeta=0 \ .
\end{equation}
This corresponds to the exponentiation of the linear combinations $\,\tfrac{1}{4} \, g_{1} + g_{3}\,$ and $\,g^{(+)}_{4} + g^{(+)}_{6}\,$ of generators in (\ref{Gener_SU3}). This sector is $\cN=1$, and its Lagrangian follows from bringing these identifications to (\ref{L_full_SU3}). An alternative construction of the G$_2$ sector that does not rely on its embedding in the SU(3)-invariant sector will be given in section \ref{sec:G2-sector}. Turning off the axion, $\chi =0$, leads to the SO(7)$_+$-invariant sector. 

The $\,\textrm{U}(3)$-invariant subsector has an additional $\,\textrm{U}(1)=\textrm{SO}(2)\,$ symmetry, with respect to the SU(3) sector, gauged by the vector $A^{1}$. This sector is thus reached by simply turning off the hypermultiplet axions
\begin{equation}
\label{U3_identification}
\tilde{\zeta} \,\,=\,\, \zeta \,\,=\,\, 0 \ ,
\end{equation}
since these are charged under that U(1), see (\ref{Cov_aB}). Together with these axions, the only other field in (\ref{SU(3)_full_content}) that needs to be turned off is the two-form $B_2 = 0$. This is forced by the third duality relation in (\ref{duality:two-form/scalar}), and is consistent with the Bianchi identities (\ref{eq:TruncBianchis_SU3}). The U(3)-invariant Lagrangian is obtained by inserting (\ref{U3_identification}) into (\ref{L_full_SU3}). It is consistent to further truncate $\chi=0$, $A^1 = \tilde{A}_1 = 0$, which leads to the SO$(6)_+$ sector. Alternatively, the U(3) sector can be further truncated by eliminating the St\"uckelberg scalar $a$, all vectors and the two-form $B^0$, thus retaining the neutral scalars $\phi$, $\varphi$, $\chi$ along with $B_1$, $C^0$, $C^1$. This truncation corresponds to the model considered in \cite{Guarino:2015jca}.

\subsection{Critical points}  
\label{subsec:CriticalPoints}

We now study the vacua of ISO$(7)_c$ supergravity with at least SU(3) invariance by analysing the critical points of the scalar potential (\ref{VSU3}). Only for the dyonic gauging $c \neq 0$, {\it i.e.}, $g m \neq 0$, does this potential have critical points or, rather, critical $\textrm{SO}(1,1) \times \textrm{U}(1)$ loci. These are the surfaces in the scalar manifold (\ref{ScalManN=2}) for which the gradient of the potential (\ref{VSU3}) vanishes. These are parameterised by the St\"uckelberg scalar $a$ and the phase $\beta$ introduced in (\ref{polar_coord}), and occur at the fixed values of the remaining scalars recorded in table~\ref{Table:SU3Points}. Abusing language, we will often refer to these critical loci simply as critical points or extrema. We have determined the residual supersymmetry and bosonic symmetry of these points within the full $\cN=8$ ISO$(7)_c$ theory. We have also calculated their scalar and vector mass spectra, both within the SU(3) sector and within the full $\cN=8$ theory. See tables \ref{Table:SU3Points} below and \ref{Table:SU3&SO4Points} in the introduction for a summary. 

All critical points in this sector are AdS. Three of them are supersymmetric: there is one point with $\cN=2$ supersymmetry and  $\textrm{SU}(3) \times \textrm{U}(1)$ bosonic symmetry that was already announced in \cite{Guarino:2015jca}; one point with $\cN=1$ supersymmetry and G$_2$ symmetry, already found in \cite{Borghese:2012qm} using the method of \cite{Dibitetto:2011gm,DallAgata:2011aa}; and we find a new point with $\cN=1$ supersymmetry and SU(3) bosonic symmetry. In addition, we find five non-supersymmetric points. Three of them were previously known, as they had already been found with the method of \cite{Dibitetto:2011gm,DallAgata:2011aa}: these are the points with SO$(7)_+$ and SO$(6)_+$ residual symmetry \cite{DallAgata:2011aa} and the point with G$_2$ symmetry \cite{Borghese:2012qm}. In addition, we numerically find two new non-supersymmetric points with SU(3) symmetry. We have appended a subscript $+$ to the SO(7) and SO(6) points to indicate that they are supported by proper (parity even) scalars, rather than (parity odd) pseudoscalars, of E$_{7(7)}/$SU(8); in fact, they are supported by dilatons only, see table \ref{Table:SU3Points}. This is also consistent with the discussion in section \ref{sec:subsecsSU3}.

\begin{table}[t!]
\renewcommand{\arraystretch}{1.5}
\scalebox{0.80}{
\begin{tabular}{cc|cccc|cc}
\noalign{\hrule height 1pt}
$\mathcal{N}$ & $\textrm{G}_{0}$ & $c^{-1/3} \, \chi$ & $c^{-1/3} \, e^{-\varphi}$ & $c^{-1/3} \, \rho$ & $c^{-1/3} \,  e^{-\phi}$ & $ g^{-2} \, c^{1/3} \, V_{0}$ & $M^2 L^2$ \\
\noalign{\hrule height 1pt}
$\mathcal{N}=2$ & $\textrm{U}(3)$ & $-\frac{1}{2}$ & $\frac{3^{1/2}}{2}$ & $0$ & $ \frac{1}{2^{1/2}}  $ & $-2^2 \, 3^{3/2} $ & $3 \pm \sqrt{17} \,,\, 2 \, , \, 2  \, , \, 2  \, , \, 0$ \\[-3mm]
&  &  &  &  &   & &  $4 \, , \,0$ \\
$\mathcal{N}=1$ & $\textrm{G}_{2}$ & $-\frac{1}{2^{7/3}} $ & $\frac{5^{1/2} \, 3^{1/2}}{2^{7/3}}$ & $-\frac{1}{2^{7/3}} $ & $\frac{5^{1/2} \, 3^{1/2}}{2^{7/3}}$ & $- \frac{2^{28/3} \, 3^{1/2}}{5^{5/2}} $ & $4 \pm \sqrt{6} \, ,\, -\tfrac{1}{6} (11\pm \sqrt{6}) \, ,\, 0 \, ,\, 0$ \\[-2mm]
&  &  &  &  &   & &  $\tfrac{1}{2}(3 \pm \sqrt{6}) $ \\
$\mathcal{N}=1$ & $\textrm{SU}(3)$ & $\frac{1}{2^2}$ & $\frac{3^{1/2} \,  5^{1/2}}{2^2}$ & $-\frac{3^{1/2}}{2^2}$ & $\frac{ 5^{1/2}}{2^2}$ & $-\frac{2^{8} \, 3^{3/2}}{5^{5/2}} $ & $4 \pm \sqrt{6} \,,\, 4 \pm \sqrt{6} \, , \, 0  \, , \, 0 $ \\[-3mm]
&  &  &  &  &   & &  $2 \, , \,6$ \\
\hline
$\mathcal{N}=0$ & $\textrm{SO}(7)_+$ & $0$ & $\frac{1}{5^{1/6}}$ & $0$ & $\frac{1}{5^{1/6}}$ & $-3 \, 5^{7/6}$ & $6 \,,\, -\tfrac{12}{5} \,,\, -\tfrac{6}{5} \,,\, -\tfrac{6}{5} \,,\, -\tfrac{6}{5} \,,\, 0$ \\[-2mm]
&  &  &  &  &   & &  $\tfrac{12}{5} \, , \,0$ \\
$\mathcal{N}=0$ & $\textrm{SO}(6)_+$ & $0$ & $2^{1/6}$ & $0$ & $ \frac{1}{2^{5/6}}  $ & $- 3 \, 2^{17/6}$ & $6 \,,\, 6 \,,\, -\tfrac{3}{4} \,,\, -\tfrac{3}{4} \,,\, 0 \,,\, 0$ \\[-3mm]
&  &  &  &  &   & &  $6 \, , \,0$ \\
$\mathcal{N}=0$ & $\textrm{G}_{2}$ & $\frac{1}{2^{4/3}}$ & $\frac{3^{1/2}}{2^{4/3}}$ & $\frac{1}{2^{4/3}}$ & $\frac{3^{1/2}}{2^{4/3}}$ & $-\frac{2^{16/3}}{3^{1/2}}$ & $6 \,,\, 6 \,,\, -1 \,,\, -1 \,,\, 0 \,,\, 0$ \\[-3mm]
&  &  &  &  &   & &  $3 \, , \,3$ \\
$\mathcal{N}=0$ & $\textrm{SU}(3)$ & $0.455$ & $0.838$ & $0.335$ & $0.601$ & $-23.457$ & $6.214 \,,\, 5.925 \,,\, 1.145 \,,\, -1.284 \,,\, 0 \,,\, 0$ \\[-3mm]
&  &  &  &  &   & &  $4.677 \, , \, 2.136$ \\
$\mathcal{N}=0$ & $\textrm{SU}(3)$ & $0.270$ & $0.733$ & $0.491$ & $0.662$ & $-23.414$ & $6.230 \,,\, 5.905 \,,\, 1.130 \,,\, -1.264 \,,\, 0 \,,\, 0$ \\[-3mm]
&  &  &  &  &   & &  $4.373 \, , \, 2.490$ \\
\noalign{\hrule height 1pt}
\end{tabular}
}
\caption{All critical points of $\cN=8$ ISO(7)-dyonically-gauged supergravity with at least SU(3) invariance. For each point we give the residual supersymmetry and bosonic symmetry within the full $\cN=8$ theory, its location, the cosmological constant and the scalar (upper) and vector (lower) masses within the SU(3) sector.}
\label{Table:SU3Points}
\end{table}

All these critical points disappear in the limits $\,c \rightarrow 0\,$ ({\it i.e.}, $m \rightarrow 0$, $g \neq 0$) and $\,c \rightarrow \infty\,$ ({\it i.e.}, $g \rightarrow 0$, $m \neq 0$), corresponding to the purely electric and purely magnetic gaugings, respectively. For the purely electric ISO(7) gauging \cite{Hull:1984yy}, in particular, we can extend the claims against critical points with at least SO$(7)_+$ symmetry \cite{Hull:1984yy} and at least G$_2$ symmetry \cite{Ahn:2002qga}: the electrically gauged ISO(7) theory does not have any critical point with at least SU(3) symmetry. In section \ref{sec:SO(4)-sector} we show that the electric gauging has no critical points with residual symmetry containing the particular SO(4) considered there. In fact, critical points in the purely electric ISO(7) gauging can be completely ruled out as follows. By an argument in \cite{DallAgata:2011aa,Dall'Agata:2014ita}, these would necessarily be Minkowski. Then, these vacua would necessarily arise from $S^6$ compactification of (massless) IIA, but this is not possible by the Maldacena-N\'u\~nez no-go theorem \cite{Maldacena:2000mw}.

It is also interesting to compare with the critical points in the SU(3)-invariant sector of the SO(8) supergravity, both electric and dyonic. The points ($\textrm{SU}(3) \times \textrm{U}(1)$, $\cN=2$), (G$_2$, $\cN=1$), and \mbox{($\textrm{SO}(7)_+$, $\cN=0$)} have direct analogs, both in the purely electric \cite{Warner:1983vz}  and the dyonic  \cite{Borghese:2012zs} SO(8) gauging. The electric SO(8) gauging also possesses a non-supersymmetric point with symmetry $\textrm{SU}(4) \sim \textrm{SO}(6)$, but in that case it is an $\textrm{SU}(4)_{-}$ point (supported by pseudoscalars), while in the dyonic ISO(7) gauging it is an SO$(6)_+$ point, as we have already noted. The \mbox{($\textrm{SO}(7)_-$, $\cN=0$)} point of the electric SO(8) gauging does not have a counterpart in the dyonic ISO(7) gauging. As discussed in \cite{deWit:2013ija} (see also \cite{Borghese:2012zs}), the $\textrm{SO}(7)_\pm$ critical points of the electric, $c =0$, SO(8) gauging become $\textrm{SO}(7)_\mp$ points at the other endpoint of the interval of the continuous, in that case, parameter $c$. A similar transition occurs for the $\textrm{SU}(4)_{-}$ point of the electric SO(8) gauging. For the dyonic ISO(7) gauging, we find that these points stay $\textrm{SO}(7)_+$ and $\textrm{SO}(6)_+$ for all non-vanishing values of the dyonically gauging parameter $c$. This is consistent with the fact that all $c \neq 0$ values are physically equivalent \cite{Dall'Agata:2014ita}. Other points in table~\ref{Table:SU3Points} have no analog in the purely electric SO(8) gauging, but do have counterparts  for dyonic SO(8). These include the (SU(3), $\cN=1$), (G$_2$, $\cN=0$) and the two (SU(3), $\cN=0$) points. Of course, the maximally supersymmetric SO(8) point of SO(8)$_c$ supergravity does not have an analog for the ISO$(7)_c$ gauging. 

We have also computed the scalar and vector masses for these critical points: see table \ref{Table:SU3Points} for the mass spectrum within the SU(3)-invariant sector and table \ref{Table:SU3&SO4Points} in the introduction for the masses within the full $\cN=8$ theory. The masses do not run with $c$, as expected, for any point. Except for the non-supersymmetric SU(3) points to be dealt with below, critical points in the ISO$(7)_c$ gauging that have analogs with the same symmetry and supersymmetry in the SO$(8)_c$ gauging, have the same mass spectra in both gaugings. This has already been noticed for the previously known points \cite{DallAgata:2011aa,Borghese:2012qm}. This happens regardless of whether those extrema are supported by scalars or pseudoscalars in either gauging: for example, the SO$(6)_+$ point of the ISO$(7)_c$ gauging and the SU$(4)_-$ point of the SO(8) gauging have the same spectrum. Such matching is possible because the masses for these points in the SO$(8)_c$ gauging do not run with $c$ either, in spite of the fact that $c$ is continous in that case. 

The situation is slightly different for the two new non-supersymmetric SU(3) points of the ISO$(7)_c$ gauging, since they have counterparts in the SO$(8)_c$ gauging whose masses do run with $c$ \cite{Borghese:2012zs}. For the first of the $\cN=0$ SU(3) points in the ISO$(7)_c$ gauging, we numerically find the following scalar masses, normalised to the radius $L$ of AdS, within the full $\cN=8$ theory 
\begin{equation}
\label{new_SU3_1_spectrum}
\begin{array}{ccrrrrrrrr}
M^2 \, L^2 &=& 6.214 \,\,(\times 1) & , &  5.925 \,\,(\times 1) & , & 1.145 \,\,(\times 1) & , & -1.284 \,\,(\times 1) & , \\
& & -1.707 \,\,(\times 12) & , & -0.860 \,\,(\times 12) & , & -1.623 \,\,(\times 8) & , & -0.159 \,\,(\times 8) & , \\
& & -1.061 \,\,(\times 6) & , & 0 \,\,(\times 20) & ,
\end{array}
\end{equation}
and vector masses
\begin{equation}
\label{new_SU3_1_spectrum_vec}
\begin{array}{ccrrrrrrrr}
M^2 \, L^2 &=& 4.677 \,\,(\times 1) & , &  2.136 \,\,(\times 1) & , & 3.184 \,\,(\times 6) & , & 2.715 \,\,(\times 6) & , \\
& & 0.150 \,\,(\times 6) & , & 0 \,\,(\times 8) & .
\end{array}
\end{equation}
The second $\cN=0$ SU(3) point in the ISO$(7)_c$ gauging has scalar masses
\begin{equation}
\label{new_SU3_2_spectrum}
\begin{array}{ccrrrrrrrr}
M^2 \, L^2 &=& 6.230 \,\,(\times 1) & , &  5.905 \,\,(\times 1) & , & 1.130 \,\,(\times 1) & , & -1.264 \,\,(\times 1) & , \\
& & -1.582 \,\,(\times 12) & , & -0.954 \,\,(\times 12) & , & -1.396 \,\,(\times 8) & , & -0.309 \,\,(\times 8) & , \\
& & -1.082 \,\,(\times 6) & , & 0 \,\,(\times 20) & ,
\end{array}
\end{equation}
and vector masses
\begin{equation}
\label{new_SU3_2_spectrum_vec}
\begin{array}{ccrrrrrrrr}
M^2 \, L^2 &=& 4.373 \,\,(\times 1) & , &  2.490 \,\,(\times 1) & , & 3.200 \,\,(\times 6) & , & 2.791 \,\,(\times 6) & , \\
& & 0.111 \,\,(\times 6) & , & 0 \,\,(\times 8) & .
\end{array}
\end{equation}
The singlets in these equations (together with two zeroes in the scalar spectra) correspond to the spectra within the SU(3) sector. The spectra (\ref{new_SU3_1_spectrum})--(\ref{new_SU3_2_spectrum_vec}) are, of course, independent of $c$. Given that these SU(3) points have counterparts in the SO$(8)_c$ gauging with $c$-dependent spectra, one may ask whether there exists a $c$ such that the spectra of the SU(3) points of that precise SO$(8)_c$ gauging coincide with the ISO$(7)_c$ spectra (\ref{new_SU3_1_spectrum})--(\ref{new_SU3_2_spectrum_vec}). If such $c$ existed, and assuming that the masses would not change in the limit, the ISO$(7)_c$ gauging could be thought of as a contraction of that particular SO$(8)_c$ gauging, like the electric, $c=0$, ISO(7) gauging \cite{Hull:1984yy} is of the electric, $c=0$, SO(8) gauging \cite{deWit:1982ig}. It turns out that such $c$ does not exist: the masses of the SU(3) points in the SO$(8)_c$ gauging do approach the values  (\ref{new_SU3_1_spectrum})--(\ref{new_SU3_2_spectrum_vec}) in the purely electric limit $c \rightarrow 0$ for which these points become unphysical.

Finally note that, among the non-supersymmetric points, only the G$_2$ point and the new SU(3) points are stable within the full $\cN=8$ theory: all of its scalar masses are above the BF bound, $\,M^2L^2 \ge -9/4\,$. Note also that the number $n_v$ of zero masses in each vector spectrum in table \ref{Table:SU3&SO4Points} corresponds to the dimension of the residual symmetry group, as it must. Denoting by $n_s$ the number of zero masses in each scalar spectrum, for all critical points except ($\cN=1$, SU(3)) and ($\cN=0$, SO$(6)_+$) it happens that $n_v + n_s = 28 \equiv$ the total number of (electric) vectors, so that all these zero-mass scalars are actually Goldstone bosons. For ($\cN=1$, SU(3)) and ($\cN=0$, SO$(6)_+$), instead, $n_v + n_s = 36$ and $n_v + n_s = 43$, respectively, so these points have 8 and 15 physical scalars of mass zero.

\section{An $\mathcal{N}=1$ truncation: the G$_2$-invariant sector}
\label{sec:G2-sector}

In section \ref{sec:subsecsSU3} we discussed how the G$_2$-invariant sector of ISO$(7)_c$ supergravity can be recovered from the SU(3) sector. Here, we give an independent characterisation of the G$_2$ sector based on the embedding
\begin{equation} \label{eqG2inSO7}
 \textrm{SO}(7) \,\, \supset \,\, \textrm{G}_{2} \ ,
\end{equation}
without first descending from SO(7) to SU(3) and then enlarging again to G$_2$. The embedding (\ref{eqG2inSO7}) is compatible with a branching $\textbf{8} \rightarrow \textbf{1} + \textbf{7}$ of the fundamental of SL(8). In terms of indices, we have $\,A \rightarrow I \oplus 8\,$ with $\,I=1, \ldots ,7$. The same branching holds for the fundamental of SU(8), ensuring that the G$_2$-invariant  sector is $\cN=1$.

\subsection{Construction and bosonic Lagrangian}
\label{sec:G2-sectorLag}

The G$_2$-invariant fields in the ISO(7) restricted duality hierarchy (\ref{eq:SL7fieldcontent4D}) include, in agreement with table \ref{Table:group_theory},  the metric $\,g_{\mu\nu}\,$, two scalars $\varphi, \chi $ and a three-form potential $C$ with four-form field strength $H_\4 = dC$. The three-form is embedded into the $\bm{28}^\prime$ three-forms $\cC^{IJ}$ of the full $\cN=8$ theory as $\cC^{IJ} = C \, \delta^{IJ}$. This sector does not contain vectors or two-forms. Only the metric and scalars enter the G$_2$-invariant Lagrangian, see (\ref{L_G2}) below.

We can find the explicit embedding of the two G$_2$-invariant scalars $\varphi, \chi $ into the coset representative $\cV$ and scalar matrix $\cM$, as we did in section \ref{subsec:construction} for the scalars of the SU(3) sector. We first identify the following G$_2$-invariant combinations of generators (\ref{Gener63}), (\ref{Gener70}) of E$_{7(7)}$,
\begin{equation}
\label{GenerG2_tunc}
\begin{array}{cclc}
g_{1} &=&  {t_{1}}^{1} + {t_{2}}^{2} + {t_{3}}^{3} + {t_{4}}^{4} + {t_{5}}^{5} + {t_{6}}^{6} + {t_{7}}^{7}  - 7 \, {t_{8}}^{8} & , \\[2mm]
g_{2} &=&  g^{(-)}_{2} + g^{(+)}_{2} = (  t_{4567} + t_{6723} + t_{2345} - t_{1357}  + t_{1346} + t_{1562} + t_{1724}  ) \\
&  &  \phantom{g^{(-)}_{4} + g^{(+)}_{4} }   \,+  
( t_{1238} + t_{1458} + t_{1678} - t_{2468} + t_{2578} + t_{4738} + t_{6358} )   & ,
\end{array}
\end{equation}
and exponentiate the Cartan $g_{1}$ and positive root $ g^{(+)}_{2}$ into a coset representative (in the SL(8) basis, see footnote \ref{FootnoteSL8})
\begin{equation}
\mathcal{V}  = e^{-12 \, \chi \, g^{(+)}_{2}} \, e^{\frac{1}{4} \, \varphi \, g_{1}} \ .
\end{equation}
The resulting scalar-dependent matrix $\,\mathcal{M}=\mathcal{V} \, \mathcal{V}^{t}\,$ is manifestly G$_2$-invariant, as its components take values along the invariant metric $\delta_{IJ}$, the associative three-form $\psi_{IJK}$ and the co-associative four-form $\tilde{\psi}_{IJKL}$ of G$_2$. See appendix \ref{app:G2sectorScalMat} for the explicit expressions. 

The bosonic Lagrangian of the G$_2$-invariant sector follows by restricting the $\cN=8$ Lagrangian (\ref{BosLag}) accordingly. In the scalar parameterisation that we are using, it reads
\begin{equation}
\label{L_G2}
\mathcal{L}  = (R - V) \, \textrm{vol}_4  + \tfrac{7}{2} \left[ d\varphi \wedge * d\varphi + e^{2 \varphi} \, d\chi \wedge * d\chi \right]  \ ,
\end{equation}
where the scalar potential simplifies from (\ref{V_generalRewrite}) to
\begin{equation} 
\label{VG2}
V = \tfrac72 \, g^2 \, e^\varphi \big( 1+ e^{2\varphi} \chi^2 \big)^2 \big( -5 + 7 e^{2\varphi} \chi^2  \big) - 7gm \,e^{7\varphi} \chi^3  + \tfrac12 m^2 e^{7\varphi}  \  .
\end{equation}
Note that the scalar kinetic terms in (\ref{L_G2}) and potential (\ref{VG2}) respectively agree with the restriction of the SU(3)-invariant kinetic terms in (\ref{L_full_SU3}) and potential (\ref{VSU3}) to the surface (\ref{G2_identification}).

The $\cN=8$ Bianchi identities (\ref{eq:TruncBianchis}) and most duality relations (\ref{H2IJDuality})--(\ref{H4Duality}) become trivial upon G$_2$-invariant truncation. The only non-trivial duality relation is that of ${H_\4 = dC}$,  coming from (\ref{H4Duality}):
\begin{equation} \label{H4DualityG2}
H_{\4} = \big[ g \, e^{\varphi } \, \big(1+ e^{2 \varphi }\, \chi ^2 \big)^2 \,  \big(5 - 7 \, e^{2 \varphi }\, \chi ^2 \big)  + m \, e^{7 \varphi } \, \chi ^3   \big] \, \textrm{vol}_4 \ .
\end{equation}
For completeness, we also record the G$_2$-invariant truncation of the duality relation (\ref{H4Duality_singlet}) for the field strength $\,\tilde{H}_{\4}\equiv \tilde{\cH}_{\4}\,$ of the singlet three-form $\,\tilde{C} \equiv \tilde{\cC}\,$ dual to the magnetic component of the embedding tensor:
\begin{equation} \label{H4MagDualityG2}
\tilde{H}_{\4} = \big[ 7 \, g  \, e^{7 \varphi } \, \chi ^3 - m \, e^{7 \varphi }  \big] \, \textrm{vol}_4 \ .
\end{equation}
From (\ref{H4DualityG2}), (\ref{H4MagDualityG2}), it is straightforward to check that these four-form field strengths and the scalar potential (\ref{VG2}) are related through
\begin{equation}
7 \, g \, H_{\4} + m \, \tilde{H}_{\4} = -2 \, V \, \textrm{vol}_{4} \ .
\end{equation}
This corresponds to the G$_2$-invariant truncation of the $\cN=8$ ISO(7) expression (\ref{H4/Potential2}). It also agrees with the restriction of the SU(3)-invariant duality relation (\ref{DualitySU3H4V}) to the surface (\ref{G2_identification}).

\subsection{Canonical $\mathcal{N}=1$ formulation}

The G$_2$-invariant sector corresponds to $\cN=1$ supergravity coupled to a chiral multiplet. The two real scalars $\varphi$, $\chi$ parameterise the upper-half plane via the first relation in (\ref{mappingN1Sugra}). For notational agreement with other $\cN=1$ sectors with chiral multiplets to be discussed in section \ref{sec:SO(4)-sector} and appendix \ref{app:Z2xSO(3)-sector}, this chiral field is denoted now by $\Phi_{1}$:
\begin{equation}
\label{def_Phi}
\Phi_{1} = - \chi + i \, e^{-\varphi} \ .
\end{equation}
In terms of this, the scalar kinetic terms in (\ref{L_G2}) can be re-expressed as 
\begin{equation} \label{kinG2can}
\tfrac{1}{2} \, \mathcal{L}^{\textrm{kin}}_{\textrm{scalar}} =  - 7 \, \frac{d \Phi_{1} \wedge *  d \bar{\Phi}_{1} }{( \Phi_{1} -\bar{\Phi}_{1})^2}  \ ,
\end{equation}
and can be seen to derive from the K\"ahler potential
\begin{equation}
\label{KahlerPhi}
K=  - 7 \log (-i (\Phi_{1}-\bar{\Phi}_{1}))   \ .
\end{equation}
Finally, the scalar potential (\ref{VG2}) is exactly recovered from the holomorphic superpotential
\begin{equation}
\label{WG2}
W =  14 \, g \, \Phi_{1}^3 + 2  \, m   \ ,
\end{equation}
using the canonical $\cN=1$ expression
\begin{equation}
\label{VN1}
\tfrac{1}{4} \, V = e^{K} \left[ K^{\Phi_{p} \bar{\Phi}_{q}} (D_{\Phi_{p}} W) (D_{\bar{\Phi}_{q}} \overline{W}) - 3 \, W \, \overline{W} \right] \ ,
\end{equation}
with $p=1$, $q=1$. Here, $K^{\Phi_{p} \bar{\Phi}_{q}}$ is the inverse of the K\"ahler metric $\,K_{\Phi_{p} \bar{\Phi}_{q}}=\partial_{\Phi_{p}}\partial_{\bar{\Phi}_{q}}K\,$ in (\ref{kinG2can}) and we have used the K\"ahler derivative $\,D_{\Phi_{p}} W \equiv \partial_{\Phi_{p}} W + (\partial_{\Phi_{p}} K )W$.

\subsection{Critical points}
\label{sec:CritPointsG2}

For $gm \neq 0$, the scalar potential (\ref{VG2}) contains three critical points: the two G$_2$ points with $\cN=1$ and $\cN=0$ and the $\cN=0$ SO$(7)_+$ point. See table \ref{Table:SU3Points} for their location in $(\varphi, \chi)$ space and table \ref{Table:SU3&SO4Points} for their spectra within the full $\cN=8$ ISO$(7)_c$ theory.

\section{An $\cN=1$ truncation: $\textrm{SO}(4)$-invariant sector}
\label{sec:SO(4)-sector}

We close the main body of the paper with a different $\cN=1$ truncation of the $\cN=8$ ISO$(7)_c$ theory: one that retains two chiral multiplets and is invariant under an SO(4) subgroup of ISO(7) embedded into the latter through the elaborate chain
\begin{equation}
\label{embedding_SO4}
\textrm{SO}(7) \, \supset \, 
 \textrm{G}_{2} \, \supset \, 
  \textrm{SO}(3)^\prime \times \textrm{SO}(4)^\prime \,  \supset \, 
   \textrm{SO}(3)_{d} \times \textrm{SO}(3)_{R}  \, \equiv \, 
   \textrm{SO}(4) \ ,
\end{equation}
with $\textrm{SO}(4)^\prime \equiv \textrm{SO}(3)_{L} \times \textrm{SO}(3)_{R}$ and $ \textrm{SO}(3)_{d}$ the diagonal subgroup of $\textrm{SO}(3)^\prime \times \textrm{SO}(3)_{L}$. The fundamental of SL(8) branches under $ \textrm{SO}(4) \equiv  \textrm{SO}(3)_{d} \times \textrm{SO}(3)_{R}$ as
\begin{equation} 
\label{8underSO4}
\textbf{8} \rightarrow (\textbf{2},\textbf{2}) + (\textbf{3},\textbf{1}) +  (\textbf{1},\textbf{1}) \ ,
\end{equation}
or, in terms of indices, $A \rightarrow  \lambda  \,\oplus \, a \, \oplus\, 8$ with $\lambda=1,3,5,7$ and $a=2,4,6$. The fundamental of SU(8) branches as in (\ref{8underSO4}) as well, and the presence of the singlet $(\textbf{1},\textbf{1})$ is responsible for the $\cN=1$ supersymmetry of this truncation. Intricate though it is, the embedding (\ref{embedding_SO4}) is very interesting: as shown in \cite{Gallerati:2014xra} using the approach of \cite{Dibitetto:2011gm,DallAgata:2011aa}, the SO$(8)_c$, SO$(7,1)_c$ and ISO$(7)_c$ $\cN=8$ gaugings have a critical point with $\cN=3$ supersymmetry and an SO(4) bosonic symmetry group that is embedded into the gauge groups as in (\ref{embedding_SO4}). Here we will give an explicit parameterisation of the sector of ISO$(7)_c$ supergravity invariant under this SO(4) and will recover the supersymmetric point along with other extrema.  See \cite{Pang:2015mra} for a recent study of this sector in SO(8)$_c$-gauged supergravity \cite{Dall'Agata:2012bb}.

\subsection{Construction and bosonic Lagrangian} 
\label{sec:SO4-sectorLag}

According to the branchings under (\ref{embedding_SO4}) recorded in table \ref{Table:group_theory}, the SO(4)-invariant truncation of the duality hierarchy (\ref{eq:SL7fieldcontent4D}) gives rise, besides the metric $\,g_{\mu \nu}\,$, to four real scalars, $(\chi, \varphi)$, $(\rho,\phi)$, one two-form coming from $\cB_I{}^J$ and two three-forms. No vectors or two-forms coming from $\cB^I$ survive the truncation. Thus, the bosonic Lagrangian of this sector contains only the metric and the four real scalars. We will focus on these fields, and will not discuss further the duality hierarchy in this sector.

The four real scalars parameterise two copies of the upper-half plane SU$(1,1)/$U(1) embedded in E$_{7(7)}/$SU(8). Like we did for the other invariant truncations, we can obtain an explicit parameterisation for the scalar geometry in this sector by exponentiating the combinations of E$_{7(7)}$ generators (\ref{Gener63}), (\ref{Gener70}) that are invariant under the SO(4) in (\ref{embedding_SO4}).  These are determined by the invariant tensors of this SO(4) (see appendix \ref{app:SO4sectorScalMat}), and can be taken as
\begin{equation}
\label{GenerSO4_tunc}
\begin{array}{cclc}
g_{1} &=&  {t_{2}}^{2} + {t_{4}}^{4} + {t_{6}}^{6} - 3 \,  {t_{8}}^{8} & , \\[2mm]

g_{2} &=&  {t_{1}}^{1} + {t_{3}}^{3} + {t_{5}}^{5} + {t_{7}}^{7} - {t_{2}}^{2} - {t_{4}}^{4} - {t_{6}}^{6} - {t_{8}}^{8} & , \\[2mm]

g_{3} &=&  g^{(-)}_{3} + g^{(+)}_{3} =  ( t_{3571} ) + ( t_{8246} ) & ,  \\[2mm]

g_{4} &=&  g^{(-)}_{4} + g^{(+)}_{4} = ( t_{4613}-t_{4657}+t_{6215}+t_{6237}+t_{2417}-t_{2435}  ) \\
&  &  \phantom{g^{(-)}_{4} + g^{(+)}_{4} }   \,+  
( t_{2578}-t_{2138}-t_{4378}-t_{4158}+t_{6358}-t_{6178} )   & .
\end{array}
\end{equation}
A coset representative on each copy of SU$(1,1)/$U(1) can then be built as
\begin{equation}
\mathcal{V}_{1}  = e^{-12 \, \chi \, g^{(+)}_{4}} \, e^{\frac{1}{2} \, \varphi \, g_{1}} 
\hspace{8mm} \textrm{ and } \hspace{8mm}
\mathcal{V}_{2}  = e^{-12 \, \rho \,  g^{(+)}_{3}} \, e^{\frac{1}{4} \, \phi \, g_{2}}  \ .
\end{equation}
Finally, the total coset representative in this sector is $\cV = \cV_1 \cV_2$, and the scalar-dependent matrix $\mathcal{M}$ is $\,\mathcal{M}=\mathcal{V} \, \mathcal{V}^{t}\,$. See appendix \ref{app:SO4sectorScalMat} for its explicit expression. 

Using this scalar parameterisation, the bosonic Lagrangian of this SO(4)-invariant sector follows from (\ref{BosLag}),
\begin{equation}
\label{L_SO4}
\mathcal{L}  = (R - V) \, \textrm{vol}_4  + \tfrac{6}{2} \left[ d\varphi \wedge * d\varphi + e^{2 \varphi} \, d\chi \wedge * d\chi \right] + \tfrac{1}{2} \left[ d\phi \wedge * d\phi + e^{2 \phi} \, d\rho \wedge * d\rho \right]  \ ,
\end{equation}
where the scalar potential (\ref{V_generalRewrite}) now reduces to 
\begin{equation}
\label{VSO4}
\begin{array}{lll}
V &=& \frac{1}{2} \, g^{2}  \, e^{-\phi } (1+e^{2 \varphi } \chi ^2)  
\left[ -24 \, e^{\varphi +\phi } - 8 \, e^{2 \phi }   +  e^{2 \varphi } \, \Big(-3+  (8 \chi ^2-3 \rho ^2) \, e^{2 \phi } \Big) \right. \\[4mm]
&+& \left.  e^{4 \varphi }  \, \chi ^2 \, \Big(  9 +  (3 \rho +4 \chi )^2  \, e^{2 \phi }   \Big)   \right]  
-     g m \, \chi ^2 \, (3 \rho + 4 \chi ) \,  e^{6 \varphi +\phi }
+  \frac{1}{2} \, m^2 \, e^{6 \varphi +\phi }  \ .
\end{array}
\end{equation}
Note that this potential depends on all four scalars in the SO(4) sector. According to the branching (\ref{embedding_SO4}), the G$_2$-invariant sector is a further subsector of the present SO(4) sector. Indeed, under the identifications
\begin{equation} \label{eq:SO4toG2}
\varphi = \phi
\hspace{8mm} \textrm{ and } \hspace{8mm}
\chi = \rho \ ,
\end{equation}
the Lagrangian (\ref{L_SO4}), (\ref{VSO4}) reduces to the G$_2$-invariant Lagrangian (\ref{L_G2}), (\ref{VG2}).

\subsection{Canonical $\mathcal{N}=1$ formulation}

We will now show that this SO(4) sector corresponds to $\cN=1$ supergravity coupled to two chiral multiplets, by casting the Lagrangian (\ref{L_SO4}), (\ref{VSO4}) in canonical $\cN=1$ form. In order to do this, we introduce the complex combinations
\begin{equation}
\label{def_Phi12}
\Phi_{1} = -\chi + i \, e^{-\varphi}
\hspace{10mm} \textrm{ and } \hspace{10mm}
\Phi_{2} =  -\rho + i \, e^{-\phi} \ 
\end{equation}
on each copy of $\,\textrm{SU}(1,1)/\textrm{U}(1)\,$. In terms of the fields (\ref{def_Phi12}), the kinetic terms in (\ref{L_SO4}) take the form
\begin{equation}
\tfrac{1}{2} \, \mathcal{L}^{\textrm{kin}}_{\textrm{scalar}} =  - 6 \, \frac{d \Phi_{1} \wedge *  d \bar{\Phi}_{1} }{( \Phi_{1} -\bar{\Phi}_{1})^2}   - \frac{d \Phi_{2}  \wedge  * d \bar{\Phi}_{2} }{( \Phi_{2} -\bar{\Phi}_{2})^2} \ ,
\end{equation}
and derive from the K\"ahler potential
\begin{equation}
\label{Kahler12}
K=  - 6 \log (-i (\Phi_{1}-\bar{\Phi}_{1}))   - \log (-i (\Phi_{2}-\bar{\Phi}_{2}))   \ .
\end{equation}
The scalar potential (\ref{VSO4}) is reproduced from the holomorphic superpotential
\begin{equation}
\label{WSO4}
W= g \, (8 \, \Phi_{1}^3 +6 \, \Phi_{1}^{2} \, \Phi_{2} ) + 2  \, m   \ ,
\end{equation}
through the canonical $\,\mathcal{N}=1\,$  expression (\ref{VN1}), now with $p=1,2$, $q=1,2$.  The simplicity of the SO(4)-invariant superpotential (\ref{WSO4}) is remarkable, given the laboured embedding (\ref{embedding_SO4}) of this SO(4) in SO(7). In comparison, the G$_2$-invariant superpotential (\ref{WG2}) is of similar simplicity, but the embedding of G$_2$ in SO(7) is straightforward. Note that the SO(4)-invariant superpotential (\ref{WSO4}) reduces to the G$_2$-invariant (\ref{WG2}) on the surface (\ref{eq:SO4toG2}), namely, when $\Phi_1 = \Phi_2$.

\subsection{Critical points}
\label{sec:CritPointSO4}

\begin{center}
\begin{table}[t]
%\begin{table}
\renewcommand{\arraystretch}{1.5}
\scalebox{0.82}{
\begin{tabular}{cc|cccc|cc}
\noalign{\hrule height 1pt}
$\mathcal{N}$ & $\textrm{G}_{0}$ & $c^{-1/3} \, \chi$ & $c^{-1/3} \, e^{-\varphi}$ & $c^{-1/3} \, \rho$ & $c^{-1/3} \,  e^{-\phi}$ & $g^{-2} \, c^{1/3} \, V_{0}$ & $M^2 L^2$ \\
\noalign{\hrule height 1pt}
$\mathcal{N}=3$ & $\textrm{SO}(4)$ & $\frac{1}{2^{4/3}}$ & $\frac{3^{1/2}}{2^{4/3}} $ & $-\frac{1}{2^{1/3}}$ & $\frac{3^{1/2}}{2^{1/3}}$ & $-\frac{2^{16/3}}{3^{1/2}} $ & $3 \,  (1 \pm \sqrt{3}) \,,\, (1 \pm \sqrt{3})$ \\
\hline
$\mathcal{N}=0$ & $\textrm{SO}(4)$ & $0.412$ & $0.651$ & $0.068$ & $1.147 $ & $-23.513$ & $6.727 \,,\, 5.287 \,,\, 0.584 \,,\, -1.586$ \\
\noalign{\hrule height 1pt}
\end{tabular}
}
\caption{Critical points of $\cN=8$ ISO(7)-dyonically-gauged supergravity with invariance equal or larger than the SO(4) subgroup of SO(7) considered in (\ref{embedding_SO4}). This list also includes the points, not shown in the table, in the G$_2$-invariant sector. 
For each point we give the residual supersymmetry and bosonic symmetry within the full $\cN=8$ theory, its location, the cosmological constant and the scalar masses within the SO(4) sector.}
\label{Table:SO4Points}
\end{table}
\end{center}

\vspace{-20pt}
 
\noindent The scalar potential (\ref{VSO4}) contains five critical points when $g m \neq 0 $, all of them AdS. See table \ref{Table:SO4Points} for a summary. Three of them occur on the surface (\ref{eq:SO4toG2}), and thus correspond to the three critical points in the G$_2$-invariant sector, see section \ref{sec:CritPointsG2}. In addition, we find two more extrema, both with symmetry SO(4). Curiously enough, both points are non-supersymmetric within this SO(4)-invariant sector but, when embedded into the full $\cN=8$ theory, one point becomes $\cN=3$ and the other one stays $\cN=0$. The reason for this peculiar behaviour of the $\cN=3$ point is that the three gravitini of the full $\cN=8$ theory that remain `massless' ({\it i.e.}, of mass $ML =1$ on the AdS vacuum) in the solution are not singlets under $ \textrm{SO}(4) \equiv  \textrm{SO}(3)_{d} \times \textrm{SO}(3)_{R}$. They instead transform as $ (\textbf{3},\textbf{1})$ and are thus  truncated out of the SO(4)-invariant sector. In more detail, the $\bm{\overline{8}}$  gravitini of the $\cN=8$ ISO$(7)_c$ theory split under the SO(4) under consideration as in (the conjugate of) (\ref{8underSO4}). The $ \textrm{SO}(4) \equiv  \textrm{SO}(3)_{d} \times \textrm{SO}(3)_{R}$-invariant sector only retains the singlet $(\textbf{1},\textbf{1})$ gravitino, while the $ (\textbf{2},\textbf{2}) + (\textbf{3},\textbf{1}) $ gravitini are truncated out. Now, this $(\textbf{1},\textbf{1})$ gravitino becomes massive ({\it i.e.}, of mass $ML >1$) at both SO(4) critical points of the scalar potential (\ref{VSO4}), thus leading to complete supersymmetry breaking within this sector for both points. An alternative way to see this is that the superpotential  (\ref{WSO4}) leads to non-vanishing F-terms, $\,D_{\Phi_{p}} W \neq 0$, for both solutions. Then, we consider these points within the full $\cN=8$ theory and analyse the mass matrix containing all $ (\textbf{2},\textbf{2}) + (\textbf{3},\textbf{1}) +  (\textbf{1},\textbf{1})$ gravitini. For one of these points we find that the $ (\textbf{2},\textbf{2})$ gravitini also become massive, but the $ (\textbf{3},\textbf{1})$ remain `massless'. This renders this point $\cN=3$ within the full $\cN=8$ theory. For the other point all gravitini become massive, giving $\cN=0$.

The $\cN=3$ point has recently been found in \cite{Gallerati:2014xra} using the method of \cite{Dibitetto:2011gm,DallAgata:2011aa}. The location of this point in scalar space given in table \ref{Table:SO4Points} above, relative to the parameterisation of section \ref{sec:SO4-sectorLag}, is new. We have also computed the scalar and vector masses about this point within the full $\cN=8$ theory, and have brought the result to table \ref{Table:SU3&SO4Points} in the introduction. Our result agrees with the spectrum reported in \cite{Gallerati:2014xra}. See that reference for the allocation of the spectrum into OSp$(4|3)$ supermultiplets. Intriguingly, the values of the potential at the ($\cN=3$, SO(4)) point and at the ($\cN=0$, G$_2$) point coincide.

We have determined numerically the position and spectrum of the non-supersymmetric SO(4) point within the full $\cN=8$ theory. The scalar masses, relative to the radius $L$ of AdS, read
\begin{equation}
\label{new_SO4_spectrum}
\begin{array}{ccrrrrrrrr}
M^2 \, L^2 &=& 6.727 \,\,(\times 1) & , &  5.287 \,\,(\times 1) & , & 0.584 \,\,(\times 1) & , & -1.586 \,\,(\times 1) & , \\
& &  -1.588  \,\,(\times 9) & , & -1.751 \,\,(\times 9) & , & 0.630 \,\,(\times 5) & , & -0.983 \,\,(\times 5) & , \\
& &  -0.730  \,\,(\times 4) & , & -1.964 \,\,(\times 4) & , & -1.176 \,\,(\times 8) & , & 0 \,\,(\times 22) & ,
\end{array}
\end{equation}
while the vector masses are 
\begin{equation}
\label{new_SO4_spectrum_vec}
\begin{array}{ccrrrrrrrr}
M^2 \, L^2 &=& 4.153 \,\,(\times 3) & , &  2.287 \,\,(\times 3) & , & 3.451 \,\,(\times 4) & , & 1.945 \,\,(\times 4) & , \\
& & 0.191 \,\,(\times 8) & , & 0 \,\,(\times 6) & .
\end{array}
\end{equation}
Note, here and for the $\cN=3$ point, the six zero masses in the vector spectrum corresponding to the six generators of the unbroken SO(4). The scalar masses are all above the BF bound, thus ensuring stability against perturbations in the full $\cN=8$ supergravity. Neither SO(4) point features flat directions. Their spectra are independent of $c$, as they must, and the points disappear from the physical scalar space in the purely electric, $c \rightarrow 0$ \cite{{Hull:1984yy}}, and purely magnetic limits.  A counterpart in the SO$(8)_c$ gauging of the non-supersymmetric SO(4) point has recently appeared in \cite{Pang:2015mra}.

%%%%%%%%%%%%%%%%%%%%%%%%%%%%%%%%%%%%
%
% Acknowledgments
%
%%%%%%%%%%%%%%%%%%%%%%%%%%%%%%%%%%%%

\section*{Acknowledgments}

We thank Daniel Jafferis for collaboration in related projects and \mbox{Franz Ciceri,} Bernard de Wit and Gianluca Inverso for discussions. AG is supported in part by the ERC Advanced Grant no. 246974, {\it Supersymmetry: a window to non-perturbative physics}.  OV is supported by the Marie Curie fellowship PIOF-GA-2012-328798, managed from the CPHT of \'Ecole Polytechnique, and partially by the Fundamental Laws Initiative at Harvard.

%%%%%%%%%%%%%%%%%%%%%%%%%%%%%%%%%%%%
%
% Appendix
%
%%%%%%%%%%%%%%%%%%%%%%%%%%%%%%%%%%%%

%\newpage

\appendix

\addtocontents{toc}{\setcounter{tocdepth}{1}}

\section{The $S^6$ as a non-geometric $T^6$}
\label{app:Z2xSO(3)-sector}

Here we analyse an $\cN=1$, $\mathbb{Z}_{2} \times \mathrm{SO}(3)$-invariant sector of $\cN=8$ ISO(7)$_{c}$ supergravity and relate it to the toroidal, non-geometric type IIA orientifold reductions of \cite{Shelton:2005cf,Aldazabal:2006up,Dibitetto:2011gm,Dibitetto:2012ia}.

\subsection{An $\,\cN=1\,$ truncation: the $\mathbb{Z}_{2} \times \mathrm{SO}(3)$-invariant sector}

The embedding of the SO(3) factor reads
\begin{equation} \label{eqSO3inSO7}
\textrm{SO}(7) \supset \textrm{SO}(6) \sim \textrm{SU}(4) \supset \textrm{SU}(3) \supset \textrm{SO}(3) \ ,
\end{equation}
while $\,\mathbb{Z}_{2}\,$ acts on the fundamental of SL(8) as
\begin{equation}
\label{Z2-symmetry}
\mathbb{Z}_{2} \,\, : \hspace{5mm} (\, 1 \,;\, 3 \,,\, 5 \,,\, 7 \,\, ; \,\, 2 \,,\, 4 \,,\, 6 \,;\, 8 \,) \,\, \rightarrow \,\, (\, -1 \,;\, -3 \,,\, -5 \,,\, -7 \,\, ; \,\, 2 \,,\, 4 \,,\, 6 \,;\, 8 \,) \ 
\end{equation}
This $\,\mathbb{Z}_{2}\,$ can be used to truncate $\,\cN=8 \rightarrow \cN=4\,$ \cite{Dibitetto:2011eu}. Taking (\ref{eqSO3inSO7}), (\ref{Z2-symmetry}) together, the fundamental of SU(8) branches under $\mathbb{Z}_{2} \times \mathrm{SO}(3)$ as $\textbf{8} \rightarrow \textbf{1}_{(-)} + \textbf{3}_{(-)} + \textbf{3}_{(+)} + \textbf{1}_{(+)}$. The truncation to the singlet sector is $\cN=1$, given the $\mathbb{Z}_{2}$-even singlet $\,\textbf{1}_{(+)}\,$ in this decomposition. This invariant sector keeps six real scalars $(\chi_{1},\varphi_{1})$, $(\chi_{2},\varphi_{2})$ and $(\chi_{3},\varphi_{3})$ along with the metric $\,g_{\mu\nu}$. We will not discuss the duality hierarchy in this sector; we only note that this $\mathbb{Z}_{2} \times \mathrm{SO}(3)$-invariant truncation does not retain vectors.

The six scalars can be grouped up into complex fields $\,\Phi_{1,2,3}\,$ taking values on three copies of the upper-half plane:
\begin{equation}
\label{Phi_SO(3)}
\Phi_{1} = -\chi_{1} + i \, e^{-\varphi_{1}}
\hspace{5mm} , \hspace{5mm}
\Phi_{2} = -\chi_{2} + i \, e^{-\varphi_{2}}
\hspace{5mm} , \hspace{5mm}
\Phi_{3} = -\chi_{3} + i \, e^{-\varphi_{3}} \ ,
\end{equation}
These scalars thus describe an $\,[\textrm{SU}(1,1)/\textrm{U}(1)]^3\,$ K\"ahler submanifold of E$_{7(7)}/$SU(8). The Lagrangian in this invariant sector can be explicitly worked out by first identifying the relevant $\mathbb{Z}_{2} \times \mathrm{SO}(3)$-invariant generators of E$_{7(7)}$,
\begin{equation}
\label{Gener_Z2xSO3}
\begin{array}{cclc}
g_{1} &=&  \phantom{-}  {t_{3}}^{3} + {t_{5}}^{5} + {t_{7}}^{7}  +  {t_{2}}^{2} + {t_{4}}^{4} + {t_{6}}^{6} - 3 \, (  {t_{1}}^{1} +  {t_{8}}^{8} )& , \\[2mm]

g_{2} &=&  {t_{1}}^{1} + {t_{3}}^{3} + {t_{5}}^{5} + {t_{7}}^{7} - {t_{2}}^{2} - {t_{4}}^{4} - {t_{6}}^{6} - {t_{8}}^{8} & , \\[2mm]

g_{3} &=&   -  {t_{3}}^{3} - {t_{5}}^{5} - {t_{7}}^{7}  +  {t_{2}}^{2} + {t_{4}}^{4} + {t_{6}}^{6} + 3 \,  ({t_{1}}^{1} -  {t_{8}}^{8})  & , \\[2mm]

g_{4} &=& g^{(-)}_{4} + g^{(+)}_{4} = (t_{4567} + t_{6723} + t_{2345}) +  (t_{1238} + t_{1458} + t_{1678})  &  , \\[2mm]

g_{5} &=& g^{(-)}_{5} + g^{(+)}_{5} =  (t_{3571}) + (t_{8246})  & ,  \\[2mm]

g_{6} &=& g^{(-)}_{6} + g^{(+)}_{6} =  (t_{1346} + t_{1562} + t_{1724}) + (t_{2578} + t_{4738} + t_{6358})  & ,

\end{array}
\end{equation}
and then exponentiating the Cartan generators and positive roots into a coset representative $\,{\mathcal{V}=\mathcal{V}_{1} \mathcal{V}_{2} \mathcal{V}_{3}}\,$, with
\begin{equation}
\label{SO(3)_exponentiation}
\begin{array}{lllllllll}
\mathcal{V}_{1} = e^{-12 \, \chi_{1} \, g^{(+)}_{4}} \, e^{\frac{1}{4} \, \varphi_{1} \, g_{1}} & , &
\mathcal{V}_{2} = e^{-12 \, \chi_{2} \, g^{(+)}_{5}} \, e^{\frac{1}{4} \, \varphi_{2} \, g_{2}} & , &
\mathcal{V}_{3} = e^{-12 \, \chi_{3} \, g^{(+)}_{6}} \, e^{\frac{1}{4} \, \varphi_{3} \, g_{3}} & .
\end{array}
\end{equation}
Plugging the resulting scalar-dependent matrix $\,\mathcal{M}=\mathcal{V} \, \mathcal{V}^{t}\,$ into (\ref{BosLag}), (\ref{V_generalRewrite}) gives rise to the bosonic Lagrangian
\begin{equation}
\label{L_Z2SO3}
\begin{array}{lll}
\mathcal{L} &=& (R - V) \, \textrm{vol}_4  + \tfrac{3}{2} \left[ d\varphi_{1} \wedge * d\varphi_{1} + e^{2 \varphi_{1}} \, d\chi_{1} \wedge * d\chi_{1} \right]    
\\[2mm] 
& + & \tfrac{1}{2} \left[ d\varphi_{2} \wedge * d\varphi_{2} + e^{2 \varphi_{2}} \, d\chi_{2} \wedge * d\chi_{2} \right] + \tfrac{3}{2} \left[ d\varphi_{3} \wedge * d\varphi_{3} + e^{2 \varphi_{3}} \, d\chi_{3} \wedge * d\chi_{3} \right] \ ,
\end{array}
\end{equation}
where the lengthy scalar potential
\begin{equation}
\label{VSO3}
\begin{array}{llll}
V &=& \frac{1}{2} \, g^2 e^{-3 \varphi _1-\varphi _2-\varphi _3} \Big[   3 \, e^{4 \varphi _1+2 \varphi _2} (e^{2 \varphi _1} \chi _1^2-1) - 6 \,  e^{4 \varphi _1+\varphi _2+\varphi _3} (e^{2 \varphi _1} \chi _1^2+3)  \\[2mm]
 & +&  3 \, e^{2 (\varphi _1+\varphi _3)} \Big( e^{2 \varphi _1} (e^{2 \varphi _1} \chi _1^2-1)  +e^{2  \varphi _2} \big(e^{4 \varphi _1} \chi _1^2 (\chi _2+2 \chi _3)^2-e^{2 \varphi _1} (2 \chi _1^2+\chi_2^2)-2\big) \Big)   \\[2mm]
 & +&  e^{4 \varphi _3} \big[  3 \, e^{2 \left(\varphi _1+\varphi _2\right)} \chi _1^2+e^{6 \varphi _1} \chi _1^2 \, \Big(e^{2
   \varphi _2} \, \big(\chi _1^2 + 3 \, (\chi _2+\chi _3) \, \chi _3 \big)^2+9 \chi _3^2\Big)   \\[2mm]
 & &    +\, 3 \, e^{4 \varphi _1} \, \Big(e^{2 \varphi _2} \, \big(\chi _1^2+\chi _3 (\chi _2+\chi _3)\big)^2+\chi _3^2\Big)+e^{2 \varphi _2}\big]  \\[2mm] 
 & -&   6 \, e^{2 \varphi _1+\varphi _2+3 \varphi _3} \big(   e^{2 \varphi _1}  ( \chi_1^2+2 \chi _3^2)+1 \big)   \Big] \\[2mm]
 &-& g m \, e^{3 \varphi _1+\varphi _2+3 \varphi _3}  \, \chi _1 \, \big(\chi _1^2+3 \, \chi _3 \, (\chi _2+\chi _3)\big) + \frac{1}{2} \, m^2 \, e^{3 \varphi _1+\varphi _2+3 \varphi _3} \ 
\end{array}
\end{equation}
depends on the six real scalars of the truncated theory. 

The SU(3), SO(4) and G$_{2}$ sectors described in the main text can be recovered as subtruncations of the $\mathbb{Z}_{2} \times \mathrm{SO}(3)$ sector. These are obtained through the identifications
\begin{equation}
\label{Field_Identifications}
\begin{array}{llllll}
\textrm{SU}(3) \textrm{ sector} &  : & \hspace{2mm} \Phi_{1} = -\chi + i \, e^{-\varphi} & , &  \Phi_{2} = \Phi_{3} = -\rho + i \, e^{-\phi} & , \\[2mm]
\textrm{SO}(4) \textrm{ sector} & : & \hspace{2mm} \Phi_{1} = \Phi_{3} = -\chi + i \, e^{-\varphi} & , &  \Phi_{2} = -\rho + i \, e^{-\phi}  & , \\[2mm]
\textrm{G}_{2} \textrm{ sector} & : & \hspace{2mm}   \Phi_{1} = \Phi_{2} =\Phi_{3} = -\chi + i \, e^{-\varphi} &  .
\end{array}
\end{equation}
The scalar potential (\ref{VSO3}) reduces on each of these three submanifolds of $[\textrm{SU}(1,1)/\textrm{U}(1)]^3$ to the scalar potentials  (\ref{VSU3}), (\ref{VSO4}) and (\ref{VG2}) of the SU(3), SO(4) and G$_2$ invariant sectors. Recovering the SU(3)-invariant scalar potential requires use of the definition (\ref{polar_coord}).

\subsection{Critical points}

All the critical points in the SU(3), SO(4) and G$_2$ sectors are also extrema of the $\mathbb{Z}_{2} \times \mathrm{SO}(3)$-invariant potential (\ref{VSO3}). In addition, a casual numerical scan yields further non-supersymmetric AdS critical points with SO(3) residual symmetry. For example, a critical point occurs at
\begin{equation}
c^{-1/3} \, \Phi_{1} = -0.554 + 0.492 \, i
\hspace{5mm} , \hspace{5mm}
c^{-1/3} \,\Phi_{2} = 0.375 \, i
\hspace{5mm} \textrm{ and } \hspace{5mm}
c^{-1/3} \,\Phi_{3} = 1.263 \, i \ ,
\end{equation}
with cosmological constant $\,g^{-2} \, c^{1/3} \, V_{0}=-27.610\,$ and scalar masses in this sector
\begin{equation}
M^2L^2 = ( \, 7.379 \,,\, 4.040 \,,\, 3.790 \,,\, -3.323 \,,\, -1.873 \,,\, -0.269 \,) \ ,
\end{equation}
normalised to the AdS radius $L$. Note the presence of an unstable mode with mass below the BF bound, $\,M^2L^2 \ge -9/4\,$.

\subsection{Canonical $\mathcal{N}=1$ formulation}

The Lagrangian (\ref{L_Z2SO3}), (\ref{VSO3}) of the  $\mathbb{Z}_{2} \times \mathrm{SO}(3)$-invariant sector of $\cN=8$ ISO$(7)_c$ supergravity can be cast in $\,\mathcal{N}=1\,$ canonical form. The relevant K\"ahler potential and superpotential are
\begin{equation}
\label{N=1SO(3)}
\begin{array}{llll}
K & = &  - 3 \, \log (-i (\Phi_{1}-\bar{\Phi}_{1})) - \log (-i (\Phi_{2}-\bar{\Phi}_{2})) - 3 \, \log (-i (\Phi_{3}-\bar{\Phi}_{3}))     & \ , \\[2mm]
W & = & g \, \left(  2 \,  \Phi_{1}^3 + 6 \, \Phi_{1}  \,   \Phi_{3}^2 + 6 \, \Phi_{1}  \,  \Phi_{2} \, \Phi_{3} \right) + 2 \, m  & \ ,
\end{array}
\end{equation}
 which give rise to the kinetic terms in (\ref{L_Z2SO3}), and to the scalar potential in (\ref{VSO3}) through the standard formula (\ref{VN1}) with $\,p=1,2,3\,$, $\,q=1,2,3\,$. The simplicity of the superpotential (\ref{N=1SO(3)}) is again in contrast with the intricacy of the scalar potential (\ref{VSO3}). By solving the F-flat conditions, $\,D_{\Phi_{p}} W=0\,$, that follow from (\ref{N=1SO(3)}), one (only) recovers the supersymmetric critical points in Table~\ref{Table:SU3Points}. The $\cN=3$ SO(4) critical point is invisible to this superpotential for reasons similar to those discussed in section \ref{sec:CritPointSO4}, but is of course an extremum of the potential (\ref{VSO3}).

\subsection{A non-geometric STU-model from $\textrm{ISO}(7)_{c}$ supergravity}

The $\,\mathcal{N}=1\,$ rewrite in (\ref{N=1SO(3)}) uncovers a connection to the non-geometric type IIA backgrounds based on toroidal $\,\mathbb{T}^{6}/(\mathbb{Z}_{2} \times \mathbb{Z}_{2})\,$ orientifold reductions  investigated in \cite{Aldazabal:2006up,Dibitetto:2012ia}. These $\mathcal{N}=1$ models have an [SU(1,1)/U(1)]$^{7}$ scalar manifold parameterised by seven complex fields $\, ( \, S \,,\, T_{1} \,,\, T_{2} \,,\, T_{3} \,,\, U_{1} \,,\, U_{2} \,,\, U_{3} \, )\,$. The moduli  $\,S\,$, $\,T_{1,2,3}\,$ and  $\,U_{1,2,3}\,$ respectively correspond to the type IIA axiodilaton, complex structure and K\"ahler moduli in the compactification.  In order to relate the non-geometric type IIA orientifold models of \cite{Aldazabal:2006up,Dibitetto:2012ia} to the $\,\mathcal{N}=1\,$ theory in (\ref{N=1SO(3)}), we further restrict to the subset of models enjoying an SO(3) \textit{plane exchange} symmetry in $\,{\mathbb{T}^{6} = \mathbb{T}^{2} \otimes \mathbb{T}^{2} \otimes \mathbb{T}^{2}\,}$. These have been referred to as \textit{isotropic} or STU-models in the literature \cite{Derendinger:2004jn}. In these STU-models the scalar manifold is reduced to [SU$(1,1)/$U(1)]$^{3}$ via the identifications $\,T\equiv T_{1}=T_{2}=T_{3}\,$ and $\,U\equiv U_{1}=U_{2}=U_{3}\,$. This results in a simplified K\"ahler potential
\begin{equation}
\label{Kahler_STU}
K_{\textrm{IIA}}=  - 3 \log (-i (U-\bar U)) - \log (-i (S-\bar S)) - 3 \log (-i (T-\bar T))    \ .
\end{equation}
This is formally the same as (\ref{N=1SO(3)}), but the fields are not yet directly identified (see below).

On the other hand, the most general flux-induced superpotential in toroidal orientifold reductions receives three types of contributions: from regular  fluxes of the type IIA form fields, from metric fluxes (if $\,\mathbb{T}^{6}\,$ is twisted) and, finally, from so-called \textit{non-geometric} fluxes. The existence of the latter has been conjectured by duality arguments strongly based on the symmetries of the straight $\,\mathbb{T}^{6}\,$ reduction \cite{Shelton:2005cf,Aldazabal:2006up,Dibitetto:2011gm}. Based on such arguments, the non-geometric fluxes are switched on directly in the four-dimensional superpotential: no reduction has been known so far that explicitly produces them from type IIA.

Now we will show that the  $\mathbb{Z}_{2} \times \mathrm{SO}(3)$-invariant sector of $\cN=8$ ISO$(7)_c$ supergravity, described by the $\cN=1$ quantities (\ref{N=1SO(3)}) corresponds, precisely, to one such non-geometric STU-model. In order to see this, we first map the scalars $\Phi_{1,2,3}$ to the scalars $S,T,U$ as $\,\Phi_{1}=-1/U\,$, $\,\Phi_{2}=S\,$ and $\,\Phi_{3}=T\,$. Plugging these identifications into (\ref{N=1SO(3)}) produces a non-standard K\"ahler potential due to the presence of $\,-U^{-1}\,$ instead of $\,U\,$. This can be taken to a standard form via a modular transformation $\,U \rightarrow -U^{-1}$. After this transformation, the K\"ahler potential  and superpotential in (\ref{N=1SO(3)}) are respectively mapped to (\ref{Kahler_STU}) and
\begin{equation}
\label{WIIA}
W_{\textrm{IIA}} = - g \, \left(  2  + 6 \,  T^2 \, U^2 + 6 \,  S \, T \, U^2 \right) + 2 \, m \, U^3   \ .
\end{equation}
This is a fluxed-induced superpotential of the type we have just reviewed. Following the flux/superpotential-couplings dictionary of \cite{Dibitetto:2011gm,Dibitetto:2012ia}, we can determine the type IIA flux-origin of each term in (\ref{WIIA}). The constant term $-2g$ descends from a regular $\hat F_{(6)}$ flux. In other words, it arises from a Freund-Rubin contribution for $\hat F_{(4)}$. The cubic coupling of $U$ is generated by the Romans mass $\hat F_\0$. From this perspective, the term $2 \, m \, U^3$ is in perfect agreement with \cite{Guarino:2015jca}, where the Romans mass $\hat F_\0$ was identified upon reduction with the magnetic coupling $m$ of dyonic ISO$(7)$ supergravity. Finally, quartic terms, like $\,T^2 U^2\,$ and $\,STU^2\,$ are of non-geometric nature in this language.

Dyonic $\cN=8$ ISO(7) supergravity and, in particular, the $\cN=1$ subsector that we are considering here, arises as a (consistent) reduction of massive type IIA on the six-sphere \cite{Guarino:2015jca,Guarino:2015vca}. Thus, this particular non-geometric model does in fact enjoy a perfectly geometric type IIA origin. It would be interesting to investigate more generally the conditions that allow for a conventional geometric interpretation of non-geometric flux reductions.

\section{The SU(3) sector and M-theory on Sasaki-Einstein} 
\label{app:MTheoryonSE7}

In this appendix we comment on the relation between the SU(3)-invariant sector of the $\cN=8$ ISO(7)$_{c}$ theory that we analysed in section \ref{sec:SU(3)-sector}, and the model of \cite{Gauntlett:2009zw}, which arises from consistent truncation of $D=11$ supergravity on any Sasaki-Einstein seven-manifold to the modes that are SU(4)-invariant under the Sasaki-Einstein SU(4)-structure. Both theories have the same field content, the same scalar manifold (\ref{ScalManN=2}) and the same  gauge group,  $\textrm{U}(1) \times \textrm{SO}(1,1)$, generated by the same hypermultiplet Killing vectors (\ref{KVs}). In both theories, U(1) is gauged electrically only and SO$(1,1)$ dyonically in their natural duality frames. Yet the theories are different: they have different scalar potentials, with different critical points. Also, they have mutually incompatible higher-dimensional origins in massive type IIA and M-theory, respectively. 

The theories turn out to differ in their embedding tensors and, in particular, in the allocation of electric and magnetic charges with respect to a common electric/magnetic duality frame. In order to see this, we first need to express both theories in the same symplectic frame. The $\textrm{Sp}(4,\mathbb{R})$ rotation
\begin{eqnarray} 
\label{SimplRottoSE7}
S^M{}_N = \left(\begin{array}{cccc}
0 	&	0	&	-1	&	0 \\
0 	&	-1	&	0	&	0 \\
1 	&	0	&	0	&	0 \\
0 	&	0	&	0	&	-1 \\
\end{array} \right)
\hspace{5mm} \textrm{ with } \hspace{5mm}
\textrm{det} \, S = 1
\hspace{5mm} \textrm{ and } \hspace{5mm}
S^{\textrm{T}} \,  \Omega \, S = \Omega \ ,
\end{eqnarray}
where $\Omega$ is given in (\ref{KahlerPot}), brings the sections $\,\hat{X}^M = (1, \tau, \tau^3, -3\tau^2)\,$ of \cite{Gauntlett:2009zw}, associated to the cubic prepotential $\,{\hat{\mathcal{F}}=-(\hat{X}^{1})^3/\hat{X}^{0}}\,$,  to the sections $\,X^{M}\,$ in (\ref{holosections}) compatible with a prepotential $\,\mathcal{F}=-2\,\sqrt{ X^{0} \, (X^{1})^3}\,$, namely, $S^M{}_N \, \hat{X}^N = X^M$. No hats were used in \cite{Gauntlett:2009zw} and the scalars $\,t = -\chi + i e^{-\varphi}\,$ here and $\,\tau = h + i e^{2U+V}\,$ there are simply identified as $\,t=\tau\,$. The symplectic rotation (\ref{SimplRottoSE7}) thus brings the theory of \cite{Gauntlett:2009zw} from the ``hatted" duality frame  to the duality frame that we are considering here for our SU(3)-invariant sector. The embedding tensor of the theory \cite{Gauntlett:2009zw} transformed into the new, common frame, \textit{i.e.} $\,(S^{-1})_M{}^N \, \hat{\Theta}_N{}^\alpha \,$, turns out to be purely magnetic. It thus differs from our dyonic $\,\Theta_M{}^\alpha \,$ in (\ref{N=2Theta}).

\section{Construction of the $\,\mathcal{N}=8\,$ $\,\textrm{ISO}(7)\,$ dyonic theory}
\label{app:ConstN=8}

In this appendix we build the family of symplectically deformed $\textrm{ISO}(7)_{c} = \textrm{SO}(7) \ltimes \mathbb{R}^{7}_{c}$ maximal gauged supergravities using the framework of the embedding tensor \cite{deWit:2007mt}. Following the same mnemonic as in \cite{Dall'Agata:2012bb}, we denote this family  ISO(7)$_{c}$ where $\,c\,$ is the electric/magnetic or symplectic deformation parameter. Importantly, when moving results to the main text, we have adopted differential form notation and rescaled the metric and the tensor fields as
\begin{equation}
g_{\mu \nu}^{\textrm{(here)}} =  2 \, g_{\mu \nu}^{\textrm{(text)}}  
\hspace{10mm} \textrm {and}  \hspace{10mm}  
\cB_{\mu \nu \, \alpha}^{\textrm{(here)}} =  2 \, \cB_{\mu \nu \, \alpha}^{\textrm{(text)}} \ .
\end{equation}
Then, the Einstein-Hilbert term, the  kinetic terms for the scalars and the scalar potential are rescaled accordingly
\begin{equation}
\mathcal{L}^{\textrm{(here)}}_{\textrm{EH}} = \tfrac{1}{2} \,  \mathcal{L}^{\textrm{(text)}}_{\textrm{EH}}
\hspace{7mm} \textrm {,}  \hspace{7mm}  
\mathcal{L}^{\textrm{kin (here)}}_{\textrm{scalar}} = \tfrac{1}{2} \, \mathcal{L}^{\textrm{kin (text)}}_{\textrm{scalar}}
\hspace{7mm} \textrm {and}  \hspace{7mm}  
V^{\textrm{(here)}} = \tfrac{1}{4} \, \, V^{\textrm{(text)}} \ .
\end{equation}

\subsection{$\textrm{E}_{7(7)}$ duality and the embedding tensor ${\Theta_{\mathbb{M}}}^{\alpha}$}

Let us start by introducing the generators of the U-duality group $\,\textrm{E}_{7(7)}\,$ of maximal supergravity in four dimensions. These are denoted $\,{[t_{\alpha}]_{\mathbb{M}}}^{\mathbb{N}}\,$ where $\,\alpha=1, \ldots,133\,$ is an adjoint index and $\,\mathbb{M}=1, \ldots , 56\,$ is a fundamental index of E$_{7(7)}$. We will use the real SL(8) basis of E$_{7(7)}$ to build the $\,56 \times 56\,$ generators $\,{[t_{\alpha}]_{\mathbb{M}}}^{\mathbb{N}}\,$. In this basis, the decomposition $\,{\textbf{56} \rightarrow \textbf{28} + \textbf{28}'}\,$ makes manifest the electric and magnetic components of an arbitrary vector $\,X_{\mathbb{M}}\,$ and translates into the index splitting $\,X_\mathbb{M} \rightarrow X_{[AB]} \oplus  \, X^{[AB]}\,$, where $\,{A=1, \ldots ,8}\,$ denotes a fundamental SL(8) index. The $\textrm{E}_{7(7)}$ generators consequentely split as $\,{t_\alpha = {t_{A}}^{B} \,\oplus \, t_{ABCD}}\,$, with $\,{t_{A}}^{A}=0\,$ and $\,t_{ABCD}=t_{[ABCD]}\,$, and correspond to a branching $\,{\textbf{133} \rightarrow \textbf{63} + \textbf{70}}\,$ under SL(8). Their matrix entries are given by\footnote{The generalised Kronecker symbols are taken to be normalised as projectors, \textit{i.e.}, $\delta_{A_{1} \ldots A_{p}}^{B_{1} \ldots B_{p}}=\pm \frac{1}{p!} \textrm{ or } 0$.}
\begin{equation}
\label{Gener63}
{[{t_{A}}^{B}]_{[CD]}}^{[EF]} = 4 \, \left( \delta_{[C}^{B} \, \delta_{D]A}^{EF} + \frac{1}{8} \, \delta_{A}^{B} \, \delta_{CD}^{EF}  \right)
\hspace{5mm} \textrm{ and } \hspace{5mm} 
{[{t_{A}}^{B}]^{[EF]}}_{[CD]} = - {[{t_{A}}^{B}]_{[CD]}}^{[EF]} \ ,
\end{equation}
for those in the $\textbf{63}$ (block-diagonal matrices) representing SL(8) generators and by
\begin{equation}
\label{Gener70}
[t_{ABCD}]_{[EF][GH]} = \frac{2}{4!} \, \epsilon_{ABCDEFGH}
\hspace{5mm} \textrm{ and } \hspace{5mm} 
[t_{ABCD}]^{[EF][GH]} = 2 \, \delta_{ABCD}^{EFGH} \ ,
\end{equation}
for those generators in the $\textbf{70}$ (off-block-diagonal matrices) completing to $\,\textrm{E}_{7(7)}\,$.

The most general gauging of a $28$-dimensional group $\,\textrm{G} \subset \textrm{SL}(8)\subset \textrm{E}_{7(7)}\,$ in maximal supergravity is encoded within an \textit{embedding tensor} ${\Theta_{\mathbb{M}}}^{\alpha}$ of the form \cite{DallAgata:2011aa}
\begin{equation}
\label{Theta_Def}
\Theta_{[AB]\,\,\,\,D}^{\phantom{[AB]}C}= 2 \, \delta_{[A}^{C} \, \theta_{B]D} 
\hspace{10mm} , \hspace{10mm} 
 {\Theta^{[AB]C}}_{D}  = 2 \, \delta^{[A}_{D} \,  \xi^{B]C} \ ,
\end{equation}
where the index $\,\alpha\,$ in $\,{\Theta_{\mathbb{M}}}^{\alpha}\,$ is restricted to the adjoint of SL(8), namely, to the generators in (\ref{Gener63}). The matrices $\,\theta\,$ and $\,\xi\,$ are symmetric and specify the gauging $\,\textrm{G}\,$ as a function of the number of negative, positive and vanishing eigenvalues. The $\,\Theta$-tensor obeys the (quadratic) constraints for a consistent gauging in maximal supergravity \cite{deWit:2007mt}
\begin{equation}
\label{QC}
\Omega^{\mathbb{MN}} \, {\Theta_{\mathbb{M}}}^{\alpha} \, {\Theta_{\mathbb{N}}}^{\beta} = 0
\hspace{10mm} \textrm{with} \hspace{10mm} 
\Omega^{\mathbb{MN}}=\left(\begin{array}{cc} 0_{28} & \mathbb{I}_{28} \\ - \mathbb{I}_{28} & 0_{28} \end{array}\right) \ ,
\end{equation}
where $\,\Omega_{\mathbb{MN}}\,$ is the Sp($56,\mathbb{R}$)-invariant matrix satisfying $\,\Omega_{\mathbb{MP}}\,\Omega^{\mathbb{MQ}}=\delta_{\mathbb{P}}^{\mathbb{Q}}\,$.

Using the form of the SL(8) generators in (\ref{Gener63}), it is possible to build an $X$-tensor\footnote{The $X$-tensor is usually decomposed as $\,{X_{\mathbb{M}\mathbb{N}}}^{\mathbb{P}}={X_{[\mathbb{M}\mathbb{N}]}}^{\mathbb{P}} + {Z^{\mathbb{P}}}_{\mathbb{M}\mathbb{N}}\,$ with 
\begin{equation}
{{Z^{\mathbb{P}}}_{\mathbb{M}\mathbb{N}} =  Z^{\mathbb{P} , \alpha} \, d_{\alpha \, \mathbb{MN}}={X_{(\mathbb{M}\mathbb{N})}}^{\mathbb{P}}\,} \ ,
\end{equation}
where $\,Z^{\mathbb{P} \, , \alpha} = \tfrac{1}{2} \, \Omega^{\mathbb{PQ}} \, {\Theta_{\mathbb{Q}}}^{\alpha}\,$ and $\,d_{\alpha \, \mathbb{MN}} \equiv {[t_{\alpha}]_{\mathbb{M}}}^{\mathbb{P}} \, \Omega_{\mathbb{NP}}\,$. The $Z$-tensor plays an important role in the tensor hierarchy of maximal supergravity \cite{deWit:2005hv,deWit:2007mt,deWit:2008ta}.}  
\begin{equation}
\label{ET}
{X_{\mathbb{M}\mathbb{N}}}^{\mathbb{P}}={\Theta_{\mathbb{M}}}^{\alpha}\,{[t_{\alpha}]_{\mathbb{N}}}^{\mathbb{P}}={{\Theta_{\mathbb{M}}}^{C}}_{D}\,{[{t_{C}}^{D}]_{\mathbb{N}}}^{\mathbb{P}} \ ,
\end{equation}
that consists of both electric $X_{[AB]}$ and magnetic $X^{[AB]}$ components often referred to as \textit{charges}. The former are given by
\begin{equation}
\label{X_electric}
{X_{[AB] [CD]}}^{[EF]} = - X_{[AB] \phantom{[EF]} [CD]}^{\phantom{[AB]} [EF]} = -8 \, \delta_{[A}^{[E} \theta_{B][C} \delta_{D]}^{F]}  \ ,
\end{equation}
whereas the latter read
\begin{equation}
\label{X_magnetic}
X^{[AB] \phantom{[CD]}[EF]}_{\phantom{[AB]}[CD]} = - {X^{[AB] [EF]}}_{[CD]} = -8 \, \delta_{[C}^{[A} \xi^{B][E} \delta_{D]}^{F]} \ .
\end{equation}
As we will see later, having magnetic charges (\ref{X_magnetic}), \textit{i.e.} $\xi \neq 0$, requires not only the introduction of magnetic vector fields $\tilde{\cA}_{[AB] \,\mu}$ in the Lagrangian but also of two-form tensor fields $\,\cB_{\mu \nu \, \alpha}\,$ in order to obtain a consistent gauge algebra \cite{deWit:2005ub}.

\subsection{Dyonic $\,\textrm{ISO}(7)_{c}\,$ gaugings}

In order to describe the family of ISO(7)$_{c}$ gaugings, it proves natural to split the index $\,{A=(I,8)}\,$ with $\,I=1, \ldots ,7\,$. The generators of $\,\textrm{ISO}(7)=\textrm{SO}(7) \ltimes \mathbb{R}^{7}\subset \textrm{SL}(8)\,$ are given by 28 linear combinations of the block-diagonal generators ${[{t_{A}}^{B}]_{\mathbb{M}}}^{\mathbb{N}}$ in (\ref{Gener63}). These are
\begin{equation}
\label{linear_comb_t}
\begin{array}{cclllcccc}
T_{\textrm{SO}(7)} &:&  T_{IJ} \, \equiv\,  2 \, {t_{[I}}^{K} \delta_{J]K}  & \hspace{10mm} \textrm{and} \hspace{10mm} &
T_{\mathbb{R}^{7}} &:&  T_{I\phantom{8}} \, \equiv\, {t_{8}}^{J} \, \delta_{JI} & ,
\end{array}
\end{equation}
comprising SO(7) generators $\,T_{IJ}=T_{[IJ]}\,$ in the $\,\textbf{21}\,$ of SO(7) plus $\,\mathbb{R}^{7}\,$ generators $\,T_{I}\,$ in the $\,\textbf{7}\,$ of SO(7). They satisfy the standard commutation relations
\begin{equation}
\label{ISO7_brackets}
\begin{array}{rlll}
[T_{IJ} , T_{KL}] &=& 4 \, \delta_{[I[K} \, T_{L]J]} & , \\[2mm]
[T_{I} , T_{KL}] &=& 2 \,  \delta_{I[K} \, T_{L]} & , \\[2mm]
[T_{I} , T_{K}] &=& 0   & ,
\end{array}
\end{equation}
which specify the structure constants of ISO(7). The completion to SL(8) requires additional generators $\,T^{(S)}_{IJ} \equiv 2 \, {t_{(I}}^{L} \delta_{J)L}\,$ and $\,T^{\perp}_{I} \equiv\,  {t_{I}}^{8} \,$ in the $\,(\textbf{1 + 27})\,$ and $\,\textbf{7}\,$ of SO(7), respectively. When embedding $\,\textrm{SO}(7) \subset \textrm{SL}(7) \subset \textrm{SL}(8)\,$, one has the generators decomposition $\,\textbf{63} \, \rightarrow \, \textbf{1} \, \oplus \, \textbf{48} \,\, ({t_{8}}^{8} \, , \, {t_{I}}^{J}) \, \oplus \, \textbf{7} \,\, ({t_{I}}^{8}) \, \oplus \, \textbf{7}' \,\, ({t_{8}}^{I})\,$. The entire set of SL(8) brackets is then given by (\ref{ISO7_brackets}) together with
\begin{equation}
\label{SL8_ext_brackets_1}
\begin{array}{rlll}
[T_{IJ} , T^{(S)}_{KL}] &=& -4 \, \delta_{[I(K} \, T^{(S)}_{L)J]} & , \\[2mm]
[T^{(S)}_{IJ} , T^{(S)}_{KL}] &=& -4 \, \delta_{(I(K} \, T_{L)J)} & , \\[2mm]
[T_{I} , T^{(S)}_{KL}] &=& 2 \,  \delta_{I(K} \, T_{L)} & , 
\end{array}
\end{equation}
and
\begin{equation}
\label{SL8_ext_brackets_2}
\begin{array}{rlll}
[T^{\perp}_{I} , T_{KL}] &=& 2 \,  \delta_{I[K} \, T^{\perp}_{L]} & , \\[2mm]
[T^{\perp}_{I} , T^{(S)}_{KL}] &=& - 2 \,  \delta_{I(K} \, T^{\perp}_{L)} & , \\[2mm]
[T^{\perp}_{I},T_{K}] &=&  \tfrac{1}{2} \, \delta_{IK} \, \displaystyle\sum_{L} T^{(S)}_{LL}  + \tfrac{1}{2} (T^{(S)}_{IK} + T_{IK}) & , \\[2mm]
[T^{\perp}_{I} , T^{\perp}_{K}] &=& 0   & .
\end{array}
\end{equation}

As found in \cite{Dall'Agata:2014ita}, there is a one-parameter family of ISO(7)$_{c}$ maximal supergravities specified by $\,\theta\,$ and $\,\xi\,$ matrices of the form
\begin{equation}
\label{theta_xi_matrices}
\theta=
\left(\begin{array}{ll}
\delta_{IJ} & 0 \\
0              & 0
\end{array}\right)
\hspace{10mm} \textrm{ and } \hspace{10mm}
\xi=
\left(\begin{array}{ll}
0_{7 \times 7} & 0 \\
0              & c
\end{array}\right) \ ,
\end{equation}
which are compatible with the constraints (\ref{QC}). However, it was also proven in \cite{Dall'Agata:2014ita}  that all the values $\,c\neq 0\,$ produce equivalent theories up to a rescaling of the gauge coupling~$g$. Upon substitution of (\ref{theta_xi_matrices}) into (\ref{Theta_Def}), the components of the embedding tensor ${\Theta_{\mathbb{M}}}^{\alpha}$ take the more explicit form 
\begin{equation}
\label{Theta-comp}
\Theta_{[IJ]\phantom{K}L}^{\phantom{[IJ]}K} = 2\, \delta_{[I}^{K} \, \delta_{J]L} 
\hspace{5mm} , \hspace{5mm}
\Theta_{[I8] \phantom{8}  K}^{\phantom{[I8]} 8} = -\delta_{I K}
\hspace{5mm} , \hspace{5mm}
{\Theta^{[IJ] K}}_{L} = 0
\hspace{5mm} , \hspace{5mm}
{\Theta^{[I8] \, 8}}_{K} = c \, \delta^{I}_{K} \ ,
\end{equation}
and the charges in (\ref{X_electric}) and (\ref{X_magnetic}) are given by
\begin{equation}
\label{X-splitting}
X_{[AB]} \rightarrow  \left\lbrace 
\begin{array}{lll}
X_{[IJ]} = T_{IJ} \\[2mm]
X_{[I8]} = - T_{I}
\end{array} \right.
\hspace{10mm} \textrm{ and } \hspace{10mm}
X^{[AB]} \rightarrow \left\lbrace 
\begin{array}{lll}
X^{[IJ]} = 0 \\[2mm]
X^{[I8]} = c \, \delta^{IJ} \, T_{J}
\end{array} \right. \ .
\end{equation}
Applying analogous decompositions for the vector fields $\,\cA^{\mathbb{M}}_{\mu}\,$, namely
\begin{equation}
\label{A-splitting}
\cA_{\mu}^{[AB]} \rightarrow  \left\lbrace 
\begin{array}{lll}
\cA_{\mu}^{[IJ]} = \cA_{\mu}^{IJ} \\[2mm]
\cA_{\mu}^{[I8]} = \cA_{\mu}^{I}
\end{array} \right.
\hspace{10mm} \textrm{ and } \hspace{10mm}
\tcA_{\mu \, [AB]} \rightarrow \left\lbrace 
\begin{array}{lll}
\tcA_{\mu \, [IJ]} = \tcA_{\mu \, IJ} \\[2mm]
\tcA_{\mu \, [I8]} = \tcA_{\mu \, I}
\end{array} \right. \ ,
\end{equation}
one finds a covariant derivative $\,D_{\mu} = \partial_{\mu} - g \, \cA_{\mu}^{\mathbb{M}} \, X_{\mathbb{M}}\,$, with $\,X_{\mathbb{M}} = {\Theta_{\mathbb{M}}}^{\alpha} \, t_{\alpha}\,$, of the form
\begin{equation}
D_{\mu} = \partial_{\mu} -  \tfrac{1}{2} \, g \,   \cA_{\mu}^{IJ} \, T_{IJ} + g \, \cA_{\mu}^{I} \, T_{I}  -  m \,  \tcA_{\mu \, J} \, \delta^{JI} \, T_{I} \ ,
\end{equation}
where $\,m \equiv gc\,$ is the magnetic parameter introduced in \cite{Guarino:2015jca}. As a result, the SO(7) rotations ($T_{IJ}$) are gauged electrically whereas the $\mathbb{R}^{7}$ translations ($T_{I}$) are gauged dyonically, in agreement with \cite{Dall'Agata:2014ita}.

\subsection{The bosonic Lagrangian}

The Lagrangian of maximal supergravity is totally determined after specifying the \mbox{$X$-tensor} ${X_{\mathbb{MN}}}^{\mathbb{P}}$ in (\ref{ET}) underlying the gauging \cite{deWit:2007mt}. Using (\ref{X_electric}), (\ref{X_magnetic}) and (\ref{theta_xi_matrices}), the set of components for the ISO(7)$_{c}$ case is given by
\begin{equation}
\label{X_components}
\begin{array}{llllll}
{X_{[IJ] [KL]}}^{[MN]} &=& - X_{[IJ] \phantom{[MN]} [KL]}^{\phantom{[IJ]}[MN]} &=& -8 \, \delta_{[I}^{[M} \delta_{J][K} \delta_{L]}^{N]} & ,  \\[4mm]
{X_{[IJ] [K8]}}^{[M8]} &=& - X_{[IJ] \phantom{[M8]} [K8]}^{\phantom{[IJ]}[M8]} &=&  -2 \, \delta_{[I}^{M} \delta_{J]K}  & , \\[3mm]
{X_{[K8] [IJ]}}^{[M8]} &=& - X_{[K8] \phantom{[M8]} [IJ]}^{\phantom{[K8]}[M8]}  &=&  -2 \, \delta_{K[I} \delta_{J]}^{M}  & ,  \\[3mm]
X^{[K8] \phantom{[IJ]}[M8]}_{\phantom{[K8]}[IJ]} &=& - X^{[K8] [M8]}_{\phantom{[K8][M8]}[IJ]}  &=&  \phantom{+}2 \, c \, \delta^{K}_{[I} \delta^{M}_{J]}  &  ,
\end{array}
\end{equation}
with all the rest vanishing. Equipped with this tensor ${X_{\mathbb{MN}}}^{\mathbb{P}}$, the bosonic Lagrangian of maximal supergravity is given by \cite{deWit:2005ub,deWit:2007mt}
\begin{equation}
\label{Lbos}
\mathcal{L}_{\textrm{bos}} = \mathcal{L}_{\textrm{EH}}  + \mathcal{L}_{\textrm{VT}} + \mathcal{L}_{\textrm{scalar}} \ ,
\end{equation}
which contains the usual Einstein-Hilbert term\footnote{We use a mostly-plus convention for the metric, \textit{i.e.} $\,e=\sqrt{-g}\,$, as well as $\,-\varepsilon_{0123}=+1=\varepsilon^{0123}\,$ for the Levi-Civita symbol.} $\mathcal{L}_{\textrm{EH}} = \tfrac{1}{2} \, e \, R$, as well as vector, tensor and scalar contributions we move on to discuss now.

\subsubsection*{The scalar Lagrangian}

The maximal supergravity multiplet contains 70 scalar fields which serve as coordinates in the coset space $\textrm{E}_{7(7)}/\textrm{SU}(8)$. Using a coset representative ${\mathcal{V}_{\mathbb{M}}}^{\underline{\mathbb{N}}}$ transforming under global $\textrm{E}_{7(7)}$ transformations from the left and local SU(8) transformations from the right, the scalar-dependent matrix $\,\mathcal{M}_{\mathbb{MN}}\,$ in (\ref{Mscalar}) is built as $\,\mathcal{M} = \mathcal{V} \, \mathcal{V}^{t}\,$. In terms of $\,\mathcal{M}\,$, the scalar sector of the theory is given by
\begin{equation}
\label{L_scalars_app}
\mathcal{L}_{\textrm{scalar}} = \mathcal{L}^{\textrm{kin}}_{\textrm{scalar}} - e \, V(\mathcal{M})  = \tfrac{1}{96} \, e \, \textrm{Tr}\left( D_{\mu}\mathcal{M} \, D^{\mu}\mathcal{M}^{-1} \right) - e \, V(\mathcal{M}) \ ,
\end{equation}
where the scalar potential induced by the gauging takes the form
\begin{equation}
\begin{array}{ccl}
\label{V_general_app}
V(\mathcal{M}) & = & \dfrac{g^{2}}{672} \left(  {X_{\mathbb{MN}}}^{\mathbb{R}}  {X_{\mathbb{PQ}}}^{\mathbb{S}} \mathcal{M}^{\mathbb{MP}} \mathcal{M}^{\mathbb{NQ}}  \mathcal{M}_{\mathbb{RS}}  +   7 {X_{\mathbb{MN}}}^{\mathbb{Q}} {X_{\mathbb{PQ}}}^{\mathbb{N}} \mathcal{M}^{\mathbb{MP}}  \right) \ .
\end{array}
\end{equation}
Here we are not providing a more explicit expression neither for $\,\mathcal{M}\,$ nor for the scalar potential (\ref{V_general_app}) when particularised to the ISO(7)$_{c}$ gaugings. However, let us make an extra remark in this case. The ISO(7)$_{c}$ gaugings involve the seven non-compact generators $\,T_{I}\,$ in (\ref{linear_comb_t}) associated to the $\,\mathbb{R}^{7}_{c}\,$ translations. This implies that, if we choose an appropriate parameterisation of the $\,\textrm{E}_{7(7)}/\textrm{SU}(8)\,$ scalar coset such that 7 out of the 70 scalars are aligned with the $\,T_{I}\,$ generators, these seven scalars will not enter the scalar potential.

\subsubsection*{The vector-tensor Lagrangian}

Neglecting fermion bilinears $\mathcal{O}^{\mu\nu}$, vector fields contribute to (\ref{Lbos}) with a kinetic and a topological term codifying generalised Chern-Simons-like terms \cite{deWit:2005ub}. This is
\begin{equation}
\label{LVT}
\mathcal{L}_{\textrm{VT}}=\mathcal{L}_{\textrm{vec}} + \mathcal{L}_{\textrm{top}} \ .
\end{equation}

The former is given by
\begin{equation}
\label{Lvec}
\mathcal{L}_{\textrm{vec}} =  \tfrac{1}{4} \, e \, \Big( \mathcal{I}_{\Lambda\Sigma} \, \mathcal{H}^{\Lambda}_{\mu \nu} \, \mathcal{H}^{\Sigma \mu \nu}  + \frac{1}{2\,e} \, \varepsilon^{\mu\nu\rho\sigma}  \, \mathcal{R}_{\Lambda\Sigma} \, \mathcal{H}^{\Lambda}_{\mu \nu} \, \mathcal{H}^{\Sigma}_{\rho \sigma}  \Big) \ ,
\end{equation}
where $\,\Lambda=1,\ldots ,28\,$ is a collective index running over the electric vectors $\,\cA_{\mu}^{\Lambda} \equiv \cA_{\mu}^{[AB]}\,$ or the magnetic ones $\,\tcA_{\mu \, \Lambda} \equiv \tcA_{\mu \, [AB]}\,$ in the decomposition $\,\cA_{\mu}^{\mathbb{M}} = (\cA_{\mu}^{\Lambda} ,\tcA_{\mu \, \Lambda})\,$ as well as over their field strengths $\,\mathcal{H}^{\mathbb{M}}_{\mu \nu}=( \mathcal{H}^{\Lambda}_{\mu \nu},\tilde{\mathcal{H}}_{ \mu \nu \, \Lambda})\,$. The symmetric matrices $\,\mathcal{R}_{\Lambda\Sigma}\,$ and $\,\mathcal{I}_{\Lambda\Sigma}\,$ in  (\ref{Lvec}) depend on the scalar fields and can be combined into a complex matrix
\begin{equation}
\label{N_Matrix_app}
\mathcal{N}_{\Lambda\Sigma} = \mathcal{R}_{\Lambda\Sigma} + i \, \mathcal{I}_{\Lambda\Sigma} \ .
\end{equation}
Note that  $\,\mathcal{I}_{\Lambda\Sigma}\,$ must be negative definite for the kinetic terms in (\ref{Lvec}) to have the correct sign. The complex matrix $\,\mathcal{N}_{\Lambda\Sigma}\,$ is related to the scalar matrix $\,\mathcal{M}_{\mathbb{MN}}\,$ in (\ref{Mscalar}) via \cite{Samtleben:2008pe}
\begin{equation}
\label{M&ReIm}
\mathcal{M}_{\mathbb{MN}}= 
\left(
\begin{array}{ll}
\mathcal{M}_{\Lambda \Sigma} & {\mathcal{M}_{\Lambda}}^{\Sigma} \\[2mm]
{\mathcal{M}^{\Lambda}}_{\Sigma} & \mathcal{M}^{\Lambda \Sigma}
\end{array}
\right)
=
\left(
\begin{array}{ll}
-(\mathcal{I}+\mathcal{R}\mathcal{I}^{-1}\mathcal{R})_{\Lambda \Sigma} & {(\mathcal{R}\mathcal{I}^{-1})_{\Lambda}}^{\Sigma} \\[2mm]
{(\mathcal{I}^{-1}\mathcal{R})^{\Lambda}}_{\Sigma} & -(\mathcal{I}^{-1})^{\Lambda \Sigma}
\end{array}
\right) \ .
\end{equation}
The field strengths $\,\mathcal{H}^{\mathbb{M}}_{\mu \nu}\,$ of the vector fields are given by
\begin{equation}
\label{H_Field_Strength}
\mathcal{H}^{\mathbb{M}}_{\mu \nu} = \mathcal{F}^{\mathbb{M}}_{\mu \nu} + g \, Z^{\mathbb{M} \, , \alpha} \, \cB_{\mu \nu \alpha}
\hspace{8mm} \textrm{ with } \hspace{5mm}
\mathcal{F}^{\mathbb{M}}_{\mu \nu}=2 \, \partial_{[\mu} \cA^{\mathbb{M}}_{\nu]} \,+\, g \, {X_{\mathbb{[NP]}}}^{\mathbb{M}} \, \cA^{\mathbb{N}}_{\mu} \, \cA^{\mathbb{P}}_{\nu} \ ,
\end{equation}
and are ``modified" in the sense that incorporate a number of \textit{auxiliary} two-form tensor fields $\,\cB_{\mu \nu \alpha}\,$ subject to suitable gauge transformations which ensure that (\ref{H_Field_Strength}) transform covariantly \cite{deWit:2005ub}. The way tensor fields enter the field strengths in (\ref{H_Field_Strength}) is dictated by
\begin{equation}
Z^{\mathbb{M} \, , \alpha} = \tfrac{1}{2} \, \Omega^{\mathbb{MN}} \, {\Theta_{\mathbb{N}}}^{\alpha}  \ ,
\end{equation}
and, using the ${\Theta_{\mathbb{M}}}^{\alpha}$ components in (\ref{Theta-comp}) for the ISO(7)$_{c}$ gaugings, one finds
\begin{equation}
\label{Z_comp}
{Z^{[IJ] \, K}}_{L} = 0 
\hspace{5mm} , \hspace{5mm} 
{Z^{[I8] \, 8}}_{K} = \tfrac{1}{2} \, c \, \delta^{I}_{K} 
\hspace{5mm} , \hspace{5mm} 
Z_{[IJ]\phantom{K} L}^{\phantom{[IJ]}K} = -\delta_{[I}^{K} \, \delta_{J]L} 
\hspace{5mm} , \hspace{5mm} 
Z_{[I8]\phantom{8}K}^{\phantom{[I8]}8} = \tfrac{1}{2} \, \delta_{IK}  \ .
\end{equation}
After using (\ref{X_components}) and (\ref{Z_comp}), the electric field strengths $\,\mathcal{H}^{\Lambda}_{\mu \nu}\,$ entering the Lagrangian (\ref{Lvec}) in the case of the dyonic $\,\textrm{ISO}(7)\,$ gaugings read
\begin{equation}
\label{H_modified_Elec}
\begin{array}{llllll}
\mathcal{H}^{IJ}_{\mu \nu} &=& \mathcal{F}^{IJ}_{\mu \nu}  \quad =  \quad 2 \, \partial_{[\mu} \cA^{IJ}_{\nu]} - 2 \, g \, \delta_{KL} \, \cA^{KI}_{[\mu}  \, \cA^{JL}_{\nu]} & ,\\[2mm]
\mathcal{H}^{I}_{\mu \nu} &=& \mathcal{F}^{I}_{\mu \nu} + \tfrac{1}{2} \, m \, \cB^{I}_{\mu \nu} \\[2mm] 
&=&  2 \, \partial_{[\mu} \cA^{I}_{\nu]} - 2 \, g \, \delta_{JK} \, \cA_{[\mu}^{IJ} \,  \cA_{\nu]}^{K}   +  m \, \cA_{[\mu}^{IJ} \, \tcA_{\nu] \, J}   + \tfrac{1}{2} \, m \, \cB^{I}_{\mu \nu} & ,
\end{array}
\end{equation}
whereas the magnetic field strengths $\,\tilde{\mathcal{H}}_{\mu \nu \, \Lambda}\,$, which do not appear in (\ref{Lvec}), take the form
\begin{equation}
\label{H_modified_Mag}
\begin{array}{llllll}
\tilde{\mathcal{H}}_{\mu \nu \,  IJ} &=& \tilde{\mathcal{F}}_{\mu \nu \,  IJ}  -   g \, {\cB_{\mu\nu \, [I}}^{K} \, \delta_{J]K} &  \\[2mm]
&=& 2 \, \partial_{[\mu} \tcA_{\nu] \, IJ} \\[2mm]
&=& g \, \delta_{KI} \, \cA^{KL}_{[\mu} \, \tcA_{\nu]\, JL}  + g \, \tcA_{[\mu \, KI} \, \cA^{KL}_{\nu]} \, \delta_{JL}  + g \, \delta_{KI} \, \cA^{K}_{[\mu} \, \tcA_{\nu]\, J}  + g \, \tcA_{[\mu \, I} \, \cA^{K}_{\nu]} \, \delta_{KJ}   \\[2mm] 
&-& 2 \, m \,   \tcA_{[\mu \, I} \, \tcA_{\nu] \, J} \,\, - \,\,   g \, {\cB_{\mu\nu \, [I}}^{K} \, \delta_{J]K} & , \\[3mm]
\tilde{\mathcal{H}}_{\mu \nu\,  I} &=& \tilde{\mathcal{F}}_{\mu \nu \, I} + \tfrac{1}{2} \, g \, \delta_{IJ} \, \cB^{J}_{\mu \nu}  \quad = \quad  2 \, \partial_{[\mu} \tcA_{\nu] \, I} - g \, \delta_{IJ} \, \cA^{JK}_{[\mu} \, \tcA_{\nu] \, K}  + \tfrac{1}{2} \, g \, \delta_{IJ} \, \cB^{J}_{\mu \nu}    & .
\end{array}
\end{equation}
Therefore, a set of seven two-form tensor fields $\,\cB^{I}_{\mu \nu} \equiv {\cB_{\mu \nu \, 8}}^{I}\,$ will enter (\ref{Lvec}) if $\,m \neq 0\,$ because of $\,\mathcal{H}^{I}_{\mu \nu}\,$ in (\ref{H_modified_Elec}).

The presence of magnetic charges and tensor fields generates the topological term in (\ref{LVT}). It was obtained in \cite{deWit:2005ub} and takes the form
\begin{equation}
\label{Ltop}
\begin{array}{llll}
\mathcal{L}_{\textrm{top}} &=&  g \, \varepsilon^{\mu\nu\rho\sigma} \, \Big[   -\tfrac{1}{8}  \,  \Theta^{\Lambda \alpha} \, \cB_{\mu \nu \, \alpha} \left( 2 \, \partial_{\rho} \tcA_{\sigma \, \Lambda} + g \, X_{\mathbb{MN}\Lambda} \, \cA_{\rho}^{\mathbb{M}} \, \cA_{\sigma}^{\mathbb{N}}  - \tfrac{1}{4} \, g \, {\Theta_{\Lambda}}^{\beta} \, \cB_{\rho \sigma \,\beta}  \right)   \\[2mm]
&  & \phantom{- g  \epsilon^{\mu\nu\rho\sigma} }  -\tfrac{1}{3} \, X_{\mathbb{MN}\Lambda} \, \cA_{\mu}^{\mathbb{M}} \, \cA_{\nu}^{\mathbb{N}}  \left(   \partial_{\rho} \cA_{\sigma}^{\Lambda} + \tfrac{1}{4} \, g \, {X_{\mathbb{PQ}}}^{\Lambda} \, \cA_{\rho}^{\mathbb{P}} \, \cA_{\sigma}^{\mathbb{Q}}    \right)  \\[2mm]
&  & \phantom{- g  \epsilon^{\mu\nu\rho\sigma} }  -\tfrac{1}{6} \, {X_{\mathbb{MN}}}^{\Lambda} \, \cA_{\mu}^{\mathbb{M}} \, \cA_{\nu}^{\mathbb{N}}  \left(   \partial_{\rho} \tcA_{\sigma \, \Lambda} + \tfrac{1}{4} \, g \, X_{\mathbb{PQ} \Lambda} \, \cA_{\rho}^{\mathbb{P}} \, \cA_{\sigma}^{\mathbb{Q}}    \right)  \Big] \ .
\end{array}
\end{equation}
Particularising again to the case of ISO(7)$_{c}$ gaugings, and using the relations
\begin{equation}
\label{X-contract}
\begin{array}{llll}
{X_{\mathbb{MN}}}^{[IJ]} \, \cA_{\mu}^{\mathbb{M}} \, \cA_{\nu}^{\mathbb{N}} & = & \delta_{KL} \, \cA_{\mu}^{IK} \, \cA_{\nu}^{JL} \,\, - \,\,  (I \leftrightarrow J)   & , \\[2mm] 
X_{\mathbb{MN} [IJ]} \, \cA_{\mu}^{\mathbb{M}} \, \cA_{\nu}^{\mathbb{N}} & = & \delta_{IL} \, \cA_{\mu}^{LK} \, \tcA_{\nu \,  JK}   + ( \delta_{IL} \, \cA_{\mu}^{L} - c \, \tcA_{\mu \, I}) \, \tcA_{\nu \, J} \,\, - \,\,  (I \leftrightarrow J)  & , \\[3mm]  
{X_{\mathbb{MN}}}^{[I8]} \, \cA_{\mu}^{\mathbb{M}} \, \cA_{\nu}^{\mathbb{N}} & = & - \delta_{KL} \, \cA_{\mu}^{IK} \, \cA_{\nu}^{L}  + ( \delta_{KL} \, \cA_{\mu}^{L} - c \, \tcA_{\mu \, K}) \, \cA_{\nu}^{IK}   & , \\[2mm]  
X_{\mathbb{MN} [I8]} \, \cA_{\mu}^{\mathbb{M}} \, \cA_{\nu}^{\mathbb{N}} & = & - \cA_{\mu}^{IK} \, \tcA_{\nu \, K} & ,
\end{array}
\end{equation}
to compute the contributions of the form $\cA\cA\partial \cA$ and $\cA\cA\cA\cA$, the topological term in (\ref{Ltop}) reduces to
\begin{equation}
\label{Ltop_ISO7}
\mathcal{L}_{\textrm{top}} = m \,  \varepsilon^{\mu \nu \rho \sigma} \, \left[ -\tfrac{1}{8} \, \cB^{I}_{\mu \nu}    \, \big( \tilde{\cH}_{\rho \sigma \, I}      -  \tfrac{1}{4} \, g \, \delta_{IJ} \, \cB^{J}_{\rho \sigma}    \big)  +   \tfrac{1}{4} \,  \tilde{\cA}_{\mu \, I} \, \tilde{\cA}_{\nu \, J}  \big(  \partial_{\rho}\cA_{\sigma}^{IJ}    + \tfrac{g}{2} \,   \cA_{\rho}^{IK} \cA_{\sigma}^{JL} \, \delta_{KL} \,\big) \right]   ,
\end{equation}
with $\,\tilde{\cH}_{\mu \nu \, I}\,$ given in (\ref{H_modified_Mag}). Notice that (\ref{Ltop_ISO7}) vanishes in the purely electric case of $\,m=0$.

\subsubsection*{Three-form potentials, Bianchi identities and representation theory}

The tensor hierarchy of maximal supergravity requires also the presence of three-form potentials $\,\cC_{\mu \nu \rho \, \alpha}{}^{\mathbb{M}}\,$ transforming in the conjugate representation to the embedding tensor \cite{deWit:2008ta}. These three-forms modify the field strengths of the tensors $\,\cB_{\mu \nu \, \alpha}\,$ in a similar manner to (\ref{H_Field_Strength}), namely  \cite{deWit:2005hv,deWit:2005ub,deWit:2008ta,deWit:2008gc},
\begin{equation}
\label{H3_Def}
\mathcal{H}_{\3 \mu\nu\rho \, \alpha} =  \mathcal{F}_{\3 \mu\nu\rho \, \alpha}   +  g \,  {Y_{\alpha , \, \mathbb{P}}}^{\beta}   \, \cC_{\mu \nu \rho \, \beta}{}^{\mathbb{P}}\ ,
\end{equation}
with
\begin{equation}
\label{F3_Def}
\mathcal{F}_{\3 \mu\nu\rho \, \alpha} = 3 \, D_{[\mu} \cB_{\nu \rho] \, \alpha} + 6 \, d_{\alpha \, \mathbb{MN}}  \, \cA_{[ \mu}^{\mathbb{M}} \, \Big(   \partial_{\nu} \cA_{\rho]}^{\mathbb{N}} + \tfrac{1}{3} \, g \,  {X_{\mathbb{RS}}}^{\mathbb{N}}   \, \cA_{\nu}^{\mathbb{R}}  \, \cA_{\rho]}^{\mathbb{S}}   \Big) \ ,
\end{equation}
and
\begin{equation}
\label{DB_Def}
D_{[\mu} \cB_{\nu \rho] \, \alpha} = \partial_{[\mu} \cB_{\nu \rho] \, \alpha} + g \, {X_{\mathbb{M} \, \alpha}}^{\beta} \, \cA_{[\mu}^{\mathbb{M}} \, \cB_{\nu \rho] \beta}   \ ,
\end{equation}
and where $\,{X_{\mathbb{M} \, \alpha}}^{\beta} = {\Theta_{\mathbb{M}}}^{\gamma} \, {[t_{\gamma}]_{\alpha}}^{\beta}\,$ and $\,d_{\alpha \, \mathbb{MN}} \equiv {[t_{\alpha}]_{\mathbb{M}}}^{\mathbb{P}} \, \Omega_{\mathbb{NP}}\,$. The $Y$-tensor in (\ref{H3_Def}) is called the \textit{intertwining} tensor and takes the form \cite{deWit:2008ta,deWit:2008gc}
\begin{equation}
\label{Y-tensor}
{Y_{\alpha , \, \mathbb{P}}}^{\beta}   = {[t_{\alpha}]_{\mathbb{P}}}^{\mathbb{Q}} \, {\Theta_{\mathbb{Q}}}^{\beta} + {X_{\mathbb{P}\, \alpha}}^{\beta}  \ .
\end{equation}
The field strengths $\,\mathcal{H}_{\3 \mu\nu\rho \, \alpha}\,$  do not enter the maximal supergravity Lagrangian in the framework of \cite{deWit:2005ub,deWit:2007mt}. Moreover, by virtue of 
\begin{equation}
Z^{\mathbb{M}, \, \alpha} \,\, {Y_{\alpha , \, \mathbb{P}}}^{\beta} = 0 \ ,
\end{equation}
the $Y$-term in (\ref{H3_Def}) vanishes upon contraction with $\,Z^{\mathbb{M, \, \alpha}}\,$ and, therefore, is not relevant for the ($Z$-projected) Bianchi identities \cite{deWit:2005ub}
\begin{equation}
\label{BI}
\begin{array}{llll}
D_{[\mu} \mathcal{H}_{\nu \rho]}^{\mathbb{M}} &=& \frac{1}{3} \, g \, Z^{\mathbb{M} , \,  \alpha} \, \mathcal{H}_{\3 \mu\nu\rho \, \alpha}  & , \\[2mm]
Z^{\mathbb{M} , \,  \alpha } \, D_{[\mu} \mathcal{H}_{\3 \nu\rho\sigma] \, \alpha} & = & 3 \, g \, {X_{\mathbb{PQ}}}^{\mathbb{M}} \, \mathcal{H}_{[\mu \nu}^{\mathbb{P}} \, \mathcal{H}_{\rho \sigma]}^{\mathbb{Q}} & ,
\end{array}
\end{equation}
with
\begin{equation}
\label{D_cov_H23}
\begin{array}{llll}
D_{[\mu} \mathcal{H}_{\nu \rho]}^{\mathbb{M}} &=&\partial_{[\mu} \mathcal{H}_{\nu \rho]}^{\mathbb{M}} \,+\, g \, {X_{\mathbb{PQ}}}^{\mathbb{M}} \, \cA_{[ \mu}^{\mathbb{P}} \, \mathcal{H}_{\nu \rho]}^{\mathbb{Q}} & , \\[2mm]
D_{[\mu} \mathcal{H}_{\3 \nu\rho\sigma] \, \alpha} &=& \partial_{[\mu} \mathcal{H}_{\3 \nu\rho\sigma] \, \alpha} \,+\, g \, {X_{\mathbb{M} \, \alpha}}^{\beta} \, \cA_{[ \mu}^{\mathbb{M}} \,  \mathcal{H}_{\3 \nu\rho\sigma] \, \beta} & .
\end{array}
\end{equation}
Using (\ref{F3_Def})--(\ref{Y-tensor}), we have obtained the expressions for the three-form field strenghts in (\ref{H3_Def}) when particularised to the dyonic ISO(7) theory. Similarly, using (\ref{D_cov_H23}), we obtained the expressions for the ($Z$-unprojected) Bianchi identities. The results have been brought to the main text. Last, and for the sake of brevity, we are not presenting here the lengthy expression for $\,\cH_{\4\mu \nu \rho \sigma \, \alpha}{}^{\mathbb{M}}\,$, which can be found in the appendix~B of \cite{Bergshoeff:2009ph}.

Let us briefly comment on the representation theory underlying the field content of the tensor hierarchy for the dyonic ISO(7) supergravities. Using the branching rules
\begin{equation}  
\label{eq:Branchings}
\begin{array}{ccccc}
\textrm{E}_{7(7)} & \supset & \textrm{SL}(8) & \supset & \textrm{SL}(7) \times \mathbb{R}_{\textrm{M}} \\[2mm]   
\textbf{56}   &  \rightarrow & \textbf{28} + \textbf{28}' & \rightarrow & ( \textbf{21}_{+2} + \textbf{7}_{-6} ) + ( \textbf{21}'_{-2} + \textbf{7}'_{+6} ) \\[2mm]
\textbf{133}   &  \rightarrow & \textbf{63} + \textbf{70} & \rightarrow & ( \textbf{1}_{0} + \textbf{48}_{0} + \textbf{7}_{+8} + \textbf{7}'_{-8} ) + ( \textbf{35}_{-4} + \textbf{35}'_{+4} ) \\[2mm]
\textbf{912}   &  \rightarrow & \textbf{36} + \textbf{36}' + \ldots & \rightarrow & ( \textbf{28}_{+2} + \textbf{7}_{-6} + \textbf{1}_{-14} ) + ( \textbf{28}'_{-2} + \textbf{7}'_{+6} + \textbf{1}_{+14} ) + \ldots
\end{array}
\end{equation}
it is possible to identify the different representations attached to the different field potentials and embedding tensor deformations in the theory. These are given by
\begin{equation}
\label{eq:Branchings_2}
\begin{array}{lcll}
\cA^{\mathbb{M}} & \rightarrow &  \tcA_{IJ} \equiv \textbf{21}_{+2} \,\,\,\, , \,\,\,\,  \tcA_{I} \equiv \textbf{7}_{-6} \,\,\,\, , \,\,\,\,  \cA^{IJ} \equiv \textbf{21}'_{-2} \,\,\,\, , \,\,\,\,  \cA^{I} \equiv\textbf{7}'_{+6} & , \\[2mm]
\cB_{\alpha} & \rightarrow &  \cB \equiv \textbf{1}_{0} \,\,\,\,\,\,\,\,  , \,\,\,\,\,\,\, {\cB_{I}}^{J} \equiv \textbf{48}_{0}  \,\,\,\,\,\,\,\, , \,\,\,\,  \cB^{I} \equiv \textbf{7}'_{-8} & , \\[2mm]
{\cC_{\alpha}}^{\mathbb{M}}& \rightarrow &  \cC^{IJ} \equiv \textbf{28}'_{-2}  \,\,\,\, , \,\,\,\,  \tilde{\cC} \equiv \textbf{1}_{-14}& ,
\end{array}
\end{equation}
whereas the embedding tensor $\,{\Theta_{\mathbb{M}}}^{\alpha}\,$ sits in the $\,\textbf{28}_{+2} \,$ ($g\, \delta_{IJ}$) and $\,\textbf{1}_{+14} \,$ ($m$). Further truncations to the different invariant sectors discussed in this paper are displayed in Table~\ref{Table:group_theory}.

\section{The scalar-dependent matrix $\mathcal{M}_{\mathbb{MN}}$}
\label{app:ScalarMatrix}

In this appendix we provide the explicit form of the symmetric, scalar-dependent matrix  
\begin{equation}
\label{M_appendix}
\mathcal{M}_{\mathbb{MN}} = 
\left(
\begin{array}{ll}
\mathcal{M}_{\Lambda \Sigma} & {\mathcal{M}_{\Lambda}}^{\Sigma} \\[2mm]
{\mathcal{M}^{\Lambda}}_{\Sigma} & \mathcal{M}^{\Lambda \Sigma}
\end{array}
\right)
=
\left(
\begin{array}{ll}
\mathcal{M}_{[AB] [CD]} & {\mathcal{M}_{[AB]}}^{[CD]} \\[2mm]
{\mathcal{M}^{[AB]}}_{[CD]} & \mathcal{M}^{[AB] [CD]}
\end{array}
\right) \ 
\end{equation}
in (\ref{Mscalar}) for the the SU(3), G$_{2}$ and SO(4) invariant sectors discussed in the main text.

\subsection{The $\textrm{SU}(3)$ sector} 
\label{app:SU3sectorScalMat}

The complexification in (\ref{SU3_complexification}) translates into an index splitting of the form $\,A \rightarrow 1 \, \oplus \, i \,  \oplus \, 8 \,$, with $\,i=2, \ldots ,7\,$ a fundamental index of SO(6). This implies a splitting of the $\textbf{28}$ (and the $\textbf{28}'$) of SL(8) of the form ${[AB] \rightarrow [ij] \, \oplus \, [1j]\, \oplus \, [i8] \, \oplus \, [18]}$. The set of SU(3)-invariant forms includes the flat metric $\,\delta_{ij}\,$, a real two-form $\,J_{ij}\,$ and a holomorphic three-form $\,\Omega_{ijk}\,$. With those index conventions, these are given by
\begin{equation}
\begin{array}{llll}
J & = & e^{2} \wedge e^{3} \, + \,  e^{4} \wedge e^{5} \, + \, e^{6} \wedge e^{7} & , \\[2mm]
\Omega & = & (e^{2} + i \,  e^{3}) \, \wedge \, (e^{4} + i \,  e^{5}) \, \wedge (e^{6} + i \,  e^{7}) & .
\end{array}
\end{equation}
These forms satisfying the orthogonality and normalisation conditions
\begin{equation}
J \wedge \Omega = 0 \
\hspace{8mm} \textrm{ and } \hspace{8mm} 
\Omega \wedge \bar{\Omega} = - \tfrac{4}{3} \, i \, J \wedge J \wedge J  \ .
\end{equation}
The scalar matrix $\mathcal{M}_{\mathbb{MN}}$ depends on the six scalars $\,(\chi,\varphi)\,$ and $\,(\phi  , a , \zeta , \tilde{\zeta})\,$ entering the coset representative in (\ref{coset_SU3}). It is useful to introduce the short-hand combinations 
\begin{equation}
X= 1+ e^{2\varphi} \chi^2
\hspace{5mm} , \hspace{5mm}
Y=1+\tfrac{1}{4} \, e^{2\phi} \, (\zeta^2+\tilde{\zeta}^2)
\hspace{5mm} , \hspace{5mm}
Z= e^{2\phi} \, a  \ ,
\end{equation}
together with
\begin{equation}
j_{1} = \zeta \, Z + \tilde{\zeta} \, Y
\hspace{10mm}  \textrm{ and }  \hspace{10mm}
j_{2} = \tilde{\zeta} \, Z - \zeta \, Y \ ,
\end{equation}
in order to present the different blocks of (\ref{M_appendix}). We now turn to do that. 
\\[-2mm]

\noindent - The block $\,\mathcal{M}^{[AB][CD]}\,$ contains the following components
\begin{equation}
\label{M_scalar_SU3_1}
\begin{array}{llll}
{\cal M}^{[18] [18]} &=&   e^{-3 \, \varphi} \, X^3& , \\[2mm]
{\cal M}^{[i8] [k8]} &=&  e^{-(2\, \phi + \varphi)} \, X \, ( Y^2 + Z^2)  \,\, \delta^{ik} & ,
\end{array}
\end{equation} 
together with
\begin{equation}
\label{M_scalar_SU3_2}
\begin{array}{llll}
{\cal M}^{[18] [kl]} &=&   e^{\varphi} \, \chi^2 \, X \,\, J^{kl}& , \\[2mm]
{\cal M}^{[i8] [kl]} &=& -\frac{1}{2} \, e^{\varphi} \, \chi \, \big[ \,\,  j_{1} \,\, (\text{Re}\Omega)^{ikl} \, + \,  j_{2} \,\, (\text{Im}\Omega)^{ikl} \,\, \big] & , \\[2mm]
{\cal M}^{[i8] [1l]} &=& - e^{-\varphi} \, X \, \big[\,\,   Z \,\, \delta^{il} \, + \, (Y-1) \,\, J^{il}    \,\, \big] & ,
\end{array}
\end{equation} 
and
\begin{equation}
\label{M_scalar_SU3_3}
\begin{array}{llll}
{\cal M}^{[1j] [1l]} &=&   e^{2 \, \phi-\varphi} \,  X \,\, \delta^{jl}& , \\[2mm]
{\cal M}^{[1j] [kl]} &=& \frac{1}{2} \, e^{2\, \phi +\varphi} \, \chi \, \big[ \,\,  \zeta \,\, (\text{Re}\Omega)^{jkl} \, + \, \tilde{\zeta} \,\, (\text{Im}\Omega)^{jkl} \,\, \big] & , \\[2mm]
{\cal M}^{[ij] [kl]} &=&  e^{\varphi} \, (X-Y) \,\, J^{ij} \,J^{kl} + 3 \, e^{\varphi} \, (Y-1) \,\, J^{[ij} \, J^{kl]} + 2 \, e^{\varphi} \, Y \,\, \delta^{k[i} \, \delta^{j]l} &  .
\end{array}
\end{equation} 

\noindent - For the block $\,{\mathcal{M}^{[AB]}}_{[CD]}\,$, the set of components is given by
\begin{equation}
\label{M_scalar_SU3_4}
\begin{array}{llll}
{{\cal M}^{[18]}}_{ [18]} &=&  - e^{3 \, \varphi} \, \chi^3& , \\[2mm]
{{\cal M}^{[i8]}}_{ [k8]} &=&  e^{\varphi} \,\chi \, \big[ \, Z  \, {J^{i}}_{k} - (Y-1)  \,\, \delta^{i}_{k} \, \big]  & ,
\end{array}
\end{equation} 
together with
\begin{equation}
\label{M_scalar_SU3_5}
\begin{array}{llll}
{{\cal M}^{[18]}}_{ [kl]} &=&  - e^{-\varphi} \, \chi \, X^2 \,\, J_{kl}& , \\[2mm]
{{\cal M}^{[i8]}}_{ [kl]} &=& \frac{1}{2} \, e^{-\varphi} \, X \, \big[ \,\,  j_{1} \,\, {(\text{Re}\Omega)^{i}}_{kl} \, + \,  j_{2} \,\, {(\text{Im}\Omega)^{i}}_{kl} \,\, \big] & , \\[2mm]
{{\cal M}^{[i8]}}_{ [1l]} &=&  e^{-2\,\phi+\varphi} \, \chi \, (   Y^2 + Z^2 ) \,\, {J^{i}}_{l}  & , \\[2mm]
{{\cal M}^{[ij]}}_{ [18]} &=&  - e^{3\,\varphi} \, \chi  \,\, J^{ij}& , \\[2mm]
{{\cal M}^{[ij]}}_{ [k8]} &=& \frac{1}{2} \, e^{2\,\phi+\varphi} \, \big[ \,\,  \tilde{\zeta} \,\, {(\text{Re}\Omega)^{ij}}_{k} \, - \,  \zeta \,\, {(\text{Im}\Omega)^{ij}}_{k} \,\, \big] & , \\[2mm]
{{\cal M}^{[1j]}}_{ [k8]} &=& - e^{2\,\phi+\varphi} \, \chi  \,\, {J^{j}}_{k}  & ,
\end{array}
\end{equation} 
and
\begin{equation}
\label{M_scalar_SU3_6}
\begin{array}{llll}
{{\cal M}^{[1j]}}_{ [1l]} &=&  - e^{\varphi} \,\chi \, \big[ \, Z \, {J^{j}}_{l} + (Y-1)  \,\, \delta^{j}_{l} \, \big]   & , \\[2mm]
{{\cal M}^{[ij] }}_{[1l]} &=& \frac{1}{2} \, e^{\varphi} \,  \big[ \,\,  j_{2} \,\, {(\text{Re}\Omega)^{ij}}_{l} \, - \, j_{1} \,\, {(\text{Im}\Omega)^{ij}}_{l} \,\, \big] & , \\[2mm]
{{\cal M}^{[1j] }}_{[kl]} &=& -\frac{1}{2} \, e^{2\, \phi -\varphi} \, X \, \big[ \,\,  \zeta \,\, {(\text{Re}\Omega)^{j}}_{kl} \, + \, \tilde{\zeta} \,\, {(\text{Im}\Omega)^{j}}_{kl} \,\, \big] & , \\[2mm]
{{\cal M}^{[ij]}}_{ [kl]} &=& e^{\varphi} \, \chi \, (Y-X) \,\, J^{ij} \, J_{kl} - 3 \, e^{\varphi} \, \chi \, Y \,\, J^{[ij} \, J^{rs]} \, \delta_{r[k} \, \delta_{l]s} - 2 \, e^{\varphi} \, \chi \, (Y-1) \,\, \delta^{ij}_{kl}  & .
\end{array}
\end{equation}

\noindent - The block $\,\mathcal{M}_{[AB][CD]}\,$ has components
\begin{equation}
\label{M_scalar_SU3_7}
\begin{array}{llll}
{\cal M}_{[18] [18]} &=&   e^{3 \, \varphi} & , \\[2mm]
{\cal M}_{[i8] [k8]} &=&  e^{2\, \phi + \varphi} \,\, \delta_{ik} & ,
\end{array}
\end{equation} 
together with
\begin{equation}
\label{M_scalar_SU3_8}
\begin{array}{llll}
{\cal M}_{[18] [kl]} &=&  e^{3\,\varphi} \, \chi^2  \,\, J_{kl}& , \\[2mm]
{\cal M}_{[i8] [kl]} &=& -\frac{1}{2} \, e^{2\,\phi +\varphi} \, \chi \, \big[ \,\,  \tilde{\zeta} \,\, (\text{Re}\Omega)_{ikl} \, - \,  \zeta \,\, (\text{Im}\Omega)_{ikl} \,\, \big] & , \\[2mm]
{\cal M}_{[i8] [1l]} &=& e^{\varphi} \,  \big[\,\,   Z \,\, \delta_{il} \, - \, (Y-1) \,\, J_{il}    \,\, \big] & ,
\end{array}
\end{equation} 
and
\begin{equation}
\label{M_scalar_SU3_9}
\begin{array}{llll}
{\cal M}_{[1j] [1l]} &=&   e^{-2 \, \phi+\varphi}  \, ( Y^2 + Z^2 ) \,\, \delta_{jl}& , \\[2mm]
{\cal M}_{[1j] [kl]} &=& -\frac{1}{2} \, e^{\varphi} \, \chi \, \big[ \,\,  j_{2} \,\, (\text{Re}\Omega)_{jkl} \, - \, j_{1} \,\, (\text{Im}\Omega)_{jkl} \,\, \big] & , \\[2mm]
{\cal M}_{[ij] [kl]} &=& e^{-\varphi} \, X \,  (X-Y) \,\, J_{ij} \, J_{kl} + 3 \, e^{-\varphi} \, X (Y-1) \,\, J_{[ij} \, J_{kl]} + 2 \, e^{-\varphi} \, X \, Y \,\, \delta_{k[i} \, \delta_{j]l} &  .
\end{array}
\end{equation} 

\noindent - Due to the symmetry of $\mathcal{M}_{\mathbb{MN}}$, the last block can be obtained as $\,{\mathcal{M}_{[AB]}}^{[CD]}={\mathcal{M}^{[CD]}}_{[AB]}\,$.
\\[-2mm]

Note that different $\textrm{SU}(3)$-invariant tensors have different $\mathbb{Z}_{2}$-parity behaviour with respect to the transformation in (\ref{Z2-symmetry}): the tensors $\delta_{ij}$ and $\textrm{Re}(\Omega)_{ijk}$ are parity-even whereas $J_{ij}$ and $\textrm{Im}(\Omega)_{ijk}$ are parity-odd. Consequently, there are parity-even and parity-odd components within $\,\mathcal{M}_{\mathbb{MN}}\,$. The latter vanish when $\,a=\zeta=0\,$ (so that $j_{2}=0$), as these scalars pair up with the parity-odd generators in (\ref{Gener_SU3}).

\subsection{The $\textrm{G}_{2}$ sector} 
\label{app:G2sectorScalMat}

The decomposition $\textbf{8} \rightarrow \textbf{7} + \textbf{1}$ of the fundamental representation of SL(8) under $\textrm{G}_{2}$ selects an index splitting of the form $\,A \rightarrow I \, \oplus \, 8 \,$ with $\,I=1,\ldots ,7\,$.  Consequently, one also has a splitting of the $\textbf{28}$ (and the $\textbf{28}'$) of the form $[AB] \rightarrow [IJ] \, \oplus \, [I8]$. The set of components of the scalar-dependent matrix  (\ref{M_appendix}) can be written in terms of the $\textrm{G}_{2}$-invariant tensors $\delta_{IJ}$ and
\begin{equation}
\begin{array}{llll}
\psi_{IJK} & = & e_{123} + e_{145} + e_{167} - e_{246} + e_{257} + e_{473} + e_{635} & , \\[2mm]
\tilde{\psi}_{IJKL} & = & e_{4567}  + e_{6723} + e_{2345} - e_{1357} + e_{1346} + e_{1562} + e_{1724}  & , \\[2mm]
\end{array}
\end{equation}
which are related by seven-dimensional Hodge duality. The scalar matrix $\mathcal{M}_{\mathbb{MN}}$ in this sector depends on two scalars $(\chi,\varphi)$. Introducing the combination $\,X= 1+ e^{2\varphi} \chi^2\,$, it  contains the following blocks:
\begin{equation}
\label{M_scalar_G2}
\begin{array}{llll}
{\cal M}^{[IJ] [KL]} &=& 2 \, e^\varphi \, X \,\, \delta^{K[I}  \, \delta^{J]L} + e^{3\varphi} \, \chi^2 \,\,\tilde{\psi}^{IJKL} & , \\[2mm]
{\cal M}^{[IJ] [K8]} &=& e^\varphi \chi^2 \, X \,\, \psi^{IJK} & , \\[2mm]
{\cal M}^{[I8][K8]} &=& e^{-3\varphi}  \, X^3 \,\, \delta^{IK} & , \\[2mm]
{\cal M}^{[IJ]}{}_{[KL]} &=& - 2 \,  e^{3\varphi} \, \chi^3 \,\, \delta^{IJ}_{KL} - e^{\varphi}  \, \chi \, X \,\, \tilde{\psi}^{IJ}{}_{KL} & , \\[2mm]
{\cal M}^{[IJ]}{}_{[K8]} &=& - e^{3\varphi} \, \chi \,\,  \psi^{IJ}{}_{K} & , \\[2mm]
{\cal M}^{[I8]}{}_{[KL]} &=& - e^{-\varphi} \, \chi \, X^2 \,\,  \psi^{I}{}_{KL} & , \\[2mm]
{\cal M}^{[I8]}{}_{[K8]} &=& - e^{3 \, \varphi} \, \chi^3 \,\,  \delta^{I}_{K} & , \\[2mm]
{\cal M}_{[IJ] [KL]} &=& 2 \,  e^{-\varphi} \, X^2 \,\, \delta_{K[I}  \, \delta_{J]L} +e^{\varphi} \, \chi^2 \, X \,\,\tilde{\psi}_{IJKL} & , \\[2mm]
{\cal M}_{[IJ] [K8]} &=& e^{3\,\varphi} \chi^2  \,\, \psi_{IJK} & , \\[2mm]
{\cal M}_{[I8][K8]} &=& e^{3\varphi}  \,\, \delta_{IK} & . 
\end{array}
\end{equation} 
The $\textrm{G}_{2}$-invariant tensors are parity-even with respect to the $\mathbb{Z}_{2}$ transformation (\ref{Z2-symmetry}). Consequently, so are the $\mathcal{M}_{\mathbb{MN}}$ components (\ref{M_scalar_G2}).

\subsection{The $\textrm{SO}(4)$ sector}
\label{app:SO4sectorScalMat}

The branching $\textbf{8} \rightarrow (\textbf{2,2}) + (\textbf{3,1}) + (\textbf{1,1})$ of the fundamental SL(8) representation  under $\textrm{SO}(4)$ determines an index splitting $\,A \rightarrow \lambda \, \oplus \,  a \, \oplus  \, 8 \,$ with $\,\lambda=1,3,5,7\,$ and $\,a=2,4,6\,$.  The splitting of the $\textbf{28}$ (and the $\textbf{28}'$) is then of the form $[AB]  \rightarrow  [\lambda \mu] \, \oplus \, [ab] \, \oplus \, [a \mu] \, \oplus \, [\lambda 8] \, \oplus \, [a 8]$. The $\textrm{SO}(4)$ sector we investigate in this work retains four scalars $\,(\chi,\varphi)\,$ and $\,(\rho,\phi)\,$. In terms of these, the independent blocks of the scalar-dependent matrix (\ref{M_appendix}) can be obtained using the invariant tensors $\,\delta_{ab}\,$, $\,\epsilon_{abc}\,$, $\,\delta_{\lambda \mu}\,$, $\,\epsilon_{\lambda \mu \nu \sigma}\,$ and the $4 \times 4$ matrices\footnote{Here we use invariant tensors $(\gamma_{2},\gamma_{4},\gamma_{6})\equiv(-2\,t_{1}^{(-)},2\,t_{2}^{(-)},2\,t_{3}^{(-)})$ with $t_{1,2,3}^{(-)}$ given in eq.(4.1) of \cite{Gallerati:2014xra}.} $\,{[\gamma_{a}]^{\lambda}}_{\mu}\,$ given by
\begin{equation}
[\gamma_{2}]=
\left(
\begin{array}{c|ccc}
0 & -1 & 0 & 0 \\
\hline
1 & 0 & 0 & 0 \\
0 & 0 & 0 & 1 \\
0 & 0 & -1 & 0
\end{array}
\right)
\hspace{1mm} , \hspace{1mm}
[\gamma_{4}]=
\left(
\begin{array}{c|ccc}
0 & 0 & -1 & 0 \\
\hline
0 & 0 & 0 & -1 \\
1 & 0 & 0 & 0 \\
0 & 1 & 0 & 0
\end{array}
\right)
\hspace{1mm} , \hspace{1mm}
[\gamma_{6}]=
\left(
\begin{array}{c|ccc}
0 & 0 & 0 & -1 \\
\hline
0 & 0 & 1 & 0 \\
0 & -1 & 0 & 0 \\
1 & 0 & 0 & 0
\end{array}
\right) \ .
\end{equation}
The above $\gamma$-matrices satisfy the anti-self-duality relations
\begin{equation}
[\gamma_{a}]_{\lambda \mu}  \,\,=\,\,  -  \tfrac{1}{2} \,\, \epsilon_{\lambda \mu \nu \sigma}  \,\, [\gamma_{a}]_{\nu \sigma} \ ,
\end{equation}
as well as the usual
\begin{equation}
\{ \gamma_{a} \, , \,  \gamma_{b}\} \,\, = \,\, -2  \,\, \mathbb{I}_{4 \times 4} 
\hspace{8mm} \textrm{ and } \hspace{8mm}
[ \gamma_{a} \, , \,  \gamma_{b} ] \,\, = \,\, -2 \, \epsilon_{abc} \, \gamma_{c} \ .
\end{equation}

\noindent As in the previous cases, we define the following combinations
\begin{equation}
X = 1 + e^{2\, \varphi} \, \chi^{2}
\hspace{5mm} , \hspace{5mm}
Y = 1 + e^{2\, \phi} \, \rho^{2} \ ,
\end{equation}
which we use to list the entries of $\mathcal{M}_{\mathbb{MN}}$.
\\[-2mm]

\noindent - The block  $\,\mathcal{M}^{[AB] [CD]}\,$ in (\ref{M_appendix}) contains the components
\begin{equation}
\begin{array}{llll}
\mathcal{M}^{[\lambda 8][\mu 8]} & = & e^{-3 \, \varphi} \, X^{3} \,\, \delta^{\lambda \mu}& , \\[2mm]
\mathcal{M}^{[a8][c8]} & = & e^{-(2\,\varphi+\phi)} \, X^2 \, Y  \,\, \delta^{ac} & , 
\end{array}
\end{equation}
together with
\begin{equation}
\begin{array}{llll}
\mathcal{M}^{[a \lambda][\mu8]} & = & e^{\varphi} \, X \, \chi^2 \,\, [\gamma^{a}]^{\lambda \mu}  & , \\[2mm]
\mathcal{M}^{[\lambda \mu][a8]} & = & e^{\phi} \, X  \, \chi \, \rho \,\,  [\gamma^{a}]^{\lambda \mu} & , \\[2mm]
\mathcal{M}^{[ab][c8]} & = & - e^{2\,\varphi - \phi} \, Y \, \chi^2  \,\, \epsilon^{abc}  & , 
\end{array}
\end{equation}
and
\begin{equation}
\begin{array}{llll}
\mathcal{M}^{[\lambda \mu][\nu \sigma]} & = & -  e^{2\, \varphi + \phi} \, \chi^2 \,\,\epsilon^{\lambda \mu \nu \sigma} \,\, + \,\, 2 \, e^{\phi} \, X \,\, \delta^{\nu [\lambda} \, \delta^{\mu] \sigma}    & , \\[2mm]
\mathcal{M}^{[ab][\lambda \mu]} & = & - e^{2\, \varphi + \phi} \, \rho \, \chi  \,\,\epsilon^{abc} \, [\gamma_{c}]^{\lambda \mu}   & , \\[2mm]
\mathcal{M}^{[a \lambda][b \mu]} & = &   e^{3\, \varphi} \, \chi^2 \,\, \epsilon^{abc} \, [\gamma_{c}]^{\lambda \mu} \, + \,  e^{\varphi} \, X \,\, \delta^{ab} \, \delta^{\lambda \mu}   & , \\[2mm]
\mathcal{M}^{[ab][cd]} & = &  2 \, e^{2\,\varphi-\phi} \, Y \,\,\delta^{c[a}\,\delta^{b]d} & . \\[2mm]
\end{array}
\end{equation}

\noindent - The block $\,{\mathcal{M}^{[AB]}}_{[CD]}\,$ contains the set of components
\begin{equation}
\begin{array}{llll}
{\mathcal{M}^{[\lambda 8]}}_{[\mu 8]} & = & - e^{3 \, \varphi} \, \chi^{3} \,\, \delta^{\lambda}_{ \mu}& , \\[2mm]
{\mathcal{M}^{[a8]]}}_{[c8]} & = & - e^{2\,\varphi+\phi} \, \chi^2 \, \rho  \,\, \delta^{a}_{c} & , 
\end{array}
\end{equation}
together with
\begin{equation}
\begin{array}{llll}
{\mathcal{M}^{[a \lambda]}}_{[\mu8]} & = & -e^{3\,\varphi} \, \chi \,\, {[\gamma^{a}]^{\lambda}}_{\mu}  & , \\[2mm]
{\mathcal{M}^{[\mu8]}}_{[a \lambda]} & = &  e^{-\varphi} \, X^2 \, \chi \,\, {[\gamma_{a}]^{\mu}}_{\lambda}  & , \\[2mm]
{\mathcal{M}^{[\lambda \mu]}}_{[a8]} & = & -  e^{2\,\varphi +\phi} \, \chi \,\,  [\gamma_{a}]^{\lambda \mu} & , \\[2mm]
{\mathcal{M}^{[a8]}}_{[\lambda \mu]} & = & - e^{-\phi} \, X  \, Y \, \chi  \,\,  [\gamma^{a}]_{\lambda \mu} & , \\[2mm]
{\mathcal{M}^{[ab]}}_{[c8]} & = &   e^{2\,\varphi + \phi} \, \rho  \,\, {\epsilon^{ab}}_{c}  & , \\[2mm]
{\mathcal{M}^{[a8]}}_{[bc]} & = &  e^{-2\,\varphi + \phi} \, \rho \, X^2  \,\, {\epsilon^{a}}_{bc}  & , 
\end{array}
\end{equation}
and
\begin{equation}
\begin{array}{llll}
{\mathcal{M}^{[\lambda \mu]}}_{[\nu \sigma]} & = & e^{\phi} \, \rho \, X \,\, {\epsilon^{\lambda \mu}}_{\nu \sigma} \,\, - \,\, 2 \,  e^{2\, \varphi +\phi} \, \rho \, \chi^2 \,\, \delta^{\lambda \mu}_{\nu \sigma}    & , \\[2mm]
{\mathcal{M}^{[ab]}}_{[\lambda \mu]} & = & e^{2\, \varphi - \phi} \, Y \, \chi  \,\,{\epsilon^{ab}}_{c} \, [\gamma^{c}]_{\lambda \mu}   & , \\[2mm]
{\mathcal{M}^{[\lambda \mu]}}_{[ab]} & = &  e^{\phi} \, X \, \chi  \,\,{\epsilon_{ab}}^{c} \, [\gamma_{c}]^{\lambda \mu}   & , \\[2mm]
{\mathcal{M}^{[a \lambda]}}_{[b \mu]} & = &  - e^{\varphi} \, X \, \chi \,\, \epsilon^{a\phantom{b}c}_{\phantom{a}b} \, {[\gamma_{c}]^{\lambda}}_{\mu} \, - \,  e^{3\,\varphi} \, \chi^3 \,\, \delta^{a}_{b} \, \delta^{\lambda}_{\mu}   & , \\[2mm]
{\mathcal{M}^{[ab]}}_{[cd]} & = & - 2 \, e^{2\,\varphi+\phi} \, \rho \, \chi^2 \,\,\delta^{ab}_{cd} & . \\[2mm]
\end{array}
\end{equation}

\noindent - The block $\,\mathcal{M}_{[AB][CD]}\,$ contains the pieces
\begin{equation}
\begin{array}{llll}
\mathcal{M}_{[\lambda 8][\mu 8]} & = & e^{3 \, \varphi} \,\, \delta_{\lambda \mu}& , \\[2mm]
\mathcal{M}_{[a8][c8]} & = & e^{2\,\varphi+\phi} \,\, \delta_{ac} & , 
\end{array}
\end{equation}
together with
\begin{equation}
\begin{array}{llll}
\mathcal{M}_{[a \lambda][\mu8]} & = & e^{3\,\varphi} \, \chi^2 \,\, [\gamma_{a}]_{\lambda \mu}  & , \\[2mm]
\mathcal{M}_{[\lambda \mu][a8]} & = & e^{2\,\varphi + \phi} \, \chi \, \rho \,\,  [\gamma_{a}]_{\lambda \mu} & , \\[2mm]
\mathcal{M}_{[ab][c8]} & = & - e^{2\,\varphi + \phi} \, \chi^2  \,\, \epsilon_{abc}  & , 
\end{array}
\end{equation}
and
\begin{equation}
\begin{array}{llll}
\mathcal{M}_{[\lambda \mu][\nu \sigma]} & = & - e^{2\, \varphi - \phi} \, Y \, \chi^2 \,\,\epsilon_{\lambda \mu \nu \sigma} \,\, + \,\,  2 \, e^{-\phi} \, Y  X \,\, \delta_{\nu [\lambda} \, \delta_{\mu] \sigma}    & , \\[2mm]
\mathcal{M}_{[ab][\lambda \mu]} & = & - e^{\phi} \, \rho \, \chi \, X  \,\,\epsilon_{abc} \, [\gamma^{c}]_{\lambda \mu}   & , \\[2mm]
\mathcal{M}_{[a \lambda][b \mu]} & = &  e^{\varphi} \, \chi^2 \, X \,\, \epsilon_{abc} \, [\gamma^{c}]_{\lambda \mu} \, + \, e^{-\varphi} \, X^2 \,\, \delta_{ab} \, \delta_{\lambda \mu}   & , \\[2mm]
\mathcal{M}_{[ab][cd]} & = &  2 \, e^{-2\,\varphi+\phi} \, X^2 \,\,\delta_{c[a}\,\delta_{b]d} & . \\[2mm]
\end{array}
\end{equation}

\noindent - The last block is obtained as $\,{\mathcal{M}_{[AB]}}^{[CD]}={\mathcal{M}^{[CD]}}_{[AB]}\,$, since $\mathcal{M}_{\mathbb{MN}}$ is symmetric. 
\\[-2mm]

The $\textrm{SO}(4)$-invariant tensors are parity-even with respect to the  $\mathbb{Z}_{2}$ transformation in (\ref{Z2-symmetry}) and so are the $\mathcal{M}_{\mathbb{MN}}$ components listed above.

%%%%%%%%%%%%%%%%%%%%%%%%%%%%%%%%%%%%
%
% Bibliography
%
%%%%%%%%%%%%%%%%%%%%%%%%%%%%%%%%%%%%

\small

\bibliography{references}

\end{document}